\documentclass[aps,prd,twocolumn,showpacs,nofootinbib,floatfix]{revtex4}
\pdfoutput=1

\usepackage{color,amsmath,amssymb,graphicx,latexsym,subfigure}
\usepackage{threeparttable}

\def\stop{{\tilde t}}     % stop
\def\squa{{\tilde q}}  % squark
\def\sbot{{\tilde b}}  % sbottom
\def\gino{{\tilde g}}     % gluino
\def\slep{{\tilde l}}     % slepton
\def\chaino{{\tilde \chi}^+}     % chargino
\def\neu2{{\tilde \chi}^0_2}     % second neutralino
\def\slashed#1{#1\!\!\!\!/}
\def\missET{\slashed E_\mathrm{T}} % missing E_T
\def\missPT{\slashed {\vec{P}}_\mathrm{T}} % missing P_T

\begin{document}
\title{Status of Dark Matter Detection}
\author{Xiao-Jun Bi}
\author{Peng-Fei Yin}
\author{Qiang Yuan}
\affiliation{Key Laboratory of Particle Astrophysics, Institute of High
Energy Physics, Chinese Academy of Sciences, Beijing 100049, China}

\begin{abstract}
The detection of dark matter has made great progresses in recent years.
We give a brief review on the status and progress in dark matter detection,
including the progresses in direct detection, collider detection at LHC and
focus on the indirect detection. The results from PAMELA, ATIC, Fermi-LAT
and relevant studies on these results are introduced. Then
we give the progress on indirect
detection of gamma rays from Fermi-LAT and ground based Cerenkov telescopes.
Finally the detection of neutrinos and constraints on the nature of dark
matter are reviewed briefly.
\end{abstract}

\pacs{12.60.Jv,14.80.Ly}
\maketitle

\section{Introduction}

\subsection{Astronomical evidence}

The standard cosmology is established in the last decade, thanks
to the precise cosmological measurements, such as the cosmic
microwave background (CMB) radiation measured by WMAP
\cite{Hinshaw:2012fq,Calabrese:2013jyk}, the distance-redshift 
relation of the Type Ia supernovae
\cite{Riess:1998cb,Perlmutter:1998np,Frieman:2008sn} and
the large scale structure (LSS) survey from SDSS
\cite{Reid:2009xm,Reid:2012sw} and 6df \cite{Beutler:2012px}. 
The energy budget in the standard cosmology consists of $4\%$ 
baryonic matter, $23\%$ dark matter (DM) and $73\%$ dark energy (DE)
\cite{Kowalski:2008ez,Komatsu:2010fb}. To unveil the mystery
of the dark side of the Universe is a fundamental problem of
modern cosmology and physics. In this review we focus on the
progress in DM detection.

Actually the existence of DM has been established for a much longer 
time. The most direct way that indicates the existence of DM is from 
the rotation curve of spiral galaxies \cite{Rubin:1980zd}. The rotation 
curve shows the rotation velocity of an object around the galaxy center 
as a function of radius $r$, which scales like $\sqrt{M(r)/r}$
with $M(r)$ the mass within the orbit $r$. The rotation curve should 
decrease as $1/\sqrt{r}$ if $r$ is beyond most of the visible part of 
the galaxy. However, most measured rotation curves keep flat at large
distances.
%The large is much larger than
%expectation estimated according to the visible mass.
The large rotation velocity implies a dark halo around the galaxy
to provide larger centripetal force that exerts on the object.

At the scale of galaxy clusters, evidence of DM is also ample. The
first evidence of DM was from the observation of the Coma cluster 
by F. Zwicky in 1930s \cite{Zwicky:1933gu}.
%The discovery of DM can be traced back to 1930s, when
%F. Zwicky studied the dynamics of galaxies in the Coma cluster.
He found unexpected large velocity dispersion of the member 
galaxies, which implied the existence of ``missing mass'' to hold 
the galaxies \cite{Zwicky:1933gu}. The observation of
X-ray emission of hot gas in the clusters can give precise measurement
of the gravitational potential felt by the gas to keep the hot gas
in hydrostatic equilibrium. % balance its thermal pressure.
Other measurement of weak lensing effect on the background galaxies 
by the clusters gives direct indication of DM component in clusters.
%The confirmation of the
%exotic gravity arised in 1970s according to the measurements of
%rotation curves of spiral galaxies \cite{Rubin:1980zd}.
%From then on, more and more astronomical observations show the
%commonly existence of the ``dark gravity'' in the Universe, and DM
%is also widely accepted as a new concept in the
%community\footnote{Note also there are some people believe that
%the ``dark gravity'' could be explained through the modification
%of the fundamental law of dynamics or gravity (e.g.,
%\cite{1983ApJ...270..365M,2004PhRvD..70h3509B})}.
Especially the bullet cluster gives strong support to the DM component 
in cluster. The Bullet cluster consists of two colliding
galaxy clusters. The X-ray image, which reflects the gas component
of the colliding system, shows obvious lag compared with the
gravitational lensing image, which traces the mass distribution
\cite{Clowe:2006eq}. It is easy to understand that the gas
is decelerated due to the viscosity, while the DM component
can pass through each other without collision. The Bullet cluster
was regarded as the most direct evidence of DM.

The existence of DM in the cosmological scale is inferred by a
global fit to the CMB, supernovae and LSS data. The WMAP data give 
the most accurate determination of the DM component in the universe 
with $\Omega_{\rm CDM}h^2=0.112\pm 0.006$ \cite{Komatsu:2010fb}, 
with $h$ the Hubble constant in unit of 100 km s$^{-1}$ Mpc$^{-1}$.
%measurements of the distance-redshift relation
%of the Type Ia supernovae, the anisotropy power spectrum of CMB
%and
%the evolution of cluster mass function \cite{2009ApJ...692.1060V}
%has reconstructed the so called concordance cosmology model
%give independent
%constraint on the matter density $\Omega_M$. All kinds of the
%cosmological measurements cross together to one point
%with $(\Omega_M,\Omega_{\Lambda})\sim(0.3,0.7)$.
% which is known as the concordance cosmology.
%According to the observation and theory of big bang
%nucleosynthesis, the baryon density is constrained to be
%$\Omega_bh^2=0.0214\pm0.002$ \cite{2003ApJS..149....1K}, which
%indicates that most of the matter density is made up of
%non-baryonic DM.

\subsection{Detection methods}

All the current evidence of DM comes from the gravitational effect by DM.
From the point of view that all the matter in the universe comes from
a big-bang a sole DM component with only gravitational interaction is hard 
to properly account for the observed DM. A popular DM candidate is the
weakly interacting massive particle (WIMP). In such scenario the WIMPs
can reach thermal equilibrium in the early universe and decouple from
the thermal equilibrium when the temperature decreases. The relic density
of WIMPs can be calculated by solving the Boltzmann equation.
A good approximate solution of the Boltzmann equation gives
\begin{equation}
\Omega_{\rm DM}h^2 \sim \frac{3\times 10^{-27} {\rm cm^3 s^{-1}}}{\langle \sigma v\rangle}\ \ ,
\label{relic}
\end{equation}
where $\Omega_{\rm DM}=\rho_{\rm DM}/\rho_0$ is the DM density over the
critical density, $h$ is the Hubble constant, $\langle \sigma v\rangle$ 
is the thermal averaged DM annihilation cross section times velocity. 
We often refer $\langle \sigma v\rangle$ as the DM annihilation cross 
section. It represents the interaction strength of the DM particles and 
the standard model (SM) particles. $\langle \sigma v\rangle = 
3\times 10^{-26} {\rm cm^3 s^{-1}}$ gives the correct relic density and 
is often taken as the benchmark value for the DM annihilation cross section.
It is found a WIMP with mass and interaction strength at the weak scale
can easily give correct relic density. If such a scenario is confirmed, 
it will become the third evidence supporting the hot big bang cosmology 
after CMB and the big bang nucleosynthesis. Probing such a decoupling 
process enables us to study the universe as early as its temperature 
was $\sim$GeV. It has become a fundamental problem to detect the DM 
particles and determine its nature in cosmology and particle physics.
WIMPs, interacting weakly with the SM particles, make it possible to 
detect the DM in experiments. A great deal of WIMP candidates have been 
proposed, such as the lightest neutralino in the supersymmetric (SUSY) 
model and the lightest Kaluza-Klein particle in the Universal extra 
dimension model (for a review see \cite{Bertone:2004pz}).

\begin{figure}[!htb]
\begin{center}
\includegraphics[width=\columnwidth]{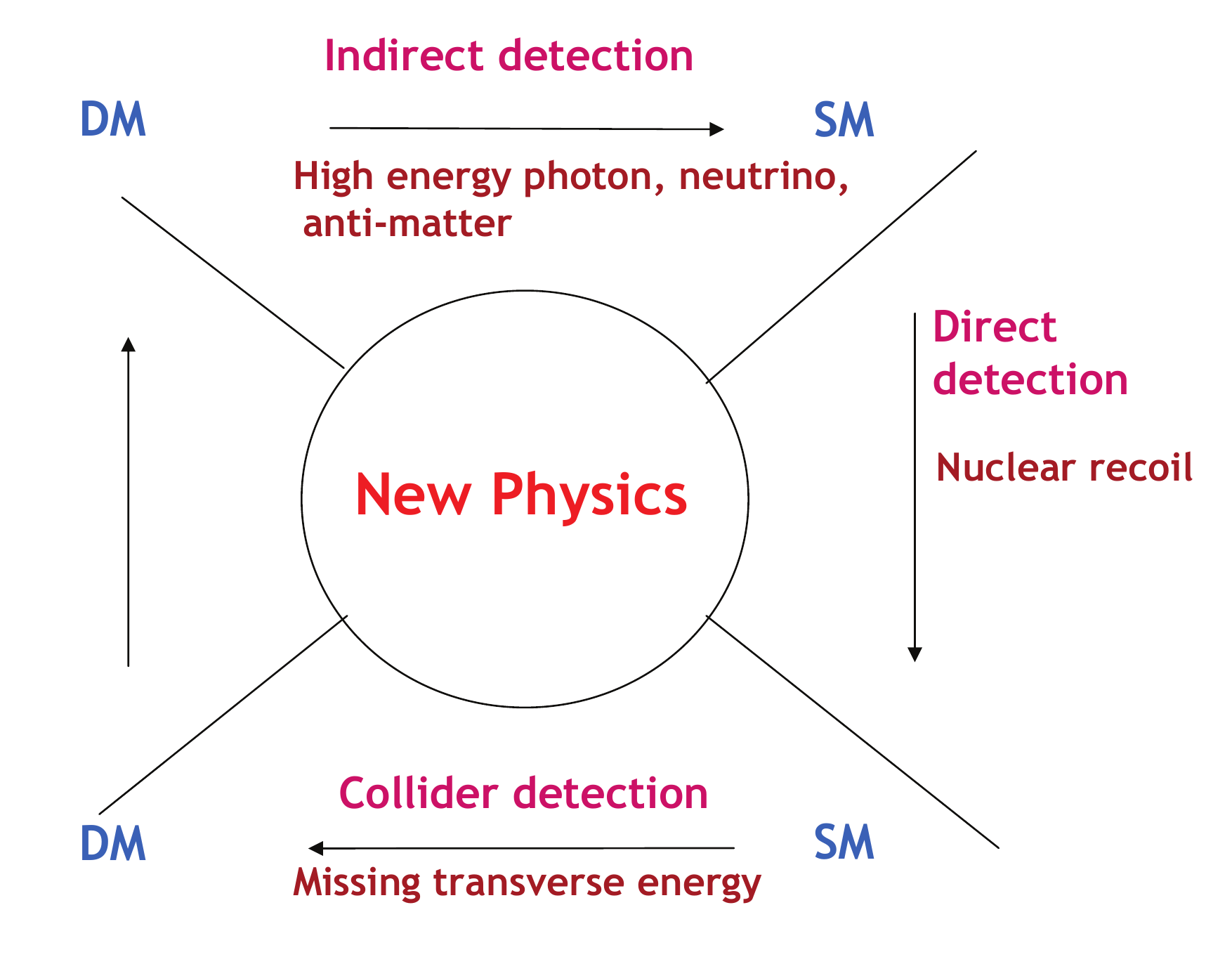}
\caption{Schematic plot to show the relation among the {\it direct
detection}, {\it indirect detection} and {\it collider detection}
of DM. The arrows indicate the direction of reaction.
\label{fig:method}}
\end{center}
\end{figure}

%Based on the
%astronomical observations, we can get the basic properties of DM
%particle, that is, neutral, stable, cold or warm, non-baryonic and
%weakly interacting. The most popular candidate is the so-called
%weakly interacting massive particle (WIMP). There is no such a
%particle candidate in the standard model of particle physics.

Fig. \ref{fig:method} shows the scheme to probe the DM particles.
To determine the nature of DM particles we have to study the
interaction between the DM and the SM particles. In general there 
are three different directions to study the interaction. One direction 
is to search for the scattering signal between DM particle and the 
detector nucleon. It is called the {\it direct detection} of DM.
The {\it indirect detection} is to detect the annihilation or
decay products of DM particles. Finally the {\it collider detection} 
is to search for the DM production process in high energy particle 
collisions. The three ways of DM detection are not independent, but 
complementary to each other.

In this review we will focus on the latest progresses of the
{\it indirect detection} of DM. The status of direct detection and 
collider detection is briefly summarized.

\section{Status of direct detection}

\subsection{Recoil event rate}

Direct detection searches for the nuclear recoil signals which are induced by the scattering of DM
particles against the target nuclei in the underground detectors \cite{Goodman:1984dc} (for reviews, see Ref. \cite{Jungman:1995df,Lewin:1995rx,Zacek:2007mi,Armengaud:2010zg,Strigari:2012gn}). For the DM with mass of $\sim O(10^2)$ GeV and
local velocity of $\sim 10^{-3} c$, the typical energy scale of the recoil signal is $O(10)$keV.
The expected differential event rate per nucleus is
\begin{equation}
\frac{d R}{d E_R}= \frac{\rho_\chi}{m_\chi} \int^{v_{max}}_{v_{min}} d^3 v \; f(\vec{v}) v \frac{d\sigma(\vec{v},E_R)}{dE_R},
\label{recespe}
\end{equation}
where $E_R$ is the nuclear recoil energy, $\rho_\chi$ is the local DM mass density, $f(\vec{v})$ is the velocity distribution of DM in the lab frame, $d\sigma/d E_R$ is the cross section of the scattering between the DM and target nucleus. Different experiment material and techniques are sensitive to search for different interactions between the DM and nucleus.

In the non-relativistic limit, the interaction between DM and the nuclei
can be divided into two classes: the spin-independent (SI) and
spin-dependent (SD). The SI interaction couples to the mass of the
detector nuclei while
the SD couples to the spin of the nuclei.
Coherent SI interaction between DM and the nuclei leads to an enhancement
of the scattering rate $\sigma \propto A^2$.
%Therefore the direct detection
%experiments usually use heavy nuclei

\subsection{Experimental results}

The key issue for direct detection is to control the background
 (for detailed discussions on backgrounds at the direct detections, see Ref. \cite{Armengaud:2010zg}). To shield the huge background from cosmic rays the
detectors are usually located in deep underground laboratory. Since the
gamma photons and electrons from the radioactive isotopes in the
surrounding rock, air and the detector apparatus will induce
electronic recoils in the detector, good shielding and high
purity of material for detector are required. Note that the
characteristics of electronic recoil events are different from
nuclear recoil signals. Many techniques have been developed to
distinguish them. Moreover the electron recoil events are often
produced in the surface of the detector. Therefore the outer part of detector
volume can be used to veto background.

Recently, more than 20 direct detection experiments worldwide are running or under construction. Three kinds of signals namely scintillation, ionization, and photon can be used to record the recoil events. Some experiments detect one kind
of signal, while some experiments can measure a combination of two
kinds of signals to discriminate the electronic and nuclear 
recoils. According to the detection technique and the detector material, 
these experiments fall into different classes, such as scintillator experiments (e.g. DAMA \cite{Bernabei:2008yi}, KIMS \cite{Lee.:2007qn}), cryogenic crystal experiments (e.g. CDMS \cite{Ahmed:2009zw}, CoGeNT \cite{Aalseth:2010vx}, CRESST \cite{Angloher:2011uu}, EDELWEISS \cite{Armengaud:2011cy}, TEXONO \cite{Lin:2007ka}, CDEX \cite{Kang:2013hda}), noble liquid experiments (e.g. XENON \cite{Aprile:2012nq}, ZEPLIN \cite{Akimov:2011tj}, PandaX \cite{Li:2012zs}), superheated liquid experiments (e.g. COUPP \cite{Behnke:2012ys}, PICASSO \cite{Archambault:2012pm}, SIMPLE \cite{Felizardo:2011uw}), etc.

Up to now, most of the direct detection experiments 
do not observe any DM induced
nuclear recoil events and set stringent constraints on the DM-nucleon
scattering cross section. However, the following experiments claim they have
observed some signal-like events. 

\begin{itemize}

\item DAMA is a NaI scintillator detector located in Gran Sasso.
DAMA collaboration reported an annual modulation effect with a high
confidence level $\sim 8.2 \sigma$ in the 2-6 keVee energy interval
\cite{Bernabei:2000qi,Bernabei:2008yi}. Such result is consistent
with expectation of DM events. Due to the Earth rotation around the Sun, the
variation of DM flux will lead to a $\sim 7\%$ anual modulation of
the scattering event rate. If DAMA result is induced by DM-nucleus
elastic SI  scattering, a kind of DM particle with mass of $\sim 10$
GeV and scatting cross section of $\sim O(10^{-40})$cm$^{-2}$ is
needed \cite{Savage:2008er,Chang:2008xa}. Such light DM can be
provided in many theoretical framework, such as SUSY
\cite{Draper:2010ew,Belli:2011kw,Kang:2010mh}, asymmetric DM
\cite{Kaplan:2009ag}, mirror DM \cite{Foot:2010hu}, etc.

\item CoGeNT is a cryogenic germanium detector with a low energy threshold. CoGeNT collaboration reported an excess of events with energy smaller than 3 keVee in 2010 \cite{Aalseth:2010vx} and an annual modulation signal with $2.8\sigma$ confidence level in 2011 \cite{Aalseth:2011wp}. Such results can be explained by a DM with mass of $\sim 10$ GeV and scatting cross section of $10^{-41}-10^{-40}$cm$^{-2}$ \cite{Fitzpatrick:2010em,Chang:2010yk} (see also \cite{Schwetz:2011xm,Fox:2011px,Farina:2011pw}) which is roughly consistent with that needed for DAMA \cite{Hooper:2010uy,Arina:2012dr}.

\item  CRESST-II is a Calcium Tungstate (CaWO$_4$) detector which measures both scintillation and phonon signals. In 2011, CRESST-II collaboration reported an excess of events in the 10-40 keV energy interval \cite{Angloher:2011uu} which is consistent with a $10-30$ GeV DM interpretation \cite{Kopp:2011yr}.

\end{itemize}

The results of DAMA, CoGeNT and CRESST-II seem inconsistent
 with the results by other experiments with higher sensitivity, such as XENON
\cite{Angle:2011th,Aprile:2012nq} and CDMS \cite{Ahmed:2010wy}, which
give null results. Therefore,
the nature of these anomalous events are still unclear. 
There are discussions about the possibility that the DAMA events are 
induced by atmospheric muon or radioactive isotopes in the literature 
\cite{Blum:2011jf,Pradler:2012qt}. If DAMA, CoGeNT and CRESST-II results 
are produced by ordinary SI DM, it means there
exist large experimental uncertainties in the other experiments
\cite{Hooper:2010uy}, which seems unacceptable. The astrophysical
uncertainties arising from DM velocity distribution are not sufficient
to relax such tensions either \cite{Schwetz:2010gv,Fox:2010bz}. Many exotic
DM models have been proposed, such as isospin violation DM
\cite{Chang:2010yk,Feng:2011vu,Gao:2011ka}, momentum dependent
scattering DM \cite{Chang:2009yt,An:2010kc}, inelastic DM
\cite{TuckerSmith:2001hy,Chang:2008gd} or a combination of them
\cite{Cline:2012ei}. These models are becoming difficult to explain
all the experiment results simultaneously with improvement of CDMS
and XENON sensitivity \cite{Aprile:2012nq} \footnote{For the
constraints from indirect detections on the isospin violation DM,
see Ref. \cite{Chen:2011vda,Kumar:2011dr,Jin:2012jn}.}. For instance,
inelastic DM model is strongly constrained by the new XENON100 results
\cite{Aprile:2011ts,Aprile:2012nq}.

\begin{figure}[!htbp]
\begin{center}
\includegraphics[width=0.9\columnwidth]{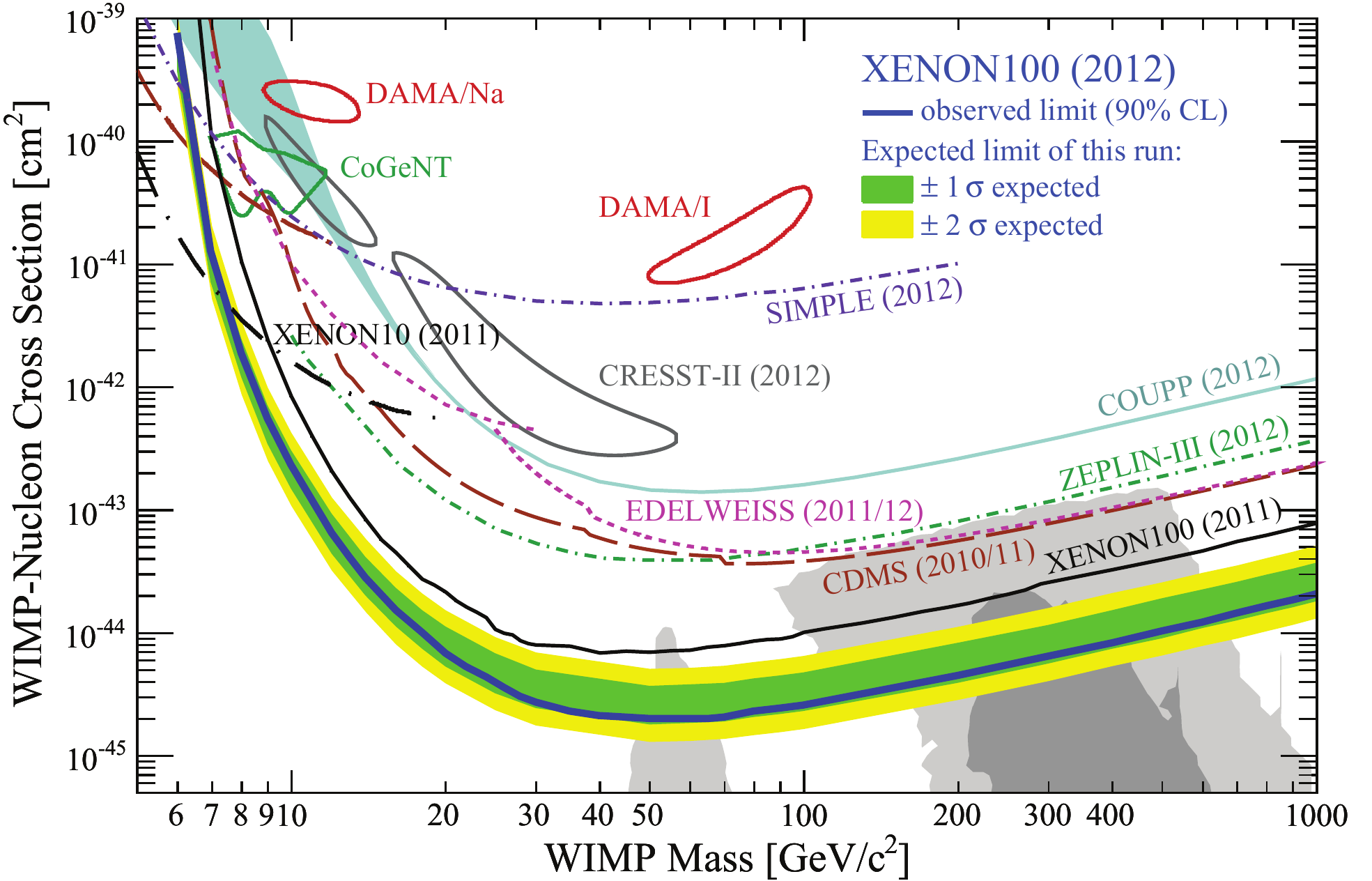}
\caption{Constraints on SI DM-nucleon scattering cross section by XENON100. For comparison, other
results from DAMA \cite{Bernabei:2008yi,Savage:2008er}, CoGeNT \cite{Aalseth:2010vx}, CRESST-II \cite{Angloher:2011uu}, CDMS \cite{Ahmed:2009zw,Ahmed:2010wy}, EDELWEISS \cite{Armengaud:2011cy}, SIMPLE \cite{Felizardo:2011uw}, COUPP \cite{Behnke:2012ys}, ZEPLIN-III \cite{Akimov:2011tj} and XENON10\cite{Angle:2011th}, are also shown, together with the preferred regions in CMSSM \cite{Strege:2011pk,Fowlie:2012im,Buchmueller:2011ab}. Figure from \cite{Aprile:2012nq}.
\label{XENONSI}}
\end{center}
\end{figure}

XENON experiment is a dual phase noble liquid detector located in Gran Sasso with simultaneous measurements of the primary scintillation (S1) and secondary ionization signals (S2). The ratio of the two kinds of signals can be 
used to discriminate the electronic and nuclear recoil events. 
The most stringent constraints on SI DM-nucleon cross section are 
set by XENON100 with the exposure of $34\times 224.6$ kg days 
\cite{Aprile:2012nq}(for the constraints on SD cross section, 
see Sec. \ref{solarneu}). For DM with mass of 55 GeV, the upper-limit
reaches $2\times10^{-45}$ cm$^2$. It has excluded some preferred
parameter regions of the CMSSM. Especially, the pure higgsino DM
is strongly disfavor due to
large expected SI cross section. 
%The co-annihilation or resonance
%enhancement is necessary to acquire correct DM relic density.

Recently many experiment collaborations
are preparing for upgrading their detectors to larger volume.
The sensitivities for DM-nucleon scattering cross section
will be improved by a magnitude of two orders in the next
five years \cite{Aprile:2012zx}.

\section{Status of collider detection}

Since the DM mass is usually assumed to be $\lesssim O(10^2)$ GeV, the DM
particles are expected to be generated
at the high energy colliders, such as Tevatron \cite{Aaltonen:2012jb},
LHC \cite{ATLAS:2012ky,deJong:2012zt} and ILC \cite{Djouadi:2007ik}.
Once produced, these particles escape the detector without energy
deposit due to their extremely weak interactions. Such signal, named ``missing
transverse energy'' (MET), can be reconstructed by the associated jets,
photons, or leptons based on momentum conservation in the plane
perpendicular to the beam pipe \footnote{In fact, the variable
reconstructed directly is the ``missing transverse momentum'' $\missPT$.
MET $\missET$ is the magnitude of $\missPT$. Since the exact energies
of initial partons are unknown, only MET is meaningful at the hadron
colliders. It is possible to reconstruct the total missing energy
at the $e^+e^-$ colliders.}. It is possible to determine the DM
mass at the colliders (see Ref.
\cite{Cheng:2007xv,Burns:2008va,Barr:2010zj,Han:2012nm} and
references therein). Moreover, searches for DM particles and MET
signals are essential to determine the mass spectra and typical
parameters of the new physics models (for some reviews, see Ref.
\cite{Baer:2008uu,Baer:2009uc}). It will reveal the origin of
electroweak symmetry breaking and the nature of new fundamental symmetries.

At the hadron colliders, the main SM backgrounds arise from the processes
which produce neutrinos, such as $Z(\to \nu \bar{\nu})$+jets,
$W(\to l \nu)$+jets, $t \bar{t}$ and single top production. Another
background is the ``fake MET''.
It arises from the QCD multi-jets due to the fact that the measurement
for jet energy has large errors. Since the MET induced by DM
is related to DM mass, large MET cut condition , e.g. $\missET > 100$ GeV,
is often adopted to reduce background.

\subsection{Direct production}

Direct production means the DM particles are produced in pair by
the collisions of high-energy SM particles. Since the DM particle
pair can not be observed, an additional energetic jet or photon
from initial state radiation is needed to trigger the event and
to reconstruct MET. Such signal is called ``mono-jet'' or ``mono-photon''
\footnote{In principle, charged lepton from W \cite{Bai:2012xg} or
Z boson \cite{Carpenter:2012rg} coming from initial state radiation
can also be used to trigger the event, it is called ``mono-lepton''
or ``mono-Z''.}. Searching for Mono-jet is more important at the
hadron colliders due to large event rate
\cite{Feng:2005gj,Bai:2010hh,Goodman:2010ku,Fox:2011pm},
while the Mono-photon signal is essential at the $e^+e^-$
colliders \cite{Birkedal:2004xn,Fox:2011fx,Dreiner:2012xm}.

To constrain the nature of DM by searching the collider monojet events
is usually finished in a model-independent way.
The effective field theory is used to
describe the interaction between the DM and SM particles
\cite{Beltran:2008xg,Cao:2009uw,Bai:2010hh,Goodman:2010ku,Zheng:2010js,Yu:2011by}.
For each interaction form, the constraints on the DM
mass and interaction coupling can be derived by the results
from collider detection, direct detection, indirect detection
and DM relic density. 
%It is helpful to explore the nature of DM
%by comparing with these constraints.

\begin{figure*}[t]
\begin{tabular}{ccc}
\includegraphics[width=0.30\textwidth]{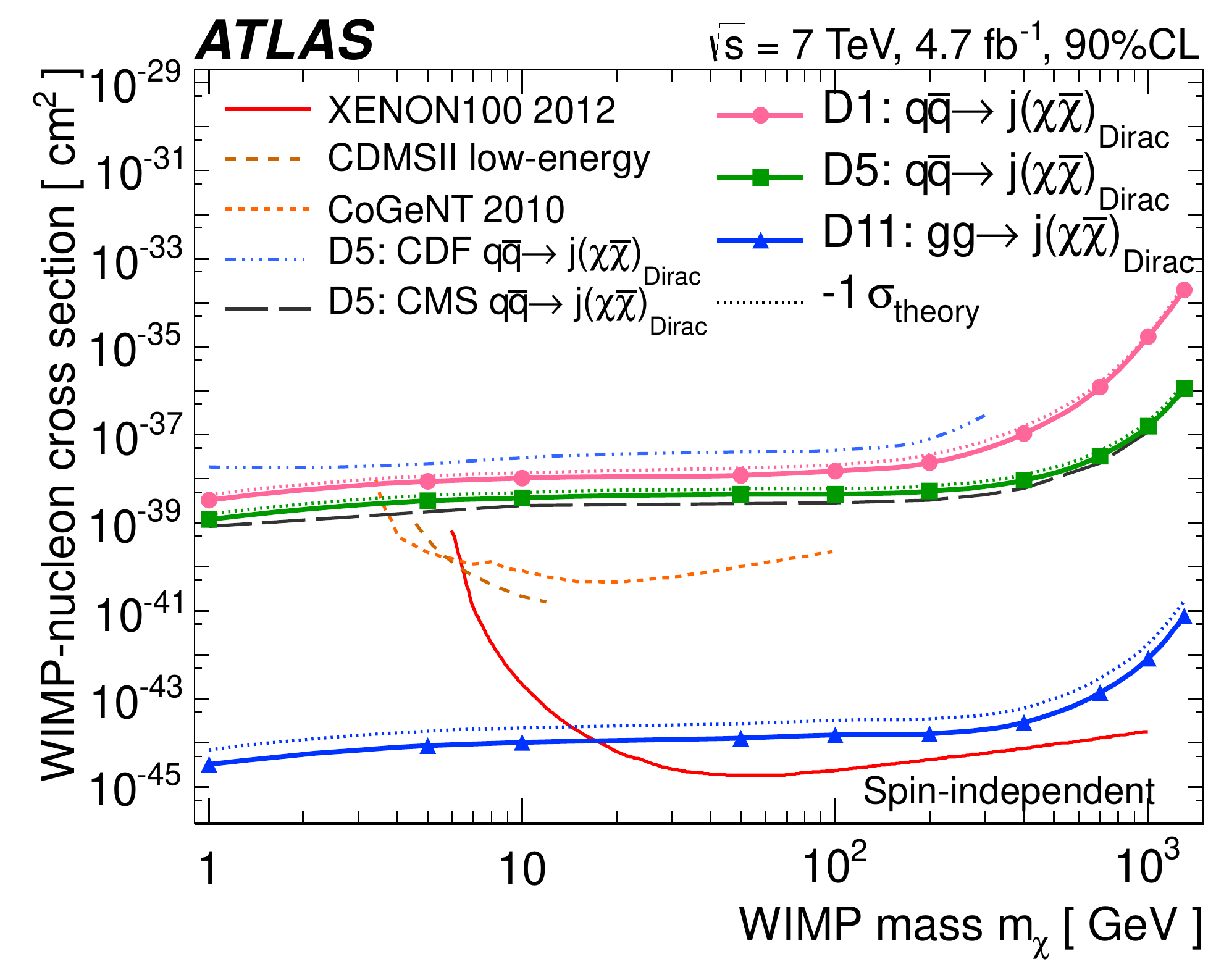} &
\includegraphics[width=0.30\textwidth]{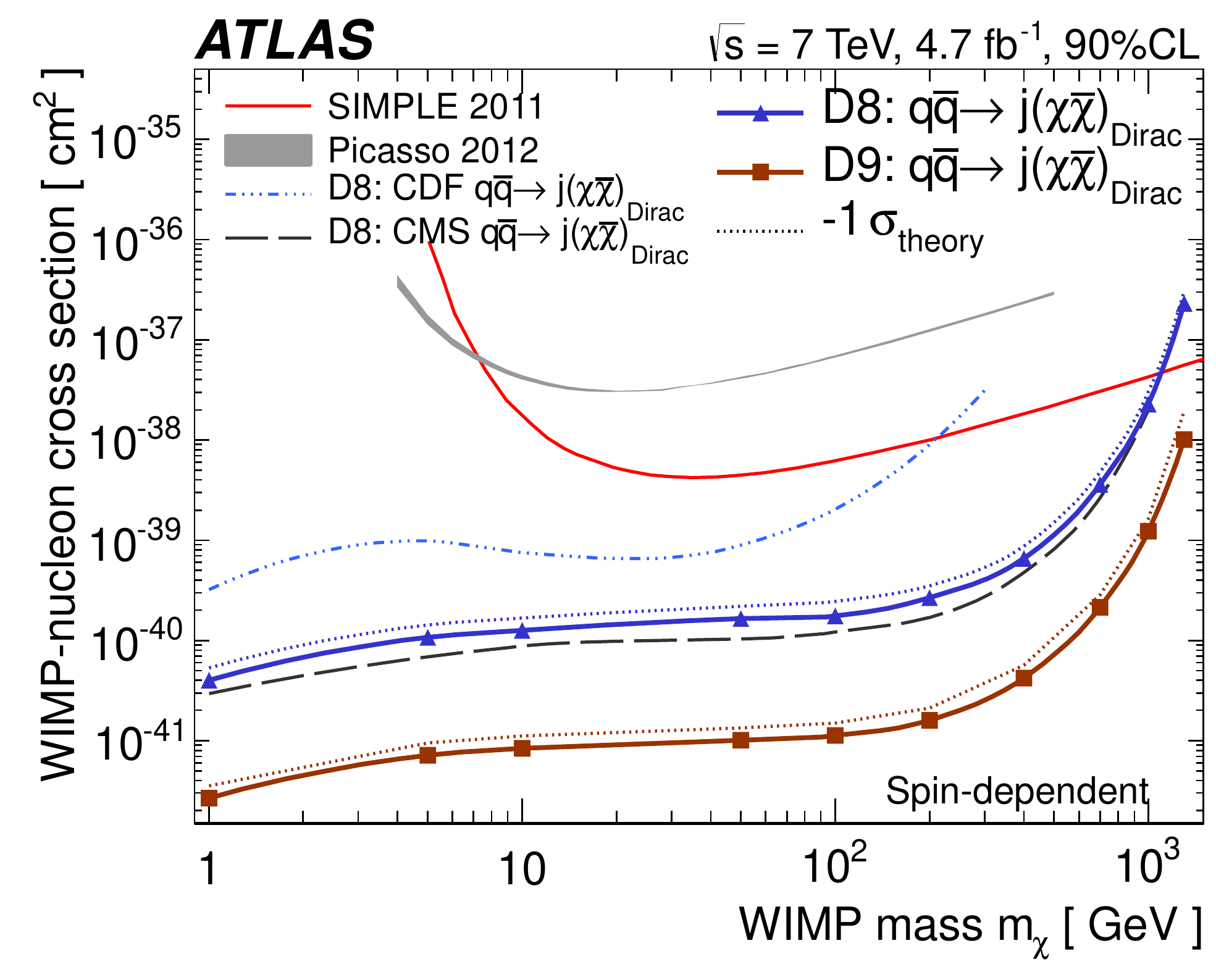} &
\includegraphics[width=0.30\textwidth]{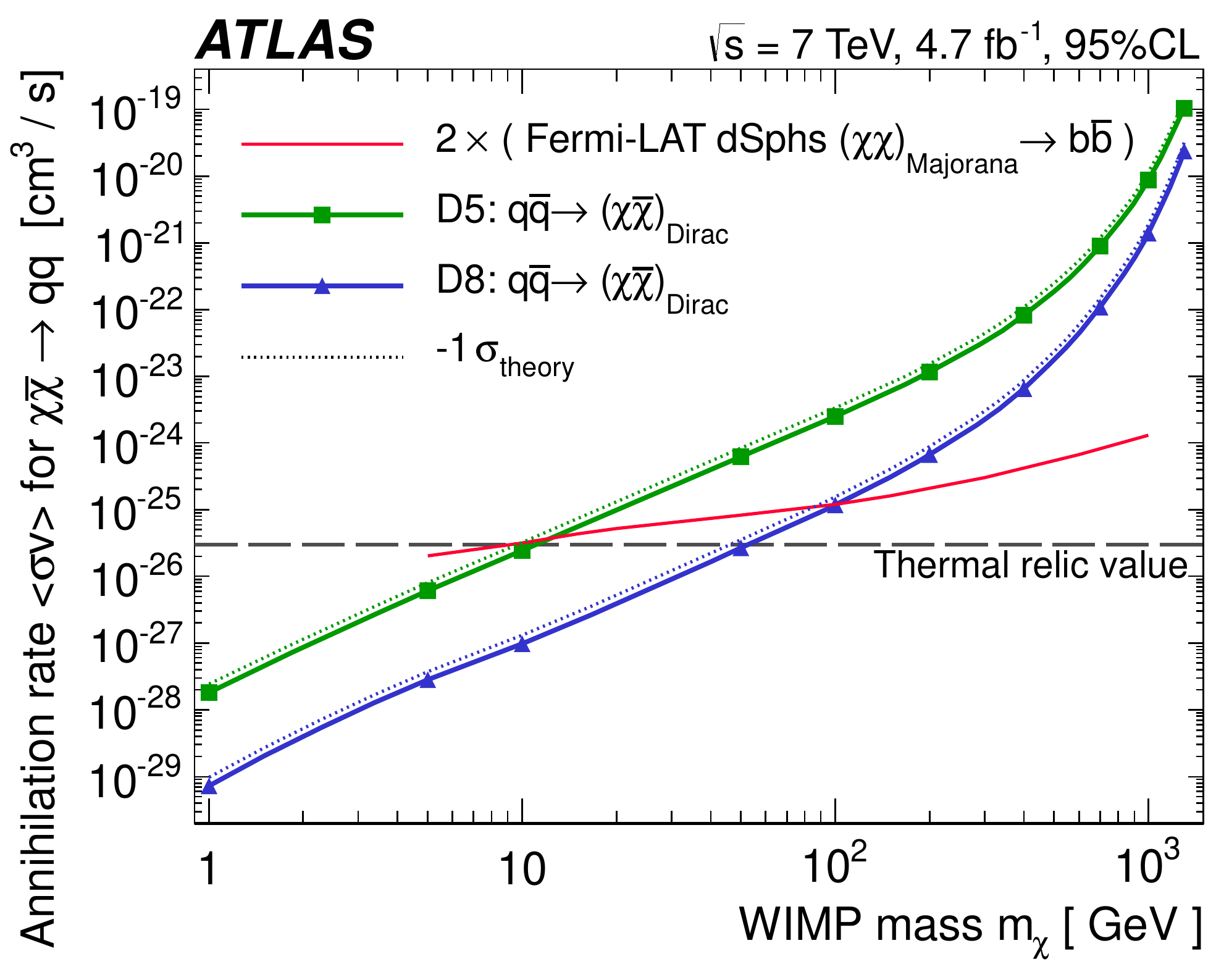}
\end{tabular}
\caption{Inferred ATLAS 90$\%$ limits on SI (left) and SD (middle) DM-nucleon scattering cross section, and ATLAS 95$\%$ limits on DM annihilation cross section (right). D1, D5, D8, D9 and D11 denote the effective interaction operators $\chi \bar{\chi} q \bar{q}$ (scalar), $\chi \gamma^\mu \bar{\chi} q \gamma_\mu \bar{q}$ (vector), $\chi \gamma^\mu \gamma^5 \bar{\chi} q \gamma_\mu \gamma_5 \bar{q}$ (axial-vector), $\chi \sigma^{\mu \nu} \bar{\chi} q \sigma_{\mu \nu} \bar{q}$ (tensor) and $\chi \bar{\chi} (G^a_{\mu \nu})^2$ (scalar) respectively \cite{Goodman:2010ku}. For comparison, the limits from XENON100 \cite{Aprile:2012nq}, CDMS \cite{Ahmed:2010wy}, CoGeNT \cite{Aalseth:2010vx}, SIMPLE \cite{Felizardo:2011uw}, PICASSO \cite{Archambault:2012pm}, CDF \cite{Aaltonen:2012jb}, CMS \cite{Chatrchyan:2012me} and Fermi-LAT \cite{Ackermann:2011wa} are also shown. Figures from \cite{ATLAS:2012ky}.
\label{ATLDMlim}}
\end{figure*}

Fig. \ref{ATLDMlim} shows the ATLAS limits on SI, SD DM-nucleon scattering and DM annihilation cross sections \cite{ATLAS:2012ky}. Four typical DM-quark interaction operators and one DM-gluon interaction operator \cite{Goodman:2010ku} are considered in the ATLAS mono-jet analysis. From Fig. \ref{ATLDMlim} we can see the DM searches at the LHC have some advantages compared with the other detections.
\begin{itemize}
\item Since the light DM has large production cross section due to the phase space and parton distribution function, LHC has good sensitivity for DM with a mass below 10 GeV. While the sensitivities of direct detection decrease quickly in this region due to the detector energy threshold.
\item Since the scalar and axial-vector operators have similar behaviors in the relativistic limit at the colliders, the LHC constraints on SI and SD DM-nucleon scattering can be comparable. For DM with a mass of $O(10)$ GeV, the LHC limits on SI scattering $\sim O(10^{-39})$ cm$^2$ are weaker than XENON limits. However, the LHC limits on SD scattering $\sim O(10^{-40})$ cm$^2$ are much better than the results from direct detections.
\item If DM interaction with gluon is significant, the production cross section of $gg \to \chi \bar{\chi}$ will be very large at the LHC due to parton distribution function. Therefore LHC has strong capability to detect such DM. The LHC constraints on SI DM-nucleon scattering induced by DM-gluon interaction can be comparable with XENON limits.
\end{itemize}

It should be noticed that the constraints derived in effective
theory is only valid under some assumptions.
%First, it often assume there are no other heavier new particles which can be efficiently produced at the colliders. This is not the case for many BSM models, such as SUSY which is discussed in the next subsection. The coupling of effective operator reflect the strength of interaction and the mass scale of mediator. It is not an arbitrary number if some theoretical requirements for UV completed model, such as perturbativity \cite{Goodman:2010ku} and unitarity \cite{Shoemaker:2011vi}, are taken into account.
The effective theory requires the
particle that mediates the interaction
between DM and SM particles be very heavy, and can be integrated
out at the collider energy scale.
If the s-channel mediator is light, the event rate will fall
with jet transverse energy as $1/p_t^2$, while the event rate
is flat with jet $p_t$ in the effective theory.
In this case the limits are not applicable \cite{Goodman:2010ku,Fox:2011fx,Goodman:2011jq}.

\subsection{Indirect production}

Indirect production means the DM particles are produced by the
cascade decays of some heavier new particles. The most important
example is the supersymmetry (SUSY) model
\footnote{The latest ATLAS and CMS SUSY
search results can be found in
https://twiki.cern.ch/twiki/bin/view/AtlasPublic/SupersymmetryPublicResults and
https://twiki.cern.ch/twiki/bin/view/CMSPublic/PhysicsResultsSUS.}.
The LHC can produce pairs of squarks and gluinos via strong interaction
processes $pp \to \squa \squa$, $\gino \gino$ and $\squa \gino$, with
large cross section. If R-parity is conserved, such particles will
decay into the lighter sparticles until the decay chain ends up in
the lightest supersymmetry particle (LSP).
The final states depend on the mass spectra and decay
mode of the sparticles. The typical
SUSY signal is usually classified
according to jet, b-jet and lepton multiplicity
\cite{Baer:1991xs,Feldman:2008hs}, such as jets+MET, 1 b-jet+jets+MET,
1 lepton+jets+MET, two opposite sign leptons+jets+MET (OS),
two same sign leptons+jets+MET (SS), etc.

\begin{figure}[!htbp]
\begin{center}
\includegraphics[width=0.45\textwidth]{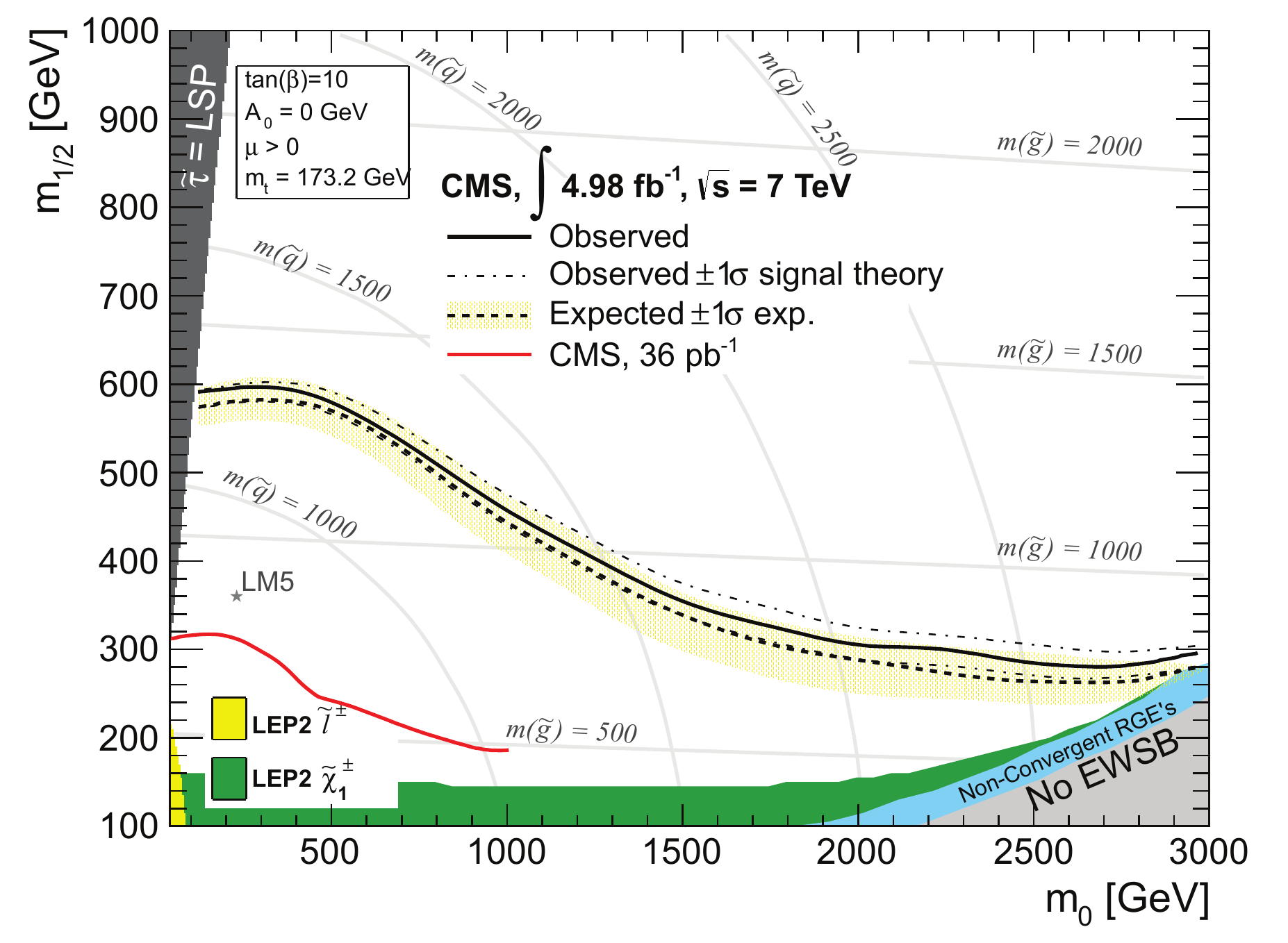}
\caption{The CMS limits in the CMSSM $m_0-m_{1/2}$ plane. The other CMSSM parameters are $\tan \beta =10$, $\mu >0$ and $A_0=0$.
The limits from CMS search with $36$ pb$^{-1}$ \cite{Collaboration:2011ida} of data and LEP \cite{LEPlim} are also shown. Figure from \cite{:2012mfa}.
\label{CMSsug}}
\end{center}
\end{figure}

If the dominated decay channels of squarks and gluino are
$\squa \to q \chi$ and $\gino \to q q' \chi$, the typical
signals are 2-4 jets+MET. If the squarks and gluino are much
heavier than neutralino, the leading jets are energetic and
the reconstructed MET is large. By choosing suitable cut
conditions, the SM backgrounds can be suppressed efficiently
\cite{:2012mfa,:2012rz}. A number of kinetic variables, such
as effective mass \cite{:2012rz}, razo \cite{Rogan:2010kb,razorcms},
$\alpha_T$ \cite{Randall:2008rw,Chatrchyan:2012wa} and $M_{T2}$
\cite{Lester:1999tx,Barr:2003rg,Chatrchyan:2012jx}, are also
helpful to discriminate signals and background. Since no excess above SM
predictions has been confirmed, the ATLAS and CMS collaborations
have set stringent constraints on the masses of gluino and the
first two generations of squarks. Fig. \ref{CMSsug} shows the
upper-limits in the CMSSM framework in the $m_0$-$m_{1/2}$ plane
for $\tan \beta=10$ and $A_0=0$ based on CMS results with
$\sqrt{s}=7$ TeV and 4.7 fb$^{-1}$ of data \cite{:2012mfa}.
For all gluino(squarks) masses, squarks(gluino) with masses
below $\sim$1200 GeV(800 GeV) have been excluded.

The mass constraints on stop and sbottom are much weaker. It is well-known that the lighter stop can be the lightest colored
sparticle due to the larger top Yukawa coupling and large mass
splitting terms in many SUSY models. The light stop is also
well-motivated by the ``naturalness'' argument
\cite{Kitano:2006gv,Papucci:2011wy}, and is consistent with
recent LHC Higgs results \cite{Kang:2012sy,Cao:2012fz}. Light
stop/sbottom can be produced by the decays of gluinos
$\gino \to t \stop / b \sbot $ which are not very heavy
as suggest in the ``natural SUSY'' framework \cite{Papucci:2011wy}.
It is called gluino-mediated stop/sbottom production. The final
states may contain many b-jets due to the processes of
$\stop \to t \chi \to bW^+\chi$, $\stop \to b \chaino$ and
$\sbot \to b \chi$ which are helpful to reduce backgrounds.
For the gluinos with masses below 1 TeV, the DM masses are
excluded up to $\sim$500 GeV (300 GeV) for
$\gino \to t\bar{t}\chi$($\gino \to b\bar{b}\chi$)
channel by the recent LHC results
\cite{deJong:2012zt,Chatrchyan:2012wa,Chatrchyan:2012jx,Aad:2012pq,ATLgtt}.

If gluino is very heavy, the main production process of
stop/sbottom is directly pair production $pp \to t\stop/ b\sbot$.
For light stop, the constraints depend on the mass splitting
between stop and neutralino, and the assumptions of stop decay
modes \cite{Papucci:2011wy,Bi:2011ha,Cao:2012rz}. The constraints
for decay mode $\stop \to t \chi$ are very stringent in the
$m_\chi-m_\stop$ plane \cite{:2012ar,:2012si}. If stop and
neutralino are almost degenerate in mass as suggested by the
``stop co-annihilation'' scenario, the dominated stop decay
mode may be flavor changing neutral current $\stop \to c\chi$.
In this case, since the charm jet from stop decay may be too
soft, an additional energetic jet is required to reconstruct
the MET \cite{Carena:2008mj} (see also \cite{Ajaib:2011hs,Yu:2012kj}
and references therein). The constraints for such signal channel are weak.

In many SUSY frameworks, sparticles in the electro-weak sector namely neutralinos, charginos and sleptons, are much lighter than colored sparticles. These sparticles can be pair produced via Drell-Yan processes at the colliders \cite{:2012ewa,:2012ku}. For the neutralino-chargino pair production $pp \to \chaino \neu2$, the final states may include three charged leptons produced by $\chaino \to \bar{l}\nu\chi$ and $\neu2 \to l\bar{l}\chi$. The SM backgrounds can be suppressed sufficiently due to the leptons with opposite sign. For the assumptions of $m_\chaino=m_{\neu2}$ and $m_\slep=0.5(m_\chi+m_{\chaino})$, the DM masses can be excluded up to 250 GeV for chargino masses below 450 GeV by the CMS results \cite{:2012ewa}.

If heavier sparticle is long-lived, it is so-called meta-stable massive particle and can be directly observed by detectors. For instance, if the mass splitting between stop (gluino) and neutralino is extremely small, stop (gluino) will form a bound state namely R-hadron in the hadronization process before its decay (see Ref. \cite{Arvanitaki:2005nq,Johansen:2010ac} and references therein). R-hadron will lose energy in the detector due to strong interaction. Another important example is long-lived stau in the GMSB scenario where the LSP and DM candidate is gravitino \cite{Giudice:1998bp}. If stau is the NLSP \footnote{If neutralino is the NLSP, the typical signals are photons+MET where photon is produced by the decay of neutralino $\chi \to \gamma \tilde{G}$ \cite{Dimopoulos:1996vz,Chatrchyan:2012vxa}.}, it may have a long lifetime due to very weak interaction with gravitino, and can be observed in the inner tracker and outer muon detector. The LHC results can exclude stop masses up to 700 GeV, and stau mass up to 300 GeV if they are meta-stable particles \cite{:2012vd,Chatrchyan:2012sp}.

%In the AMSB model \cite{Randall:1998uk,Giudice:1998xp}, the dominated components of the lightest chargino and neutralino are charged and neutral wino respectively. Since they are quai-degenerate in mass, the lifetime of chargino is is expected to be typically a fraction of a nanosecond. Some charginos will leave tracks in the inner tracker and decay into soft pions which may be missed. Such "disappearing" track can be used to set constraints on the mass and lifetime of the chargino \cite{ATLAS:2012jp}.

\section{Status of indirect detection -- charged particles}

\subsection{Introduction}

The indirect detection searches for the DM annihilation or decay products,
including $\gamma$-rays, neutrinos and charged anti-particles such as
positrons and antiprotons. Since the interstellar space is filled with
magnetic fields the charged particles are deflected when propagating in
the interstellar space. The source information will get lost and 
therefore we can only resolve the possible signals of DM in the energy
spectra of charged particles. On the contrary the $\gamma$-rays
and neutrinos can trace back to the sources. We can search for such 
signals at the directions where the DM density is expected to be high.

There are two kinds of $\gamma$-ray spectra can be generated from DM
annihilation or decay. The DM particles can annihilate/decay into two 
photons directly. The photon energy equals approximately to the mass 
(or half mass for decaying DM) of the DM particle since the DM moves
non-relativistically today. Such spectrum is monoenergetic, and is 
usually thought to be the smoking gun of DM signal, since there is no 
astrophysical process that can produce such kind of spectrum. 
But such a process is in general highly suppressed and hard to be 
detected because the DM particles are neutral and can not couple with 
photons directly. The process can occur through a loop Feynman diagram 
that DM first annihilate into two virtual charged particles and then 
the the virtual charged particles annihilate into two real photons.
The DM particles can also annihilate into quarks, gauge bosons and 
so on, which induce continuous $\gamma$-ray spectrum by cascade decays.
The continuous $\gamma$-rays have much larger flux and easier to be 
detected. However, it does not have distinctive features from the
astrophysical background $\gamma$-rays.

Right now there are many cosmic ray (CR) experiments dedicated to
look for the DM annihilation signals. To avoid the shield of the
atmosphere the instruments are better to be placed in space.
The satellite based detector PAMELA and the international space
station (ISS) detector AMS02 two most important experiments for 
charged particle detection. Both detectors are magnetic spectrometers 
that have magnetic field to identify the charge of the incident 
particles. PAMELA was launched in 2006 and many important results 
have been published. We will give detailed discussion on the PAMELA 
results in the following. AMS02 was launched in 2011 and the data
taking and analysis are on-going. The first physical result of AMS02
will be released soon in this year. It is expected AMS02 will 
improve the PAMELA results essentially as it has much larger aperture 
than PAMELA. 

The most sensitive $\gamma$-ray detector in space is the satellite 
based Fermi, which can detect $\gamma$-rays from $20$ MeV to 
$\sim 300$ GeV. The detailed summary of the Fermi results on DM 
detection will be presented in the next section. The ground based 
image atmospheric Cerenkov telescopes (IACT) detect very high energy 
(VHE) $\gamma$-rays with energy greater than $\sim 100$ GeV. With 
the rapid development of the IACT technology the VHE $\gamma$-ray
astronomy develops quickly in recent years. We will also describe
the status of DM searches with IACTs breifly in the next section.

\subsection{Experimental status}

\begin{figure}[!htb]
\begin{center}
\includegraphics[width=0.9\columnwidth]{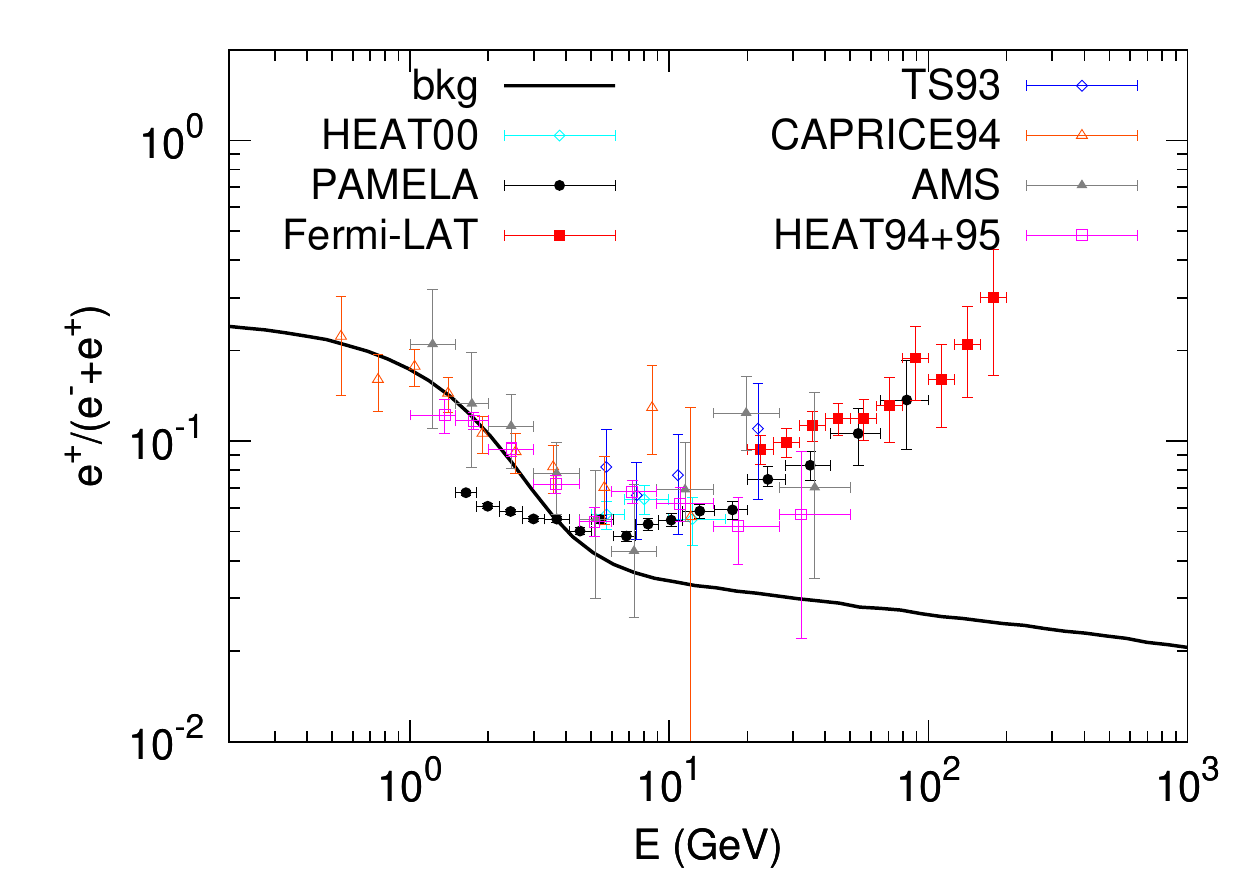}
\includegraphics[width=0.9\columnwidth]{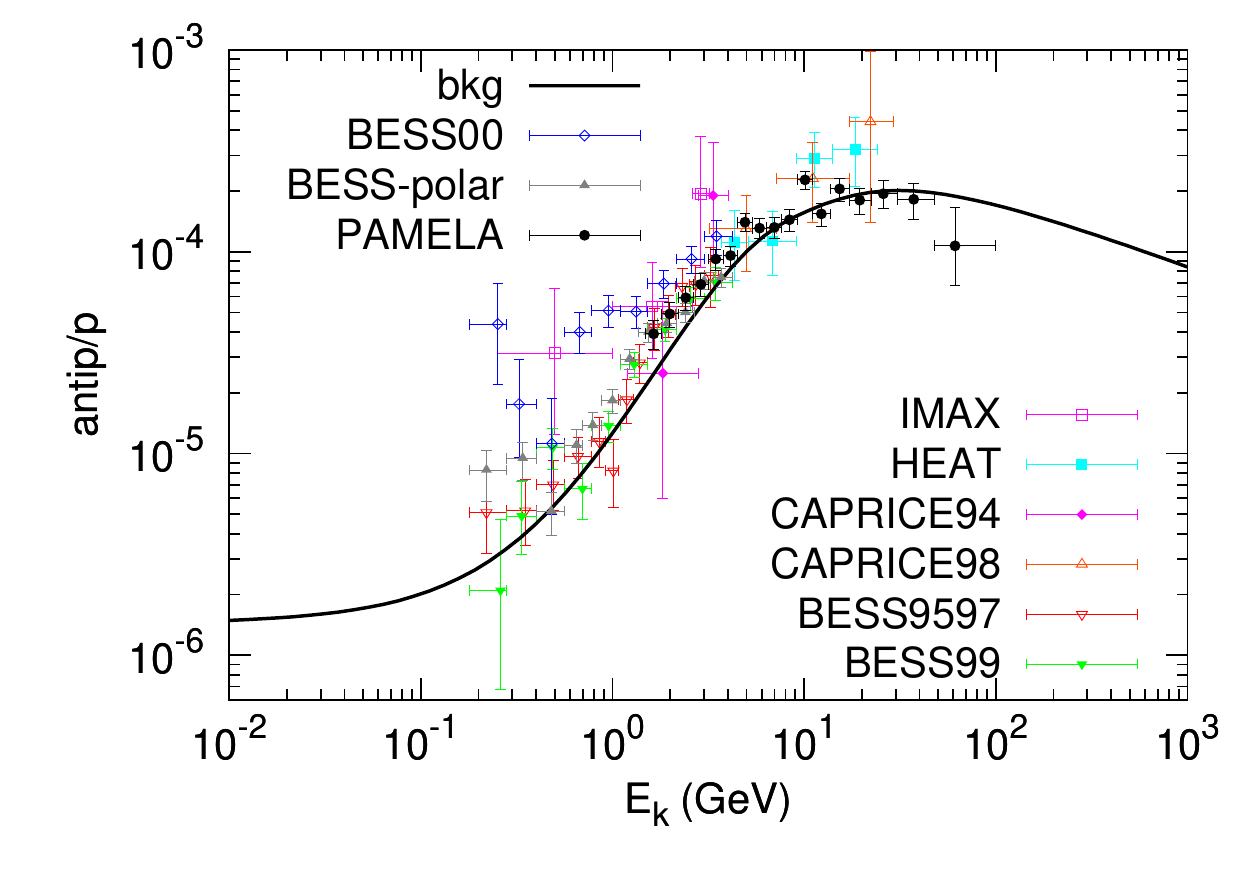}
\caption{Observational data of the positron fraction
$\phi({e^+})/(\phi({e^-})+\phi({e^+}))$ (upper) and
antiproton-proton ratio $\phi({\bar{p}})/\phi(p)$ (lower)
respectively. Lines in these figures are the expectations based on
conventional CR propagation models. The data in the figure are from:
positron fraction --- TS93 \cite{Golden:1995sq}, 
CAPRICE94 \cite{Boezio:2000zz}, AMS \cite{Aguilar:2007yf}, HEAT
\cite{Barwick:1997ig,Coutu:2001jy}, PAMELA \cite{Adriani:2008zr} 
and Fermi-LAT \cite{FermiLAT:2011ab}; antiproton-proton ratio --- 
IMAX \cite{Mitchell:1996bi}, HEAT \cite{Beach:2001ub},
CAPRICE94 \cite{Boezio:1997ec}, CAPRICE98 \cite{Boezio:2001ac}, 
BESS95+97 \cite{Orito:1999re}, BESS99 \cite{Asaoka:2001fv}, 
BESS00 \cite{Asaoka:2001fv}, BESS-polar \cite{Abe:2008sh} and 
PAMELA \cite{Adriani:2008zq}. 
\label{data:pamela}}
\end{center}
\end{figure}

The most interesting result on DM indirect detection in the recent
years comes from PAMELA, which observed obvious positron excess in
the cosmic rays (CRs) \cite{Adriani:2008zr}. The upper panel of
Fig. \ref{data:pamela} shows the positron fraction
$\phi(e^+)/(\phi(e^-)+\phi(e^+))$ measured by PAMELA and several 
previous experiments. The black curve shows the expectation of the 
positron fraction from the conventional CR propagation model. 
In the conventional model there are no primary positrons, and the
positrons are secondary products through the interactions of CRs
and the interstellar medium (ISM) during the propagation. 
Here the expected positron fraction is calculated with the GALPROP 
package \cite{Strong:1998pw}. The propagation parameters used are 
listed in Table \ref{table:prop}. The conventional propagation model 
can reproduce most of the observed CR data on the Earth. 
For the calculated curves, a solar modulation under the force field 
approximation \cite{Gleeson:1968zza} is applied with modulation 
potential $500$ MV. Since below $\sim 10$ GeV the flux is affected 
by the solar modulation effect, we will pay more attention on the 
high energy end. It can seen that above $\sim 10$ GeV the positron 
ratio shows an obvious excess beyond the expected background from 
CR physics.

In fact, the early HEAT \cite{Barwick:1997ig} and AMS \cite{Aguilar:2007yf} 
data have shown the hints of positron excess. The PAMELA data confirmed
this excess with high significance \cite{Adriani:2008zr}. The rise of the 
positron fraction for energies higher than $\sim10$ GeV means that the
positron spectrum is even harder than the electron spectrum and
cannot be understood easily in the CR background model.

\begin{table}[htb]
\centering \caption{Conventional GALPROP model parameters}
\begin{tabular}{cccccccc}
\hline \hline
  $z_h$ & $D_0$ & $\delta$ & $\rho_0$ & $v_A$ & $\gamma_{e^-}^{\ \ a}$ & $\gamma_{\rm nuc}^{\ \ \ b}$ & \vspace{0mm}\\
  (kpc) & ($10^{28}$ cm$^2$ s$^{-1}$) &  & (GV) & (km s$^{-1}$) & $\gamma_1/\gamma_2$ & $\gamma_1/\gamma_2$ \\
\hline
  $4$ & $5.5$ & $0.34$ & $4$ & $32$ & $1.60/2.62$ & $1.91/2.39$ \\
  \hline
  \hline
\label{table:prop}
\end{tabular}\\
$^a$Below/above break rigidity $4$ GV.\\
$^b$Below/above break rigidity $11$ GV.
\end{table}

At the same time PAMELA also reported the antiproton-to-proton ratio 
in CRs \cite{Adriani:2008zq}. The lower panel of Fig. \ref{data:pamela} 
shows the antiproton-to-proton ratio observed by PAMELA as well as
earlier experiments. It shows that the data are well consistent with 
the expectation of the conventional CR propagation model. The old BESS 
data are also consistent with background.

\begin{figure}[!htb]
\begin{center}
\includegraphics[width=0.9\columnwidth]{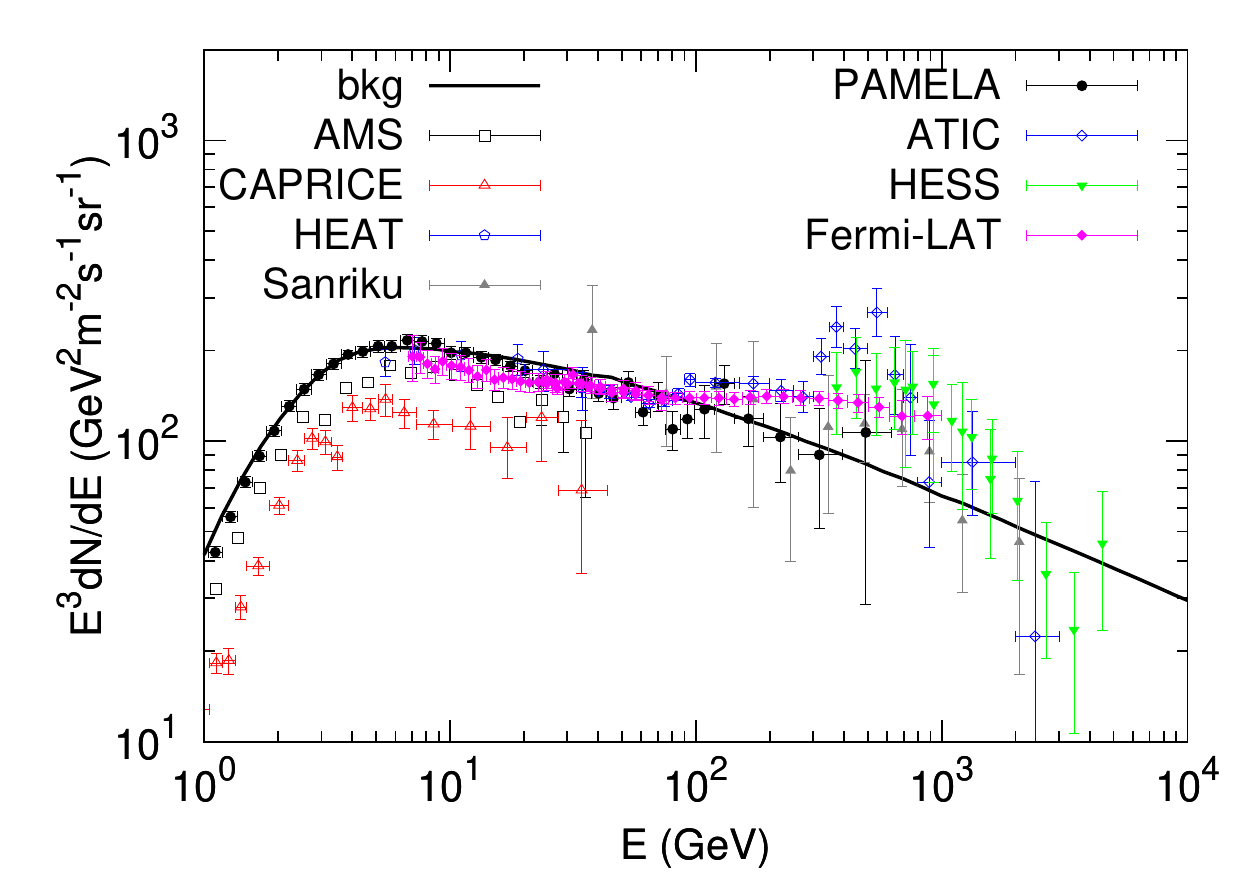}
\caption{The electron spectrum. Line in the figure is the expectation 
based on the conventional CR propagation model. Note that the low energy
part of the data are pure electrons, while the high energy data are the 
sum of electrons and positrons. The data in the figure are from 
AMS \cite{Alcaraz:2000bf}, CAPRICE \cite{Boezio:2000zz}, 
HEAT \cite{Barwick:1997kh}, Sanriku \cite{Kobayashi:1999he}, 
PAMELA \cite{Adriani:2011xv}, ATIC \cite{Chang:2008aa}, 
HESS \cite{Aharonian:2008aa,Aharonian:2009ah} and 
Fermi-LAT \cite{Abdo:2009zk}. \label{data:electron}}
\end{center}
\end{figure}

Soon after PAMELA released the positron fraction result the balloon-based
experiment ATIC published the total electron plus positron spectrum
up to several TeV \cite{Chang:2008aa}. The ATIC data show a peak
between $300$ and $800$ GeV, together with a sharp falling above
$800$ GeV \cite{Chang:2008aa}. Later Fermi-LAT also measured the total 
electron spectrum with much larger statistics. Fermi-LAT data give a 
smooth spectrum with power-law $\sim E^{-3}$ in $20-1000$ GeV, without 
peak structure as ATIC measured \cite{Abdo:2009zk}. The ground-based
Cerenkov telescope HESS also measured the electron spectrum, which
is similar to that from Fermi for $E\lesssim 1$ TeV 
\cite{Aharonian:2009ah}. Above $\sim 1$ TeV HESS found a
softening of the electron spectrum which is consistent with ATIC
data \cite{Aharonian:2008aa}. The observational results are compiled
in Fig. \ref{data:electron}. The line in Fig. \ref{data:electron}
is the expected background contribution of the electrons, which
is determined according to the low energy data.

It should be remarked here that the ATIC and Fermi data of electron 
spectrum are not consistent with each other. The ATIC data show sharp 
feature at $\sim 600$ GeV, while the Fermi data show a smooth spectrum 
consistent with a power law. This is the present largest uncertainty 
to discuss the origin of the positron excess. 

One may expect that the hard spectrum as shown for example by the 
Fermi data could be accounted for by assuming a harder injection
spectrum of the background electrons. However, the positron excess
in this case will become more significant \cite{Grasso:2009ma}. 
Therefore we can conclude that in general it is difficult to 
reproduce the observed data of both the positron fraction and the 
electron spectrum under the traditional CR background frame. The 
data indicate it is most probably that there exists new source(s) 
of primary electrons and positrons near the solar system.

Those results have stimulated a huge enthusiasm to study the
possible origins of the positron and electron excesses. In general
all the works can be divided into two classes: the astrophysical
origin, such as the nearby pulsar(s) which emit positron/electrons; 
the exotic origin including DM annihilation or decay. In the 
following we will give a brief description of the relevant studies.

\subsection{Explanations}

In the conventional propagation model of Galactic CRs, the source 
population is often assumed to be one single type and its distribution 
is usually adopted to be continuous and smooth. The positrons are 
produced through CR nuclei interacting with the ISM when propagating 
in the Milky Way. As shown above such a scenario fails to explain the 
observed positron fraction and electron spectra. To account for the 
observational data, modifications of the conventional scenario of 
production and/or propagation of CR electrons and positrons are 
necessary, through either changing the background model or invoking 
new sources of $e^{\pm}$. The sources of $e^{\pm}$ generally include: 
1) secondary production of hadronic cosmic rays interacting with
ISM, 2) pair production of photon-photon or photon-magnetic field 
interactions, 3) pair production of photon-nuclei interactions, 
and 4) DM annihilation or decay \cite{Fan:2010yq}.
The two categories of models, astrophysical and DM scenarios, are
described in detail respectively.

\subsubsection{Astrophysics origins}

The first point needs to be clarified is that whether such observational
results indeed are ``excesses''. This depends on the understanding
of the background contributions to both the positrons and electrons.
The minimal opinion is that there might be no ``excess'' at all.
It was found that there were very large uncertainties of the
theoretical expectation of CR positron flux from the primary fluxes
of protons and Helium, propagation and hadronic interaction
\cite{Delahaye:2008ua}. Therefore it should be more careful to claim
an ``excess'' and judge the amplitude of the ``excess'' given such
large uncertainties. However, it will be difficult to explain the
rising behavior of the positron fraction with only the uncertainties,
given the fact that the total $e^++e^-$ spectrum is as hard as
$\sim E^{-3}$ \cite{Serpico:2008te,Serpico:2011wg}. Through the
likelihood analysis with scanning over wide ranges of possible
uncertain parameters, significant tension between the $e^{\pm}$
related data and CR nuclei data was found, which implied the ``excess''
of the $e^{\pm}$ \cite{Auchettl:2011wi}.

A less minimal opinion is that the continuous distribution of the
CR sources might break down, especially for high energy $e^{\pm}$
which have limited propagation range \cite{Shaviv:2009bu}. The
inhomogeneity of supernova remnants (SNRs) leads to distinct features
of the primary electron spectrum and may give a rising behavior of
the positron fraction. However, the fit to the total electron spectra
is poor \cite{Shaviv:2009bu}. Furthermore the result of the positron
fraction keeps rising up to $200$ GeV, and the positron spectrum is
harder than $E^{-3}$ as revealed by Fermi-LAT \cite{FermiLAT:2011ab}
also disfavor such a scenario with modification of the primary electron
spectrum only. Finally it was pointed out that the assumptions of the
source distribution in \cite{Shaviv:2009bu} were too extreme
\cite{Stockton:2011ks}.

An alternative scenario without resorting to exotic sources of $e^{\pm}$
is proposed in \cite{Stawarz:2009ig}, where Klein-Nishina suppression of
the electron cooling was employed to produce a relatively flat electron
spectrum as measured by Fermi-LAT. The PAMELA positron fraction,
however, can not be explained with the average parameters of the
ISM. To overcome this issue, extremely large values of the starlight
intensity and gas density were needed \cite{Stawarz:2009ig}.

\begin{figure*}[!htb]
\centering
\includegraphics[width=0.9\columnwidth]{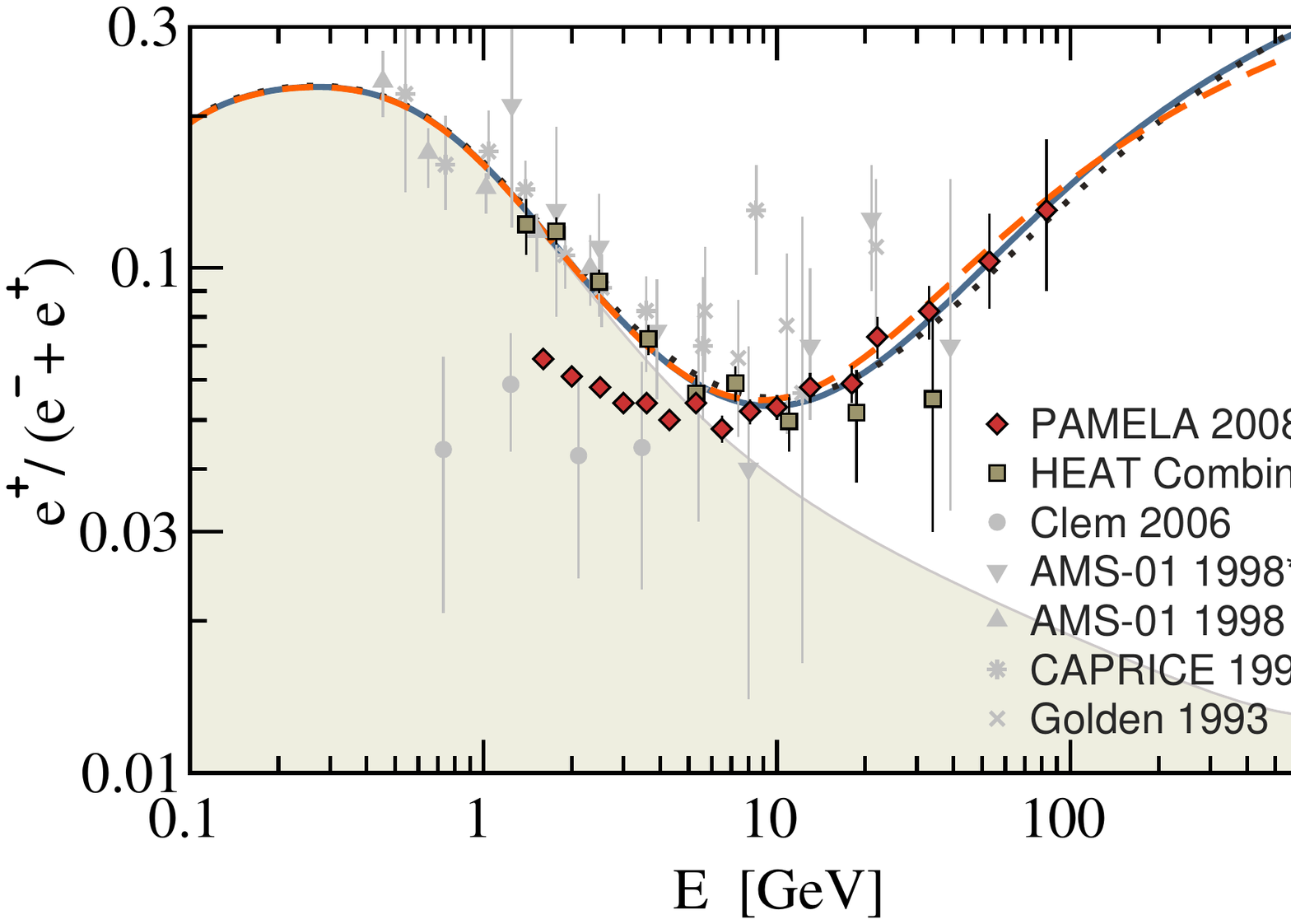} \hspace{3mm}
\includegraphics[width=0.9\columnwidth]{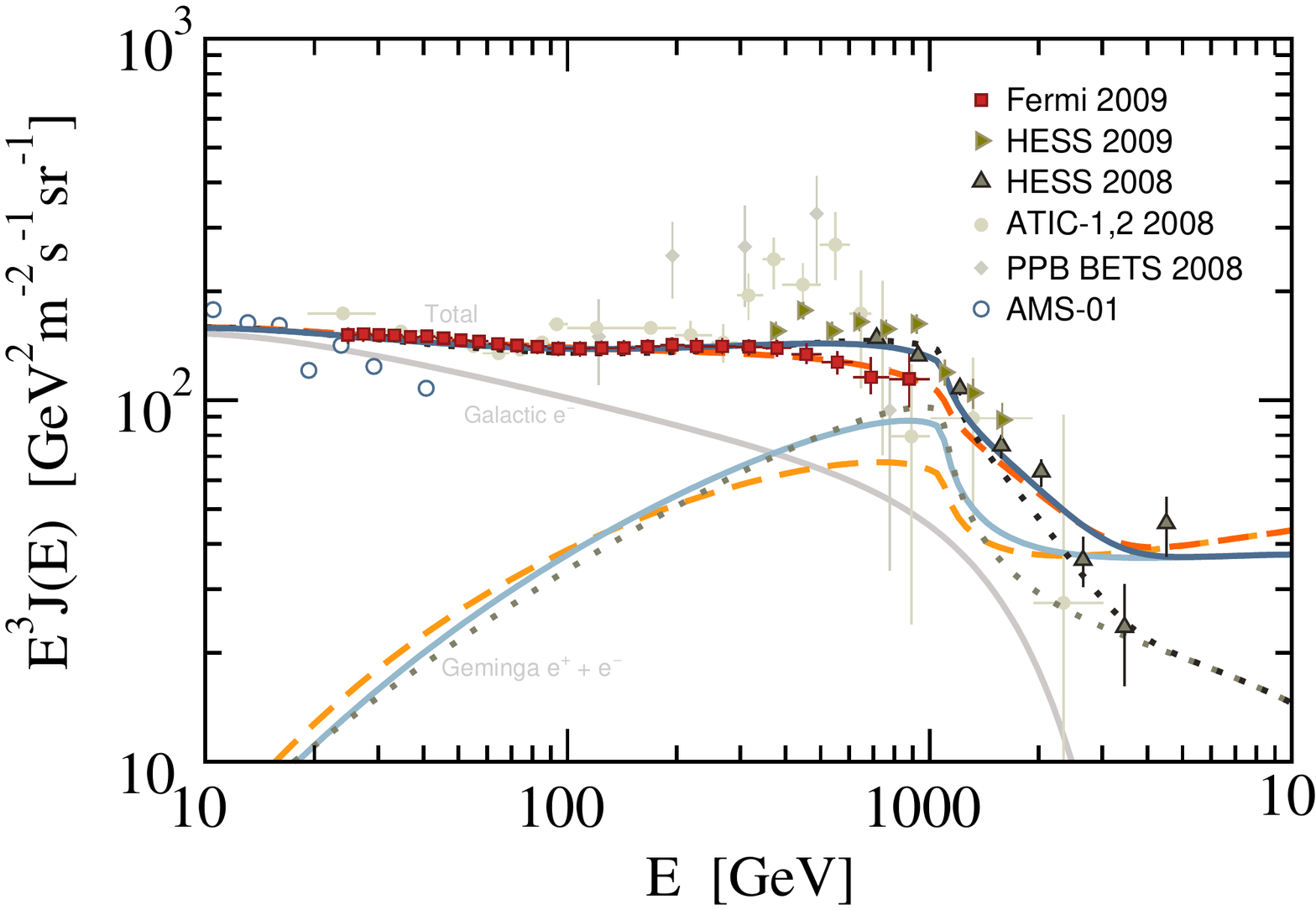}
\caption{The positron fraction (left) and total $(e^++e^-)$ spectra
of the background plus Geminga-like pulsar contribution. From
\cite{Yuksel:2008rf}.
}
\label{fig:pulsar}
\end{figure*}

In summary we can conclude that the current data may still favor the
existence of a population of ``primary'' positrons. There were many
astrophysical factories being proposed to produce the high energy
electrons/positrons, of which the pulsars are the most widely
discussed (e.g., \cite{Yuksel:2008rf,Hooper:2008kg,Profumo:2008ms,
Malyshev:2009tw,Grasso:2009ma,Kawanaka:2009dk,Heyl:2010md}.
The idea of pulsars as the accelerators of high energy electrons/positrons
is actually quite old \cite{Arons:1981xv,Harding:1987xv,Zhang:2001bj,
Grimani:2007ae}. The high energy $\gamma$-ray emission of pulsars, 
especially the recently discovered very high energy emission above 
$100$ GeV \cite{Aliu:2011zi,Aleksic:2011np} directly supports
the particle acceleration of pulsars. The strong magnetic field enables
the photon-pair cascade occur, makes pulsars natural candidate of
positron factory. There are indeed some very nearby pulsars, such
Geminga at 0.16 kpc, PSR B0656+14 at 0.29 kpc and Vela pulsar at
0.29 kpc. Fig. \ref{fig:pulsar} shows an example that a Geminga-like
pulsar together with the background can explain both the positron
fraction and total electron spectra \cite{Yuksel:2008rf}.
To fit the data is easy, but to identify which pulsars contribute
to the CR leptons is very difficult due to the diffusive propagation
of the charged particles. It was discussed that the anisotropy and
precise energy spectra of the electrons/positrons might help to
identify the sources \cite{Hooper:2008kg,Profumo:2008ms,Cernuda:2009kk,
Malyshev:2009tw}.

Other astrophysical sources of the high energy $e^{\pm}$ include
secondary $e^{\pm}$ production inside the SNRs \cite{Blasi:2009hv,
Fujita:2009wk,Ahlers:2009ae}, photo-nuclei pair production of very
young SNRs \cite{Hu:2009zzb}, supernova explosion of massive stars
\cite{Biermann:2009qi}$, \gamma$-ray burst \cite{Ioka:2008cv}, white
dwarf ``pulsars'' \cite{Kashiyama:2010ui}, pulsar wind nebula
\cite{Kamae:2010ad} and so on. For the scenario of secondary $e^{\pm}$
production inside the SNRs, the expected secondary-to-primary ratio
such as $\bar{p}/p$ \cite{Blasi:2009bd} and B/C \cite{Mertsch:2009ph}
will show distinct rising behavior at high energies, and can be
tested in future with high precision data.

\subsubsection{Dark matter}

What is more exciting is that the PAMELA result might be the long-waited
DM signal. A lot of works discussed the possibilites that the positron 
excess comes from DM annihilation or decay. If the positron/electron 
excesses are due to DM annihilation or decay it gives clear indication 
of the nature of DM particles. Firstly, since only positron/electron 
excesses are observed while the antiproton-to-proton ratio is 
consistent with the CR background prediction, the DM should couple 
dominantly with leptons. Secondly, the large amount of positrons
requires very large annihilation/decay rate of DM. For annihilating
DM scenario it requires some non-trivial enhancement mechanisms to get 
large annihilation cross section. We discuss the first property
in the following and leave the discussion of the second point in the
next subsection.

\begin{figure}[!htb]
\begin{center}
\includegraphics[width=0.9\columnwidth]{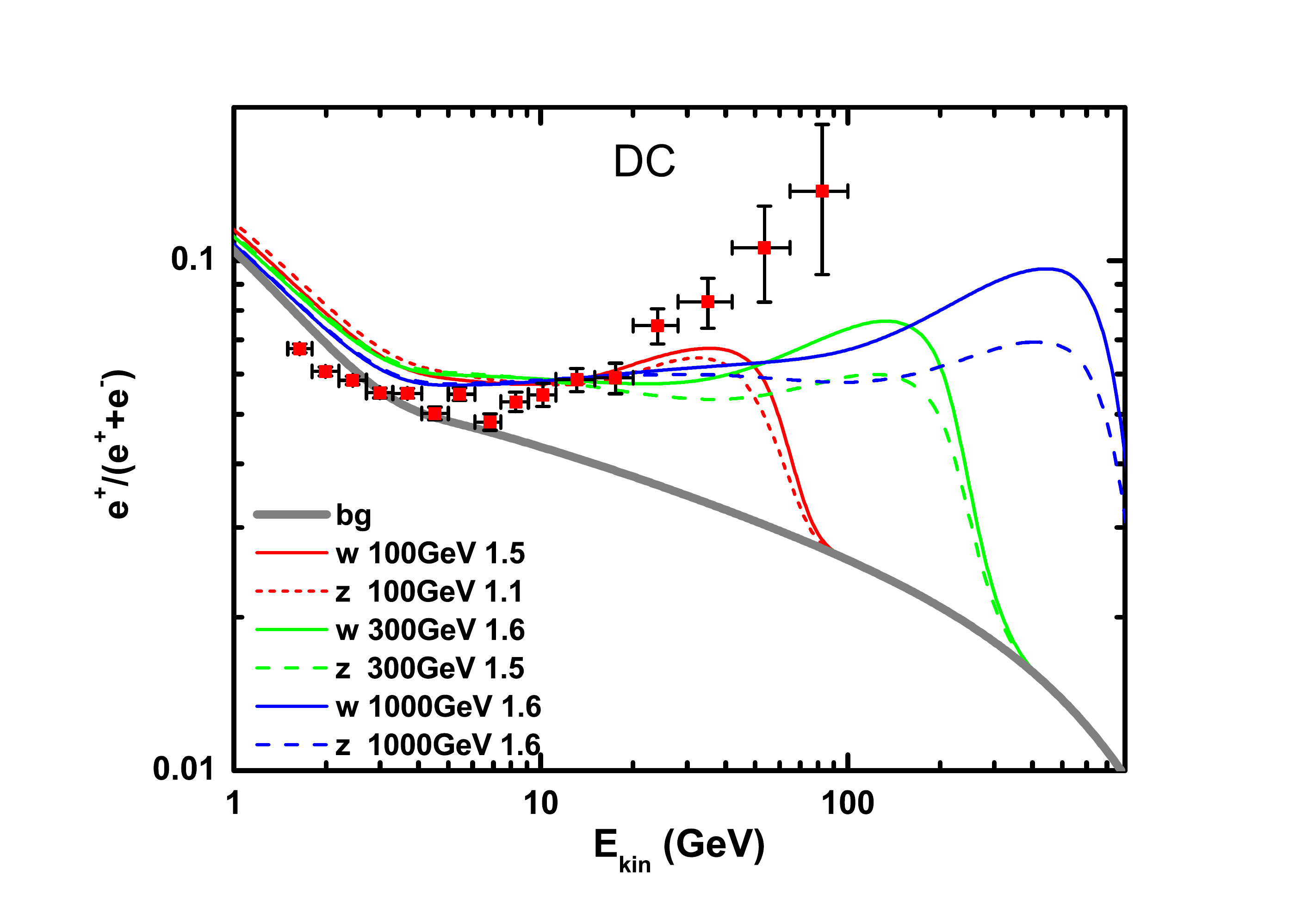}
\includegraphics[width=0.9\columnwidth]{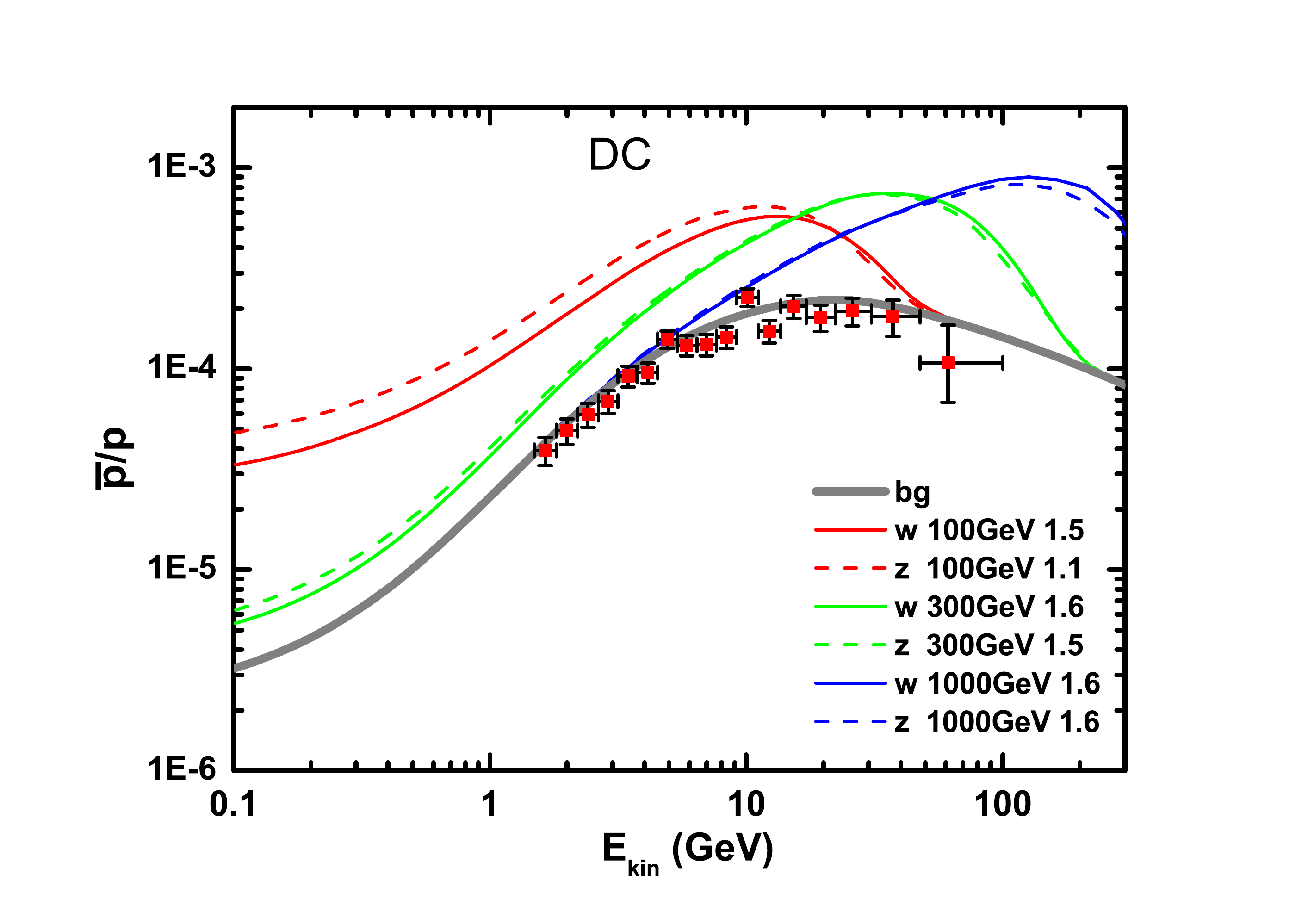}
\caption{ The positron fraction (top) and antiproton/proton ratio
(bottom) from DM decaying into gauge boson pairs. The grey lines
are the background expectation from the CR propagation model. The
numbers label the energies of the gauge bosons and the lifetimes
of the DM. From \cite{Yin:2008bs}.
\label{gaugeboson}}
\end{center}
\end{figure}

\begin{figure}[!htb]
\begin{center}
\includegraphics[width=0.9\columnwidth]{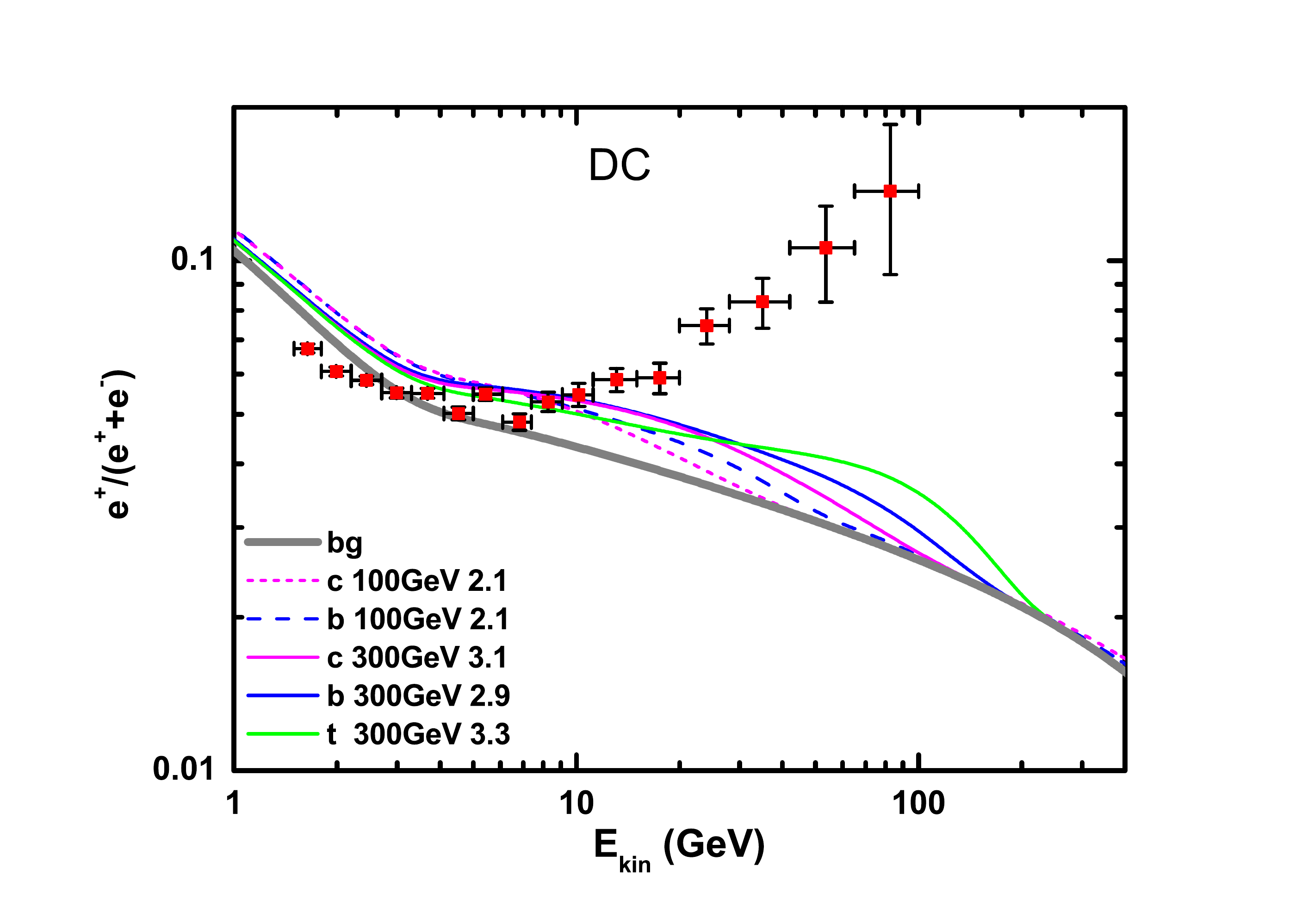}
\includegraphics[width=0.9\columnwidth]{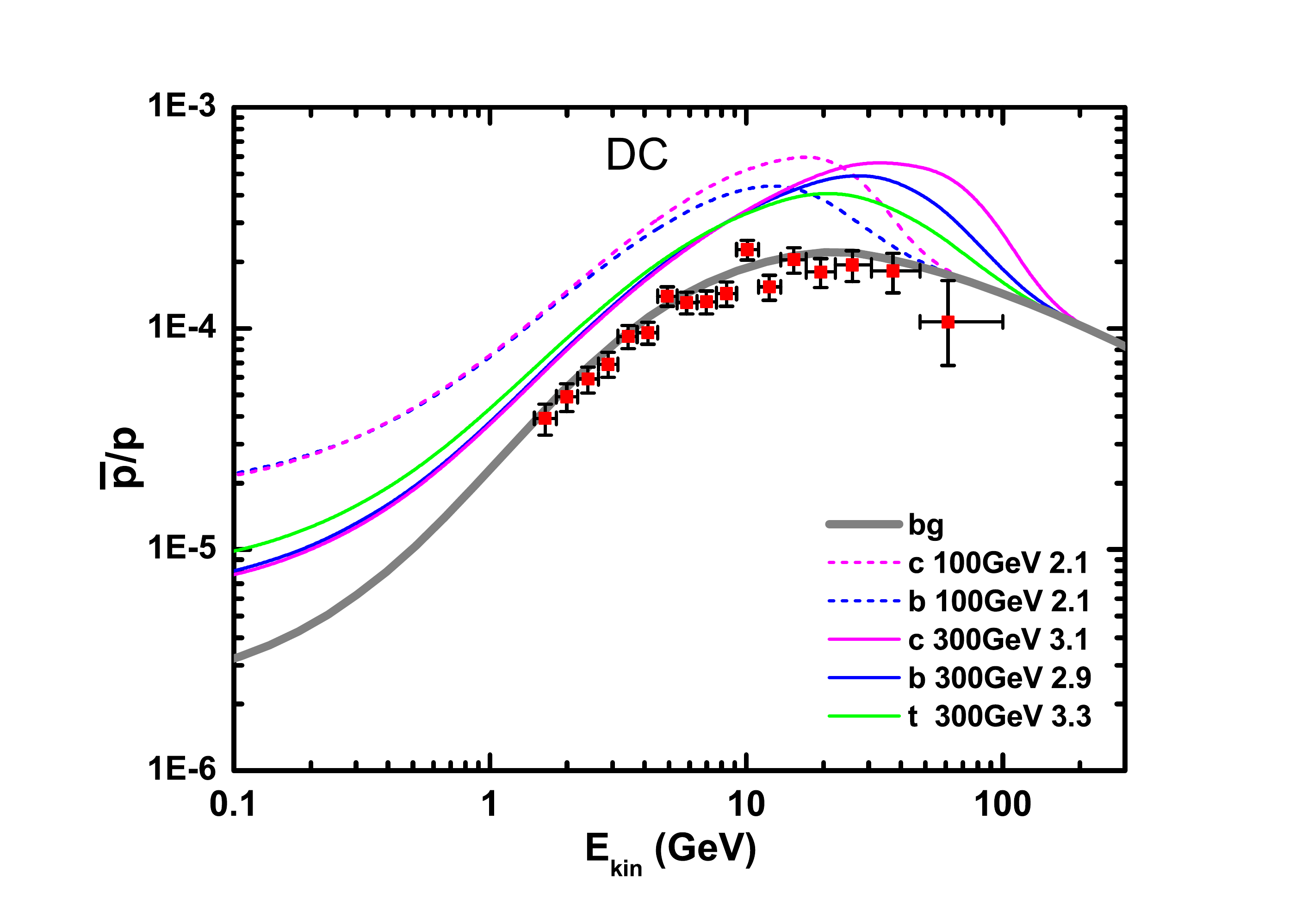}
\caption{Same as Fig. \ref{gaugeboson} but for DM decaying into
quark pairs. From \cite{Yin:2008bs}.
\label{quark}}
\end{center}
\end{figure}

\begin{figure}[!htb]
\begin{center}
\includegraphics[width=0.9\columnwidth]{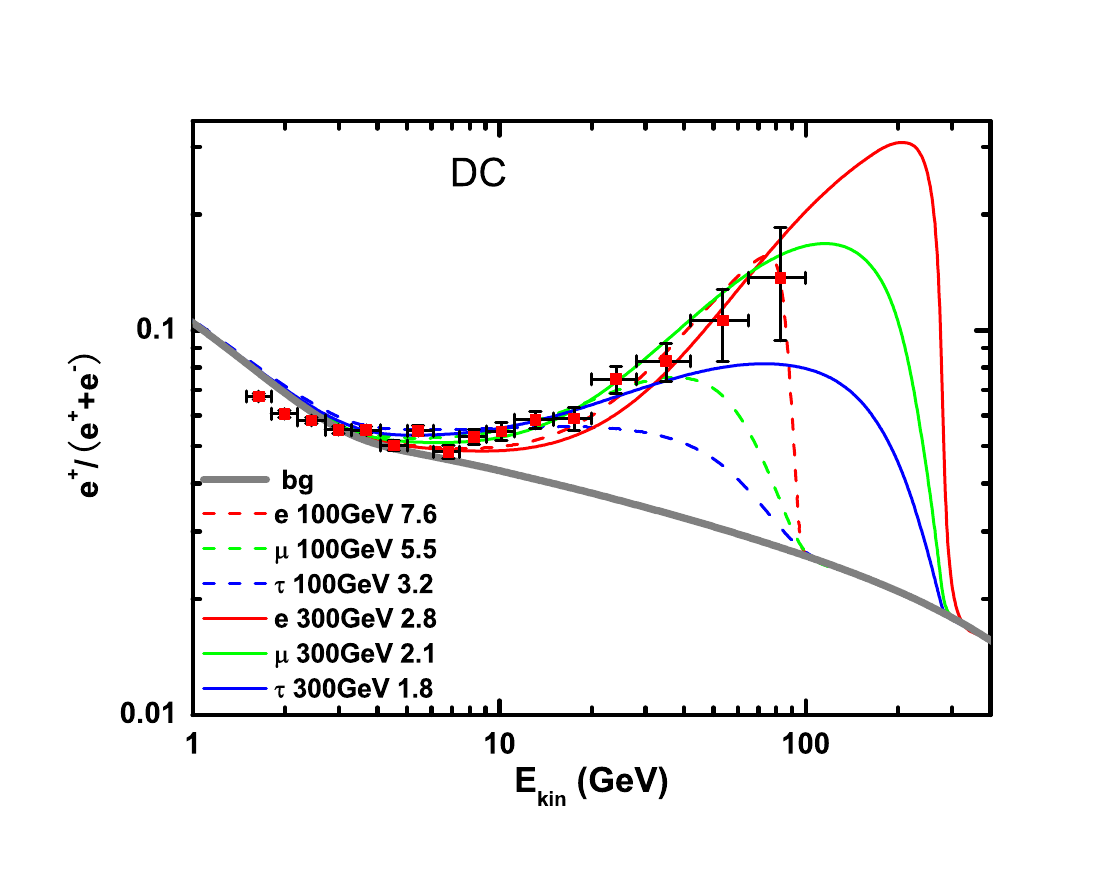}
\caption{Positron fraction for DM decaying into lepton pairs.
From \cite{Yin:2008bs}.
\label{lepton}}
\end{center}
\end{figure}

Soon after PAMELA reported the new result about the positron
fraction, the DM was proposed as a possible positron source to
explain the data (e.g., \cite{Bergstrom:2008gr,Barger:2008su}).
In \cite{Yin:2008bs} we give a careful study of the DM scenario
to explain the positron excess. We first assume DM 
decay\footnote{The propagated positron spectra at the Earth from
the annihilation scenario and the decay scenario have little 
difference. Only in the region like the Galactic center the two 
scenarios show difference. We will discuss the difference in 
Subsection \ref{diff}.} into gauge bosons, and the positrons/electrons 
are then generated from decay of the gauge bosons. Fig. \ref{gaugeboson} 
shows the calculated positron fraction and antiproton-to-proton ratio 
for different energies of the gauge bosons. It is shown that
the positron spectrum from gauge boson decay is too soft to explain
PAMELA data, even the gauge boson energy is as high as $1$ TeV.
Especially, the gauge boson channel is problematic for the
antiproton spectrum. They give several times larger antiproton-to-proton
ratio than the PAMELA data. Therefore DM decaying or annihilating to 
gauge bosons are strongly disfavored by the antiproton data.
Similarly we show the case for DM decay into quarks in Fig.\ref{quark}.
Positrons are produced after hadronization of quarks via the decay of 
charged pions. The positrons are too soft and can not account for the 
excess of positrons above $\sim 10$ GeV. Furthermore, quark hadronization 
produces too many antiprotons, which are several times larger than the
PAMELA data. Fig. \ref{lepton} shows the case for lepton channel.
It shows that in such case it is easy to account for the PAMELA data 
by assuming a proper DM mass and life time. Therefore we can conclude 
that the only possible channel to explain the observed positron excess
is DM decay or annihilate into leptons. Similar conclusions were aslo
shown in Refs. \cite{Cirelli:2008pk,Donato:2008jk}.

Therefore the first important implication for the DM scenario to explain 
the PAMELA data is that DM has to couple dominantly with leptons.

\subsection{Mechanisms to enhance DM annihilation rate}

Another important issue of the DM model to explain the electron/positron
excesses is the high production rate of electrons/positrons. For DM
annihilation scneario, it means a very large boost factor (BF), the 
ratio between the required cross section and the benchmark value 
$3\times 10^{-26}{\rm cm^3s^{-1}}$, at the order $\sim 10^3$
\cite{Bergstrom:2008gr,Cirelli:2008pk}. That is to say to explain the 
positron excess the DM annihilation rate needs to be about thousand 
times larger than that to give correct relic density of DM.

There are plenty of papers in the literature to study possible
ways to enhance the annihilation rate. The key point here is how
to reconcile the large annihilation rate with the much smaller
rate at the early time when DM decoupling. We summarize the
possible ways in the following.

\subsubsection{Substructures}

Since the DM annihilation rate is proportional to the density
square, the DM substructures are expected to enhance the
annihilation rate and give a BF. There have been a lot
of careful study on the effect of substructures on charged CRs,
based on the N-body simulation results. However, the studies show
that the enhancement effects due to substructures is generally
negligible \cite{Lavalle:1900wn}. The largest BF with the most 
extreme configurations of DM substructures is $O(10)$, which is 
much lower than that needed to explain the PAMELA electron/positron 
excesses. Considering that the DM velocity in the substructures is 
smaller than that in the smooth halos, the BF including 
the Sommerfeld effect is also calculated in \cite{Yuan:2009bb}. 
Even in this case the BF is still very small.

It is also proposed that a nearby massive substructure might be
able to provide enough BF to explain the data
\cite{Hooper:2008kv}. However, the search in the simulation
results shows a very low probability ($\sim 10^{-5}$) of the
existence of such a clump \cite{Brun:2009aj}. Therefore it
is very difficult to provide a large enough BF to account for the
PAMELA electron/positron excesses through DM substructures.

\subsubsection{Non-thermal DM}

The benchmark value of DM annihilation cross section  $~3\times
10^{-26}$cm$^3$s$^{-1}$ is acquired under the assumption that DM
is generated thermally in the early universe. If DM is produced by
some non-thermal mechanisms, its annihilation cross section is not
constrained and can be larger or smaller than the benchmark value.

%In the SUSY models, if neutralino is higssino or wino dominated,
%the annihilation cross section can be large as $O(10^{-25})\sim
%O(10^{-24})$cm$^3$s$^{-1}$ due to SM weak interaction. If
%neutralinos are generated from the decays of much heavier
%gravitinos, which are produced in the reheating after inflation
%significantly, the DM relic density and "gravitino problem" can be
%explained simultaneously \cite{Moroi:1999zb}.

The non-thermal mechanism is first proposed in Refs.
\cite{Jeannerot:1999yn,Lin:2000qq}, in which DM is generated by
the decays of topological defects, such as cosmic string.
%After typical DM freeze-out temperature $T_\chi\sim$O(GeV), DM 
%particles do not annihilate into SM particles rapidly. The dominant
%non-thermal DM relic is contributed by the cosmic string loops
%decaying at the temperature $~T_\chi$. Therefore the BBN results
%will not be destroyed by the DM production process. The value of
%DM relic density depends on the detailed cosmic string model and
%the spectra of new physics particles.
As PAMELA released the new result the non-thermal mechanism is
adopted to explain positron excess. By choosing suitable
parameters, the correct relic density and positron flux can be
explained simultaneously \cite{Bi:2009am}.

It is interesting to note that the non-thermal WIMP from cosmic
string decay is more energetic than ordinary WIMP in the early
universe, and can be treated as a kind of warm DM. Therefore the
free streaming length of non-thermal WIMP may be large. It will
lead to distinct predictions of DM substructure and indirect
searches from cold WIMP \cite{Yuan:2012wj}.

\subsubsection{Breit-Wigner enhancement}

If two DM particles annihilate into SM particles through an $s$-channel 
process, there is a resonance if the intermediate particle mass is close 
to $2m_{\chi}$. This is called the Breit-Wigner enhancement.
It is well-known that resonance effect can enhance the DM annihilation 
cross section significantly \cite{Griest:1990kh,Gondolo:1990dk}.
In the Breit-Wigner enhancement scenario \cite{Feldman:2008xs,
Ibe:2008ye,Guo:2009aj,Bi:2009uj}, the DM annihilation process is 
assumed to be $\chi \bar{\chi} \to R \to f\bar{f}$,
where $R$ is a narrow Breit-Wigner resonance with mass
$M=\sqrt{4m^2(1-\delta)}$ and decay width
$\Gamma=M\gamma$ with $|\delta|, \gamma \ll 1$.
The annihilation cross section is proportional to
\begin{equation}
\sigma v \propto \frac{1}{(\delta+v^2)^2+\gamma^2},
\label{BW1}
\end{equation}
where $v$ is the velocity of two DM particles. From Eq. (\ref{BW1}),
it can be found that DM particles with smaller $v$ in the halo have 
larger annihilation cross section than those with larger $v$ in the 
early universe.

%The freeze-out process begins at $x_f \sim O(10)$ ($x \equiv m_{DM}/T \sim v^{-2}$ is related to DM temperature), and continues until the temperature of $x_b \sim max[\delta, \gamma]^{-1}$ at which
%the DM annihilation cross section enhancement is saturated. The final value of reduced DM number density $Y_\infty$($ Y \equiv n_{DM}/s$) is proportional to $x_{b}/\langle \sigma v \rangle_0$ where $\langle\sigma v \rangle_0$ is the DM annihilation cross section at low temperature limit. While $Y_\infty$ is proportional to $x_f/\langle \sigma v \rangle$ for the ordinary S-wave DM annihilation models with a constant cross section $\langle\sigma v \rangle$. The enhancement factor can be achieved as $S \sim x_b/x_f \sim max[\delta,\gamma]^{-1}/O(10)$ \cite{Ibe:2008ye}.
%In order to acquire a large enhancement factor as $O(10^3)$, the
%parameters $\delta$ should be as small as $O(10^{-5})$
%\cite{Ibe:2008ye,Bi:2009uj}. A small decay width $\Gamma$ can be
%achieved via the very weak interactions between resonance and SM
%particles. The tiny difference between the resonance mass
% and twice the DM mass requires much more fine-tuning.

Therefore the Breit-Wigner effect may play a role of BF to explain
the PAMELA data \cite{Ibe:2008ye,Guo:2009aj}. However, to give
a large BF required by the electron/positron data at $O(10^3)$,
fine tuning of the parameters $\delta, \gamma$ as small as $O(10^{-5})$
is needed \cite{Ibe:2008ye,Guo:2009aj,Bi:2009uj}.

\begin{figure}[t]
\includegraphics[width=0.9\columnwidth]{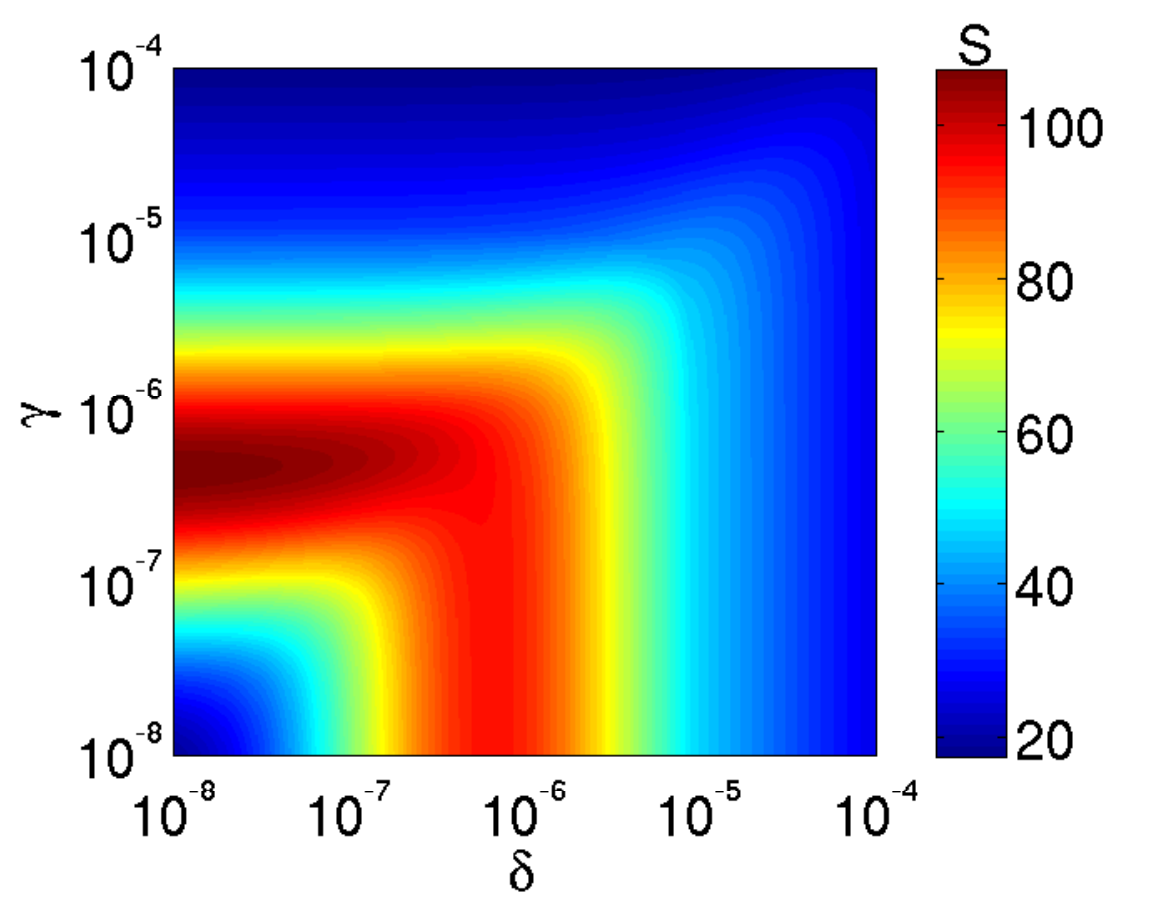}
\caption{Numerical illustration of the BF $S$ on the $\gamma-\delta$ 
parameter plane. From \cite{Bi:2011qm}.}
\label{boost1}
\end{figure}

Later another important effect, the kinetic decoupling, is discussed
\cite{Bi:2011qm}. Since after kinetic decoupling the DM particles can not
get any momentum exchange with the thermal bath, its velocity decreases
more rapidly. As $v$ decreases its annihilation rate is enhanced due to
the Breit-Wigner effect and further reduce the relic density of
DM significantly. This finally reduce the BF, as shown in Fig.
\ref{boost1}. The largest enhancement factor $S$ is at $O(10^2)$ for 
the DM with a mass of 1 TeV, as shown in Fig. \ref{boost1}. Therefore 
it is difficult to explain the anomalous positron excesses and give 
the correct DM relic density simultaneously in the minimal Breit-Wigner 
enhancement model.

%Note that the elastic scattering between DM and leptons $\chi f \to \chi f $ via the t-channel is suppressed by the heavy propagator $R$ with $\sim m_R^4 \sim 1/16m_\chi^4$, DM particles are difficult to keep in kinetic equilibrium by momentum exchange with the leptons after chemical decoupling at the temperature of O(GeV) \cite{Bi:2011qm}. Since the velocity of DM decrease much faster (as "matter") after kinetic decoupling than that in kinetic equilibrium (as "radiation") \cite{Bringmann:2006mu}, the Breit-Wigner enhancement after the kinetic decoupling will reduce the DM relic density significantly. Some new interactions to keep DM in the kinetic equilibrium until low temperature are needed to interpret the positron excess and correct DM relic density simultaneously.

%Therefore the DM annihilation cross section needs to be smaller to explain the correct DM relic density and the enhancement %factor $\sim O(10^3)$ is not easy to achieve. However,

\subsubsection{Sommerfeld enhancement}

If there is a long range attractive force between two DM particles, 
the cross section will enhance at low momentum, known as the Sommerfeld 
effect. A new light boson $\phi$ with $m_{\phi} \ll m_\chi$ is usually 
assumed to mediate the ``long range'' interaction between DM particles.
This non-perturbative quantum effect arises from the contributions
of ladder diagrams due to the exchange of new bosons between two
incoming DM particles in the annihilation process.
It leads to a velocity dependent annihilation cross section. When
the relative velocity of the two particles is high enough,
the Sommerfeld enhancement effect becomes negligible. Therefore at the
early stage of the Universe, the cross section keeps to be a
relatively low value which can give the right relic density.
When the DM particles cool down significantly today, a larger
annihilation cross section can be obtained. There are several
works employ the Sommerfeld effect to explain the large BF as
implied by the CR lepton data (e.g.
\cite{Cirelli:2008pk,Pospelov:2008jd,ArkaniHamed:2008qn,Lattanzi:2008qa}).

In the literature, the Sommerfeld enhancement has been discussed
for  $\chi \chi \to W^+ W^-$ if $W^\pm$ is light enough compared
to heavy DM with mass of several TeV \cite{Hisano:2004ds,Cirelli:2008pk}.
However, the $W$ boson may over-produce antiprotons and be conflict
with the PAMELA data. If the mediator $\phi$ boson is light enough
instead, e.g. $m_\phi \leq O(1)$GeV, the production of antiprotons 
will be kinematically suppressed 
\cite{Pospelov:2008jd,ArkaniHamed:2008qn,Cholis:2008qq,Nomura:2008ru}.
Therefore the existence of light boson interpreters the PAMELA
results elegantly.

However, further analysis finds that the enhancement of cross
section due to the Sommerfeld effect would also affect the thermal
history of DM in the early Universe.
The calculation of DM relic density with Sommerfeld enhancement
depends on several issues, such as the temperature of kinematic
decoupling \cite{Zavala:2009mi,Dent:2009bv,Feng:2010zp} and
the efficiency of self-interactions for persevering thermal
velocity distribution \cite{Feng:2010zp}. Detailed calculation
shows there is a tension between large enhancement factor and
correct DM relic density \cite{Feng:2010zp,Feng:2009hw}. In
order to obtain correct DM relic density for $m_\chi \sim$1
TeV and $m_\phi \sim 1$GeV, the maximal value of enhancement
factor $S$ for $\chi \chi \to \phi \phi\to \mu^+\mu^-\mu^+\mu^-$
is only $\sim O(10^2)$ which is smaller than the required value
$O(10^3)$. In \cite{Slatyer:2011kg} the decay channel of $\phi \to e^+e^-$
is considered and a special configuration of ``dark sector'' is assumed,
which can relax the constraint on the maximal BF.

The light mediator, a scalar or a vector boson, can interact with
leptons via the small mixing with the Higgs boson or gauge boson
in the SM. This kind of interactions is called ``dark force''
in the literature. The small mixing can be achieved by intergrading
out some heavy fields which interact with both $\phi$ boson and SM
boson, and induce small mass of $\phi$ as $m_\phi \sim 10^{-3} m_\chi$
naturally \cite{Baumgart:2009tn,Cheung:2009qd}. The light mediators
may be produced by the collisions between electrons and/or protons,
and can be tested at the ``fixed target'' experiments
\cite{Reece:2009un,Bjorken:2009mm,Freytsis:2009bh},
low-energy electron-positron colliders
\cite{Essig:2009nc,Yin:2009mc,Li:2009wz} or high-energy
hadron colliders \cite{Bai:2009it,Cheung:2009su}.

\subsubsection{Decaying dark matter}

The final way to solve the discrepancy between the early annihilation
rate and today's lepton generation rate is to assume that leptons are
generated by DM decay. Therefore the decay process dominates over the
annihilation process today. BF is not needed in such scenario obviously.

If there exists some tiny symmetry violation, the DM particles may
decay very slowly to SM particles. Since the decay of DM should not 
reduce the abundance of DM in the universe significantly, the 
lifetime of DM should be much longer than the age of the universe
$\sim 10^{17}$ s. If DM particles can decay into electrons or
muons via some leptonic interactions, the flux of positron
excess can be interpreted by the decaying DM with lifetime of
$\sim O(10^{26})$ s \cite{Chen:2008dh,Yin:2008bs,Ishiwata:2008cv,
Ibarra:2008jk,Chen:2008qs,Nardi:2008ix}.
Such long lifetime of DM can be derived naturally by some high
energy scale suppressed operators \cite{Arvanitaki:2008hq}.
For instance, if a DM particle with a mass of 1 TeV decays
via dimension 6 operators, its lifetime would be
$\tau \sim 8 \pi \Lambda^4_{GUT}/m_\chi^5 \sim 3\times 10^{27}$ s.
The signatures of high energy photons \cite{Meade:2009iu,Papucci:2009gd,
Chen:2009uq,Ibarra:2009nw,Zhang:2009ut,Huang:2011xr,Cirelli:2012ut} and
neutrinos \cite{Hisano:2008ah,Covi:2009xn} from decaying DM have
been widely studied in the literature.

Note that the fluxes of high energy photons and neutrinos induced by 
decaying DM and annihilating DM are proportional to the DM number 
density and the square of DM number density, respectively. Such features
can be used to distinguish different DM scenarios.

\subsection{Discrimination between astrophysical and dark matter scenarios
\label{diff}}

As we discussed above, all the explanations to the positron/electron
excesses fall into two categories: the astrophysical origin and the DM origin.
An important question is if we can discriminate the two kinds of origins.

First we should state that by the local observation it is impossible to
distinguish the decaying and annihilation DM scenarios \cite{Liu:2009sq}.
This is easy to understand. The final states of the decay and annihilation 
are the same. The only difference comes from the source distribution of 
positrons from DM annihilation and decay. However, the observed high 
energy electron/positrons should come from local region near the solar 
system. This makes the two cases indistinguishable.

\begin{figure*}[!htb]
\begin{center}
\includegraphics[width=0.9\columnwidth]{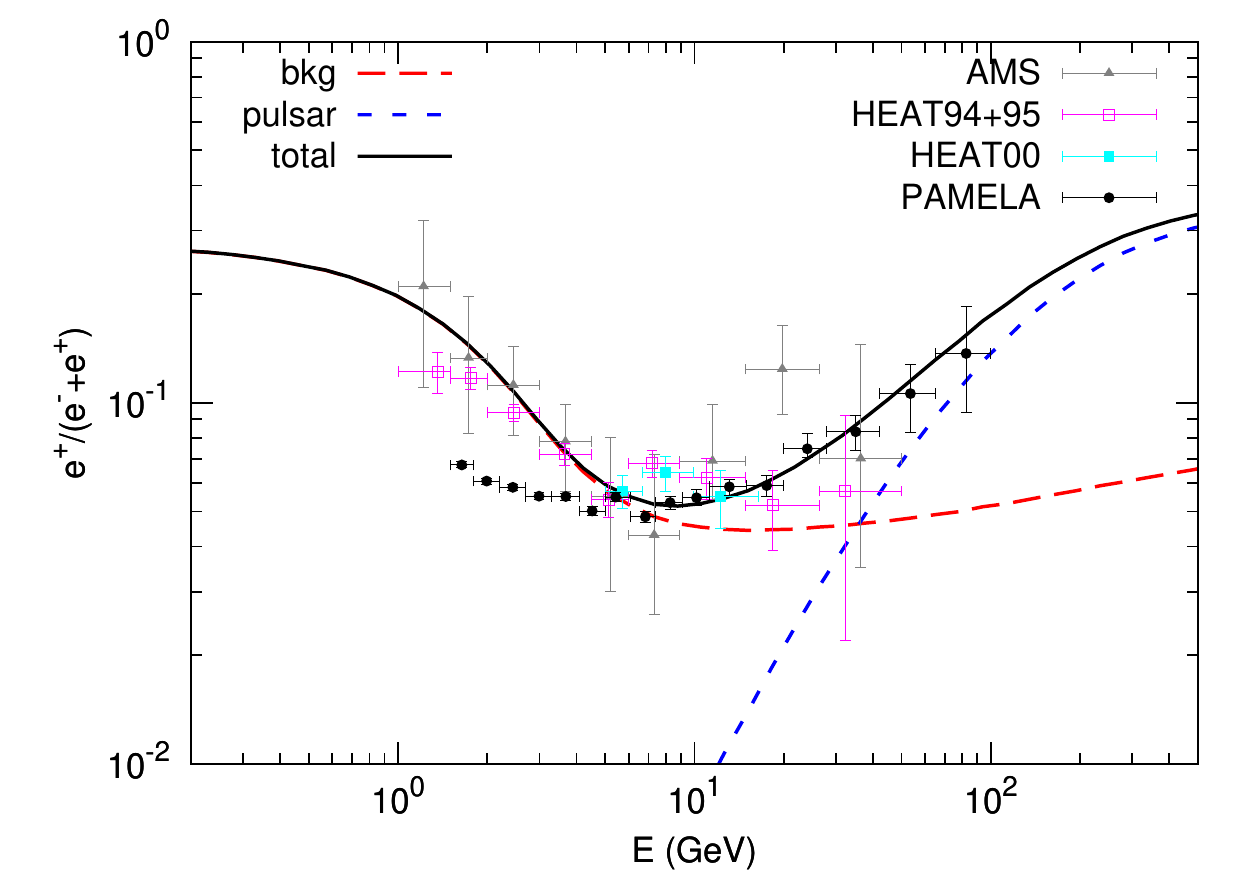}
\includegraphics[width=0.9\columnwidth]{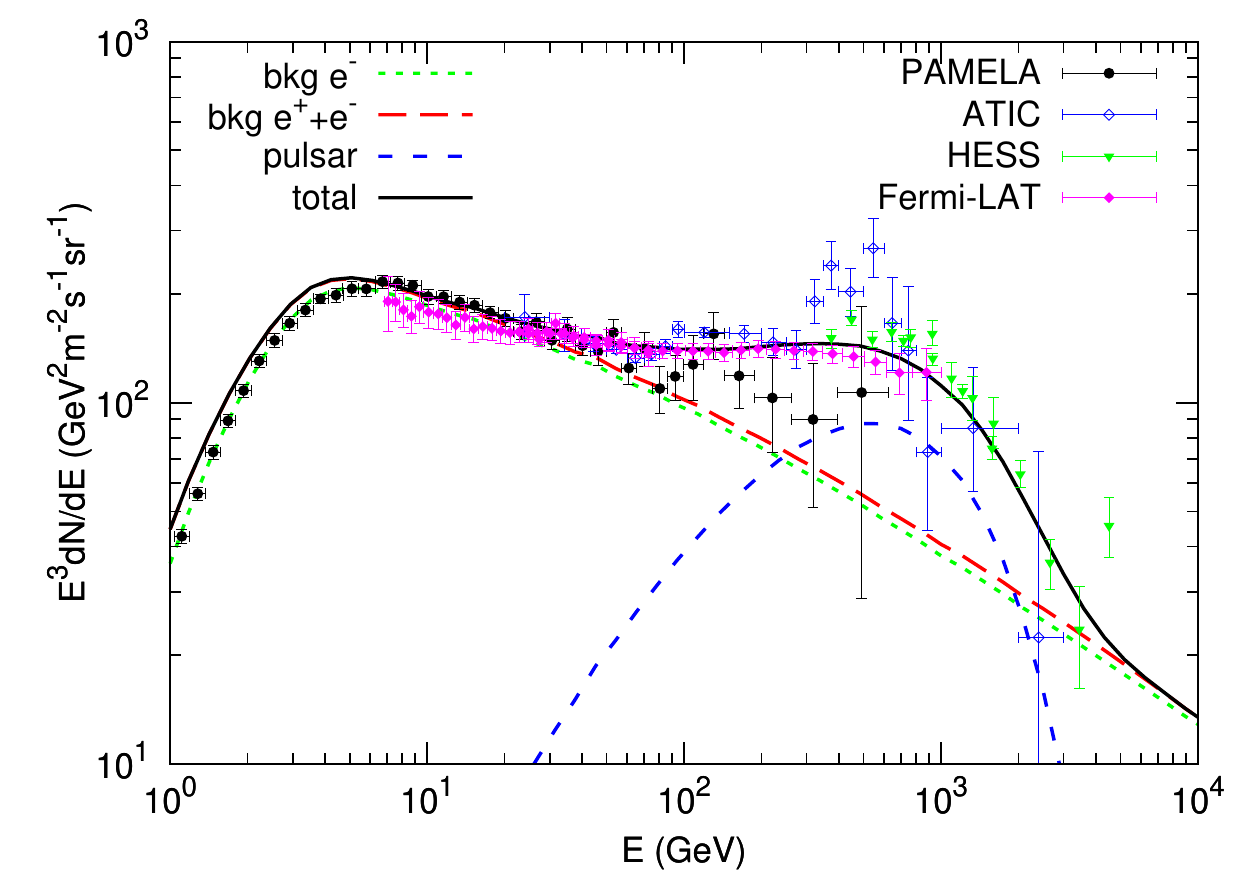}
\includegraphics[width=0.92\columnwidth]{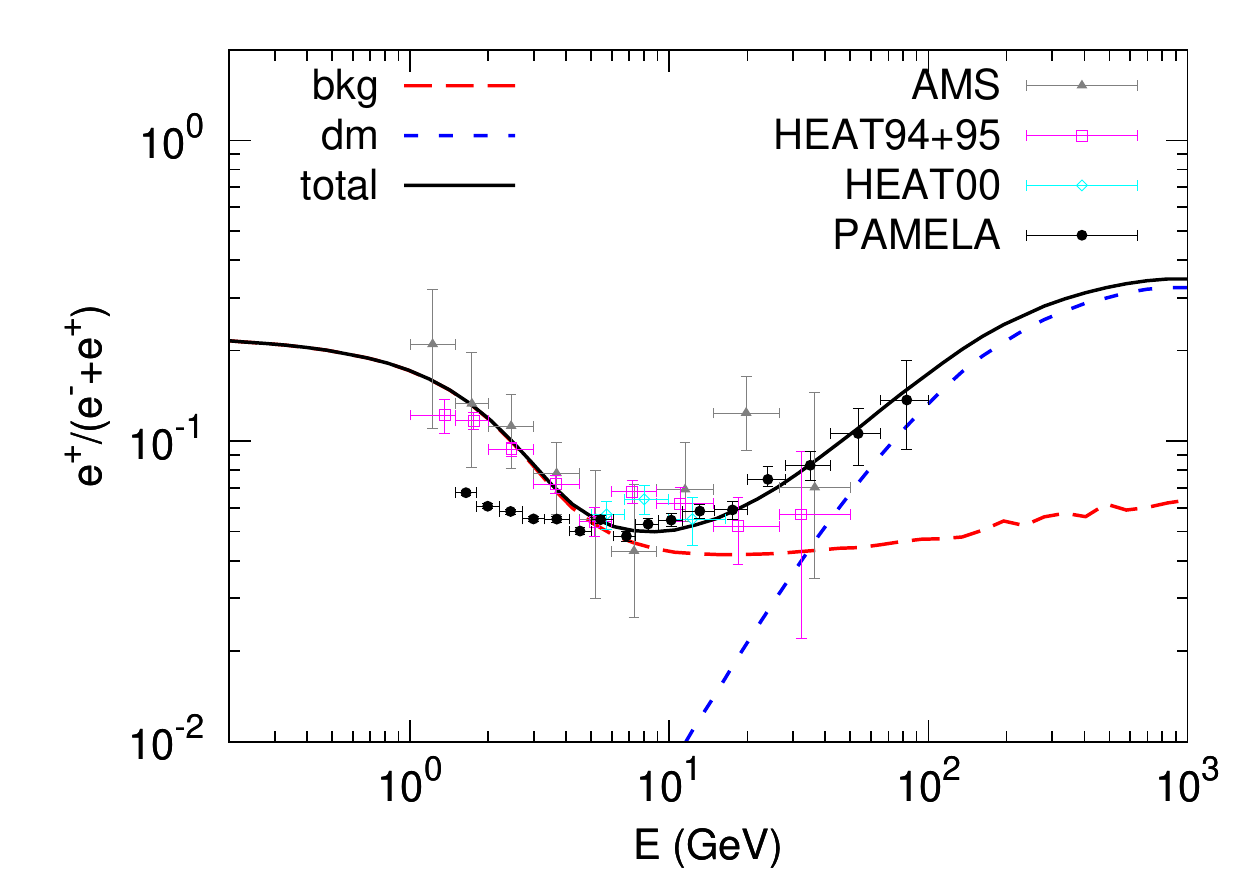}
\includegraphics[width=0.92\columnwidth]{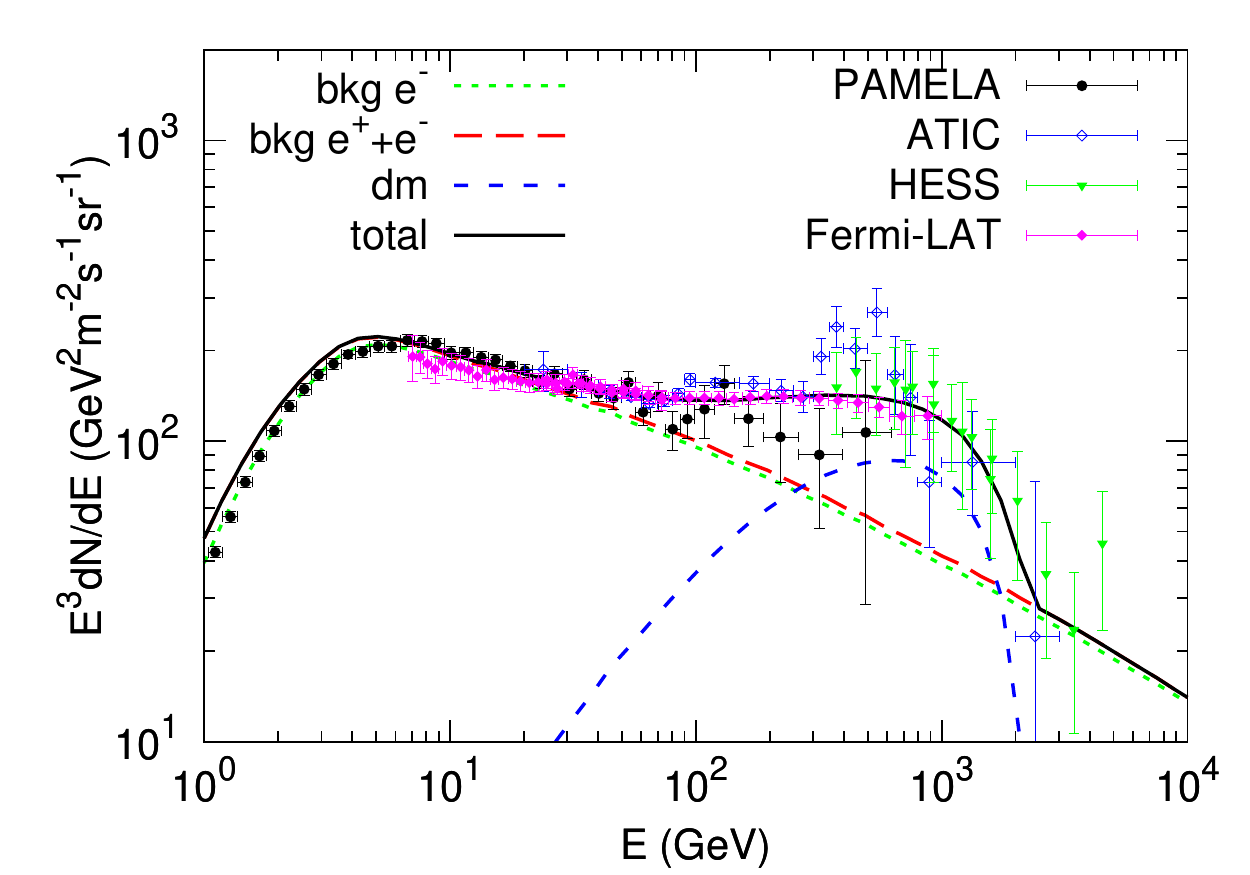}
\caption{Best-fit positron fraction (left) and electron spectra (right)
for the scenarios with pulsars (top panels) and DM annihilation (bottom)
as the extra sources of positrons and electrons. From \cite{Liu:2011re}.
\label{fig:epm_psr_dm}}
\end{center}
\end{figure*}

\begin{table*}
\begin{threeparttable}
\centering
%\tiny
\caption{Fitting parameters with $1\sigma$ uncertainties or $2\sigma$
limits. Note that for the ``bkg'' case the reduced $\chi^2$ is too large
that the uncertainties of the parameters should not be statistically
meaningful. From \cite{Liu:2011re}.}
\begin{tabular}{cccc}
\hline
\hline
& bkg & bkg+pulsar & bkg+DM \\
\hline
$\gamma_1$ & $<1.532 (95\% C.L.)$ & $<1.619 (95\% C.L.)$ & $<1.610 (95\% C.L.)$\\
\hline
$\gamma_2$ & $2.557\pm0.007$ & $2.712\pm0.014$ & $2.706\pm0.013$\\
\hline
$\log(A_{\rm bkg})$\tnote{1} & $-8.959\pm0.003$ & $-8.997\pm0.007$ & $-8.997\pm0.006$\\
\hline
$E_{\rm br} ({\rm GeV})$ & $3.599_{-0.112}^{+0.123}$ & $4.254_{-0.287}^{+0.278}$  & $ 4.283_{-0.259}^{+0.246}$\\
\hline
$\phi ({\rm GV})$ & $0.324\pm0.016$ & $0.383\pm0.042$ & $0.371\pm0.037$\\
\hline
$c_{e^+}$ & $1.462\pm0.035$ & $1.438_{-0.079}^{+0.076}$ & $1.394\pm0.053$ \\
\hline
$c_{\bar p}$ & $1.194\pm0.039$ & $1.225\pm0.043$ & $1.210\pm0.045$ \\
\hline
$\log(A_{\rm psr})$\tnote{1} & --- & $-27.923_{-0.537}^{+0.534}$ & ---\\
\hline
$\alpha$ & --- & $1.284\pm0.104$  & --- \\
\hline
$E_c ({\rm TeV})$ & --- & $0.861_{-0.164}^{+0.170}$ & --- \\
\hline
$m_\chi ({\rm TeV})$ & --- & --- & $2.341_{-0.391}^{+0.492}$ \\
\hline
$\log[\sigma v ({\rm cm^3 s^{-1}})]$ & --- & --- & $-22.34\pm0.13$\\
\hline
$B_e$ & --- & --- & $<0.379(95\%C.L.)$\\
\hline
$B_\mu$ & --- & --- & $<0.334(95\%C.L.)$\\
\hline
$B_\tau$ & --- & --- & $0.713_{-0.152}^{+0.141}$\\
\hline
$B_u$ & --- & --- &$<0.005(95\%C.L.)$\\
\hline
$\chi^2/d.o.f$ &  $3.390$ & $1.047$ & $1.078$\\
\hline
\end{tabular}
\begin{tablenotes}
\footnotesize
\item[1]In unit of ${\rm cm^{-2} sr^{-1} s^{-1} MeV^{-1}}$.
\end{tablenotes}
\label{table:par}
\end{threeparttable}
\end{table*}

In \cite{Liu:2011re} we try to see which scenario of the positrons 
between the astrophysical source and the DM is more favored by the
present data. We assume pulsars as the typical astrophysical sources 
contributing to the positron excess, and compare the goodness of fit
to the data with the DM scenario through a global fit method. 
The contribution of pulsars is parametrized with three new parameters,
the power law index $\alpha$, the cutoff energy $E_c$ and the normalization
$A_{\rm psr}$. For the DM case we fit the mass $m_\chi$, cross section 
$\langle \sigma v \rangle$, and 4 branching ratios to leptons and quarks. 
The data include the PAMELA positron fraction, the PAMELA electron 
spectrum \cite{Adriani:2011xv}, Fermi-LAT total electron spectrum 
\cite{Abdo:2009zk}, and HESS total electron spectrum 
\cite{Aharonian:2008aa,Aharonian:2009ah}. 
The fitting parameters are given in Table \ref{table:par} and the best
fitting results compared with the data are shown in Fig. 
\ref{fig:epm_psr_dm}. We see that when including either pulsar or DM
the fitting is improved essentially compared with the pure background
fitting. However, the $\chi^2/{\rm d.o.f.}$ for the pulsar and DM 
scenarios are almost the same. That is to say we can not distinguish 
the two scenarios according to the present electron/positron data.

\begin{figure}
\includegraphics[width=0.9\columnwidth]{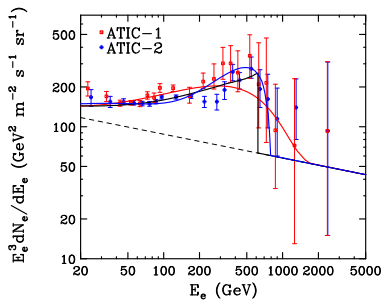}
\caption{\label{atic_fit}
The spectrum predicted from three possible sources: a nearby pulsar (red),
annihilation to $W^+W^-$ from $800$ GeV DM (blue), and annihilation of 
$620$ GeV Kaluza-Klein DM (which annihilates to $e^+e^-$, $\mu^+\mu^-$, 
and $\tau^+\tau^-$ with branching ratios 20\% each, black).
From \cite{Hall:2008qu}.}
\end{figure}

There are proposals to distinguish these two scenarios with future
experiments. In \cite{Hall:2008qu} the authors considered three cases 
to fit the ATIC data: the nearby pulsars, DM annihilation into $W$ boson 
pair and Kaluza-Klein DM that produce a sharp cutoff. The three models 
give different electron spectra, especially around the cutoff, as 
shown in Fig. \ref{atic_fit}. They proposed to distinguish the different 
scenarios by the IACTs, such as HESS, VERITAS and MAGIC.

In \cite{Malyshev:2009tw} the authors studied the pulsar contribution to 
the electron/positron spectrum. They pointed out that at higher energies 
the spectrum is dominated by a few young nearby pulsars, which therefore 
induces wiggle-like features. If the electron/positron spectrum can be 
measured with much higher precision and the wiggle-like features are 
detected it will strongly favor the pulsar origin of the positron/electron 
excesses.

At present the most important experiment for DM indirect detection is
certainly the AMS02 \cite{ams02}. The AMS02 has accumulated data for
nearly two years. The AMS02 will measure the CR spectra with much 
higher precision than the present data, which will reduce the 
uncertainties of background model significantly. It is possible
that AMS02 will make breakthroughs in the indirect detection of DM. 
In \cite{Pato:2010im,Palmonari:2011zz} the potential of AMS02 to 
measure the positron spectrum and the possibility to distinguish 
different kinds of the extra sources were discussed.

A new satellite experiment has been proposed in China, which is called
Dark Matter Particle Explorer (DAMPE) \cite{Chang:DAMPE2011}. DAMPE is
an electromagnetic calorimeter dedicated to measure the electron+positron 
with energies from about 5 GeV up to about 10 TeV. With the large aperture 
of $0.3$ cm$^2$ sr, DAMPE could measure the total electron+positron 
spectra with high precision. This will test the prediction of the 
wiggle-like features in \cite{Malyshev:2009tw}. An important goal of 
DAMPE is to measure the shape of the cutoff of electron spectrum and 
try to distinguish different scenarios to explain the positron/electron 
excesses.

Another widely used way to discriminate different models is the photon
emission \cite{Borriello:2009fa,Barger:2009yt,Cirelli:2009vg,Regis:2009md,
Ishiwata:2009dk,Zhang:2009pr,Ibarra:2009dr,Cholis:2009gv,Zhang:2009kp,
Bi:2009de,Harding:2009ye,Pohl:2009qt,Yuan:2009xq,Yuan:2010gn}.
In \cite{Zhang:2008tb} we discuss the difference of the synchrotron and 
inverse Compton (IC) radiation from the Galactic center region for
the pulsar, DM annihilation and DM decay scenarios. The key point is 
that the three scenarios have very different spatial distribution and 
the largest difference is at the Galactic center. All the scenarios 
can explain well the local observation by adjusting the parameters. 
However, once they are normalized at the position of the Earth they 
should show great difference at the Galactic center. Since the 
electrons/positrons can not propagate to the Earth due to fast energy loss 
we propose to observe the difference by their synchrotron and IC radiation.

%We adopt the Einasto form \cite{Einasto:1965bs} for DM density profile
%\begin{equation}
%\rho(r)=\rho_0\exp\left[-\frac{2}{\alpha}\left(\frac{r^{\alpha}-
%R^{\alpha}_{\odot}}{r^{\alpha}_{-2}}\right)\right],
%\end{equation}
%with $\alpha=0.2$, $r_{-2}=25$ kpc and the local DM density
%$\rho_{\odot}=0.3$ GeV cm$^{-3}$ \cite{Cholis:2008vb}
%at $r=R_{\odot}\equiv8.5\,$kpc. We know the decay DM follows $\rho$ 
%while the annihilation DM follows $\rho^2$. The spatial distribution 
%of pulsars is parametrized as \cite{Zhang:2001bj}
%\begin{equation}
%f(R,z)\propto\left(\frac{R}{R_{\odot}}\right)^{a}\,
%\exp\left[{-\frac{b(R-R_{\odot})}{R_{\odot}}}\right]
%\exp\left({-\frac{|z|}{z_s}}\right),
%\end{equation}
%where $R$ is the Galactocentric radius, and $z$ is the distance
%away from the Galactic plane. The distribution parameters
%taken as $a=1.0, b=1.8$ and $z_s\sim 0.2$ kpc.
%We notice that the pulsars distribution is concentrated at the
%Galactic Plane and tend to zero at the GC. This is totally different
%from the DM, where the density tend to very large at the GC.

\begin{figure}[!htb]
\centering
\includegraphics[width=0.9\columnwidth]{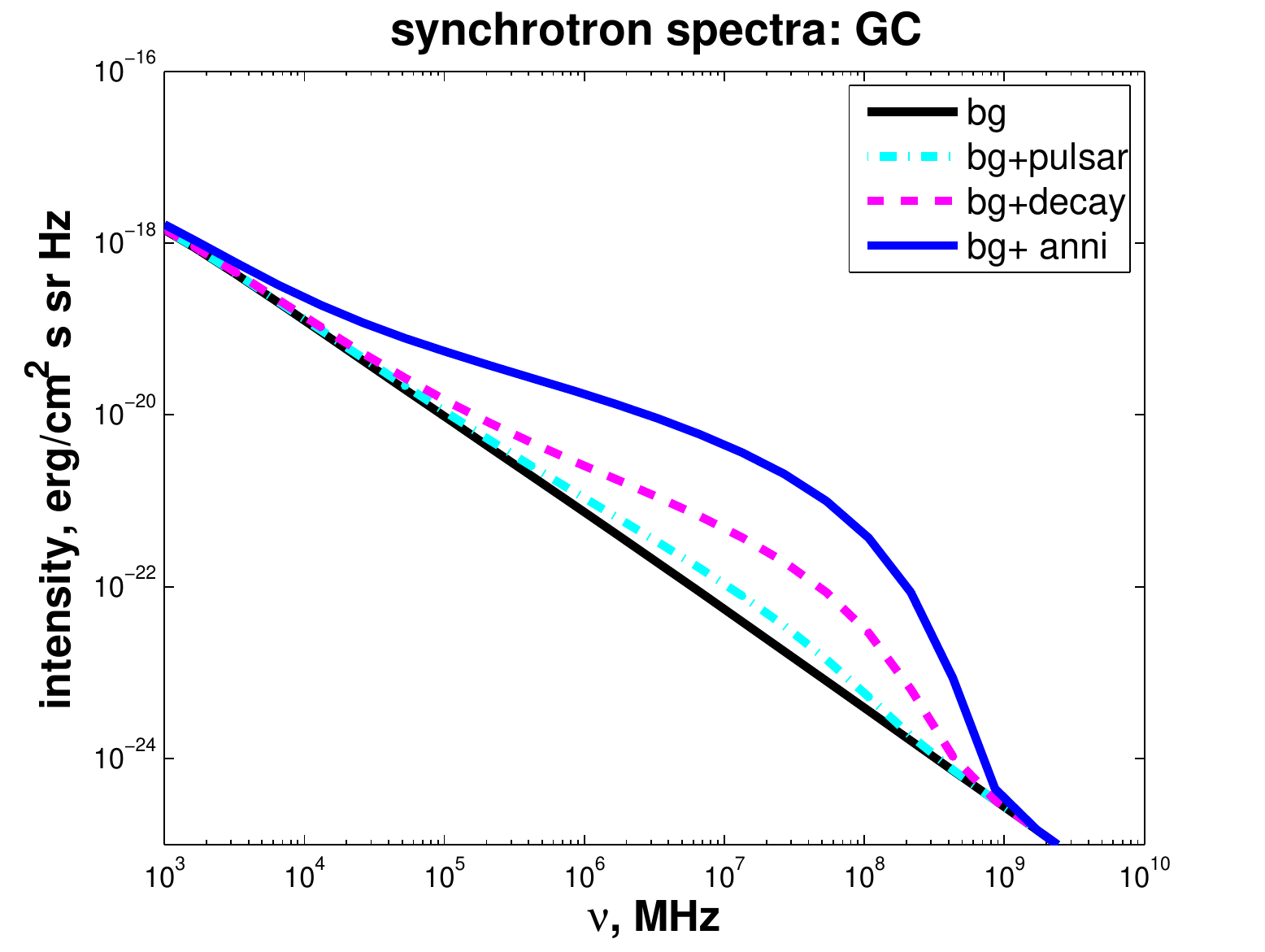}
\includegraphics[width=0.9\columnwidth]{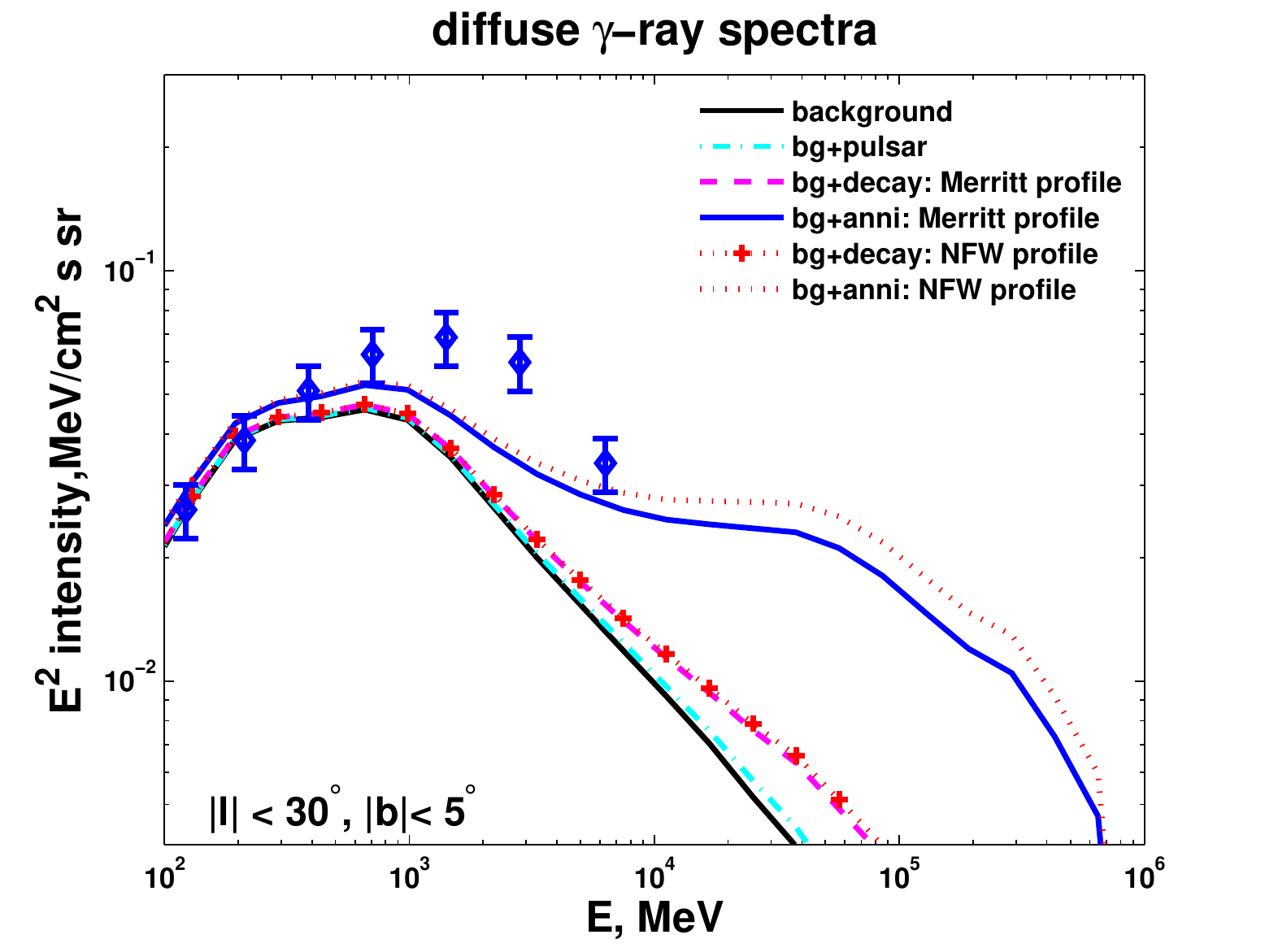}
\caption{Upper: The average synchrotron spectra of the three scenarios 
within a bin size of $20^\circ\times20^\circ$ around Galactic center;
Lower: the contributions to the diffuse $\gamma$-ray spectra in the
region $|l| <30^\circ, |b| < 5^\circ$ for the three scenarios, compared 
with the EGRET data \cite{Hunger:1997we}. From \cite{Zhang:2008tb}.
} \label{spectra}
\end{figure}

%After adjusting the parameters to fit the PAMELA, ATIC or Fermi data
%we calculated the synchrotron and IC spectra from the GC region.
Fig. \ref{spectra} shows the calculated synchrotron emission (top) 
and IC $\gamma$-ray spectrum (bottom) in the inner Galaxy. In this
calculation the DM density profile is adopted to be the Einasto
profile \cite{Einasto:1965bs}. We can see that the three scenarios 
indeed show great difference at the inner galaxy region. Here we take 
a large region around the Galactic center is mainly due to the large 
uncertainties of the DM density profile. We also considered the cases 
for NFW or cored isothermal profiles. We find even for the cored 
isothermal profile the radiation spectra still show distinguishable 
differences.

\section{Status of indirect detection -- gamma rays}

Gamma-ray photons are better than the charged CRs for the indirect search
of DM due to the simple propagation. By means of $\gamma$-rays we can
trace back to the sites where DM concentrates and map the distribution
of DM. The energy spectrum of $\gamma$-rays is less affected during the
propagation (except for high redshift sources) and could directly reflect
the properties of DM particles. Furthermore, the photons can enlarge the
detection range of DM significantly compared with the charged CRs which
are almost limited in the Milky Way.

There are in general two ways to produce photons from the DM: the
{\it primary} emission radiated directly from the DM annihilation/decay
or from the decay of the final state particles, and the {\it secondary}
emission produced through the inverse Compton scattering, synchrotron
radiation or bremsstrahlung radiation of the DM-induced particles
(mainly electrons and positrons). For WIMPs, the photon emission is
mainly at the $\gamma$-ray band, with some kind of {\it secondary}
emission such as the synchrotron radiation covering from X-ray to radio
bands. In this review we focus on the recent progress in $\gamma$-ray
search of DM.

Fermi space telescope is one of the most important facilities in operation
for the $\gamma$-ray detection. The sensitivity of searching for DM with
Fermi is up to now the highest for the general WIMP models. There
are also ground-based very high energy (VHE) $\gamma$-ray detectors
such as the imaging atmospheric Cerenkov telescopes (e.g., HESS, VERITAS
and MAGIC) and air shower array detectors (e.g., ARGO-YBJ), which could
be specifically powerful for heavy DM ($m_{\chi}\sim$TeV).

\subsection{Fermi}

\subsubsection{Dwarf galaxies}

The Milky Way dwarf galaxies are ideal laboratories for the indirect
detection of DM. The dwarf galaxies are DM dominated with mass-to-light
ratio of the order $10^2-10^3$ \cite{Wolf:2009tu}. The lack of gas
content makes it free of $\gamma$-ray emission, resulting in a clean
target for the search of DM signal. There are many works using the
Fermi-LAT data to search for DM signal or constrain the DM model
parameters \cite{Abdo:2010ex,Ackermann:2011wa,GeringerSameth:2011iw,
Cholis:2012am,Baushev:2012ke,Mazziotta:2012ux}.

\begin{figure}[!htb]
\centering
\includegraphics[width=\columnwidth]{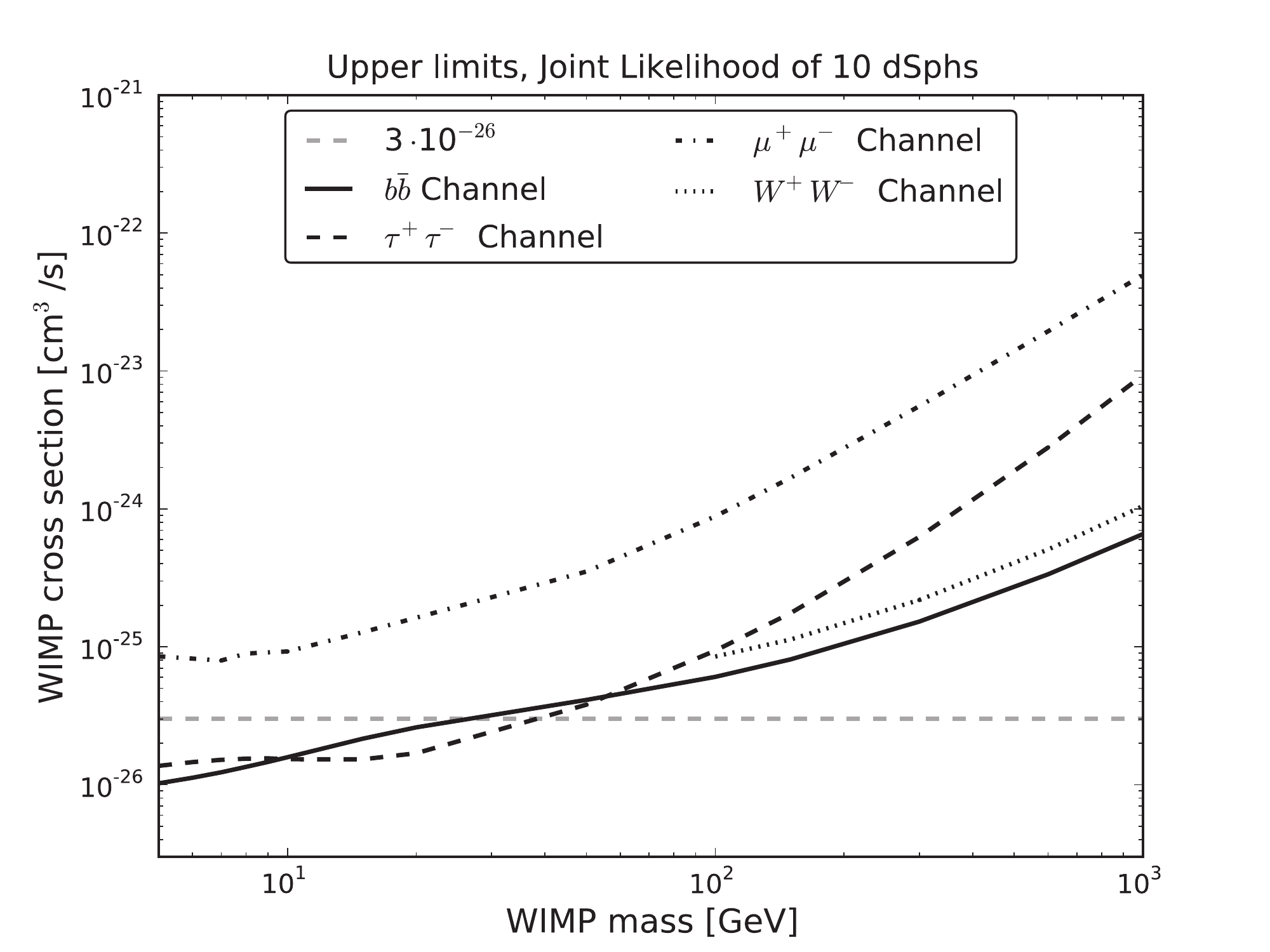}
\caption{Fermi 95\% confidence level upper limits on WIMP annihilation
cross section for $b\bar{b}$, $W^+W^-$, $\mu^+\mu^-$ and $\tau^+\tau^-$
channels which induce continuous $\gamma$-ray spectra. From \cite{Ackermann:2011wa}.
}
\label{fig:dwarf}
\end{figure}

Using the first $11$ month data of dwarf galaxies, the Fermi-LAT
collaboration found no indication of $\gamma$-ray emission from these
objects and strong constraints on the DM annihilation cross section were
given \cite{Abdo:2010ex}. With accumulation of the exposure, the limits
were significantly improved, through additionally a joint analysis
of many dwarf galaxies by Fermi-LAT collaboration and others
\cite{Ackermann:2011wa,GeringerSameth:2011iw}. Fig. \ref{fig:dwarf}
shows the $95\%$ confidence level upper limits on the cross section
$\langle\sigma v\rangle$ for selected annihilation channels derived in
\cite{Ackermann:2011wa}. In such an analysis, the uncertainty of the
DM density distribution in individual dwarf galaxy is also taken into
account in the likelihood fitting. It is shown that for DM with
mass less than $\sim30$ GeV and annihilation channels $b\bar{b}$ and
$\tau^+\tau^-$, the generic cross section of WIMPs which were thermally
produced in the early Universe and match the correct relic density,
$\langle\sigma v\rangle\sim3\times10^{-26}$ cm$^3$ s$^{-1}$, can be
ruled out by the $\gamma$-ray observations.

In \cite{Tsai:2012cs} an update of the constraints to the 4-year Fermi-LAT
data was given. In this analysis, a DM model-independent likelihood map,
on the ``$E_{\rm bin}-$flux'' plane, of the Fermi-LAT data was derived
through binning the data in multiple energy bins. Then the total
likelihood of any ``signal'' spectrum can be easily obtained with the
likelihood map. This method was tested to give consistent results with
that done using the Fermi Scientific
Tool\footnote{http://fermi.gsfc.nasa.gov/ssc/data/analysis/software/}
(same as in \cite{Ackermann:2011wa}). It was found that the 4-year
result \cite{Tsai:2012cs} was comparable with the 2-year one
\cite{Ackermann:2011wa} shown above for $m_\chi\lesssim20$ GeV and
even weaker by a factor of $2-3$ for higher mass DM particles.
Similar conclusion was also reported recently in the Fermi international
Symposium \cite{Drlica-Wagner:2012}. Such a result could be due to the
statistical fluctuations and different event classifications in those
analyses \cite{Drlica-Wagner:2012}. In addition the photon yield spectrum
also have a factor of $2-3$ difference between different simulation
codes \cite{Corcella:2000bw,Sjostrand:2006za,Sjostrand:2007gs}.
Finally, the substructures in the dwarf galaxies, although generally
not important, would contribute a factor of several to the uncertainty
based on the numerical simulation \cite{Gao:2011rf}. Therefore we should
keep in mind that the uncertainty of the constraints from $\gamma$-ray
observations of dwarf galaxies, in spite that it is less affected by the
DM density profile \cite{Charbonnier:2011ft}, is still a factor of
$\sim10$.

\subsubsection{Galaxy clusters}

Galaxy clusters are another good type of targets to search for DM
signal. There is currently no $\gamma$-ray emission found from galaxy
clusters, and stringent limits on the $\gamma$-ray emission were set by
Fermi-LAT \citep{Ackermann:2010qj}. Due to the low background emission and
the large amount of DM, galaxy clusters can also give effective constraints
on the DM model parameters \cite{Ackermann:2010rg,Yuan:2010gn,Dugger:2010ys,
Ke:2011xw,Huang:2011xr,Zimmer:2011vy,Ando:2012vu,Combet:2012tt,Nezri:2012tu,
Han:2012uw}.

\begin{figure}[!htb]
\centering
\includegraphics[width=\columnwidth]{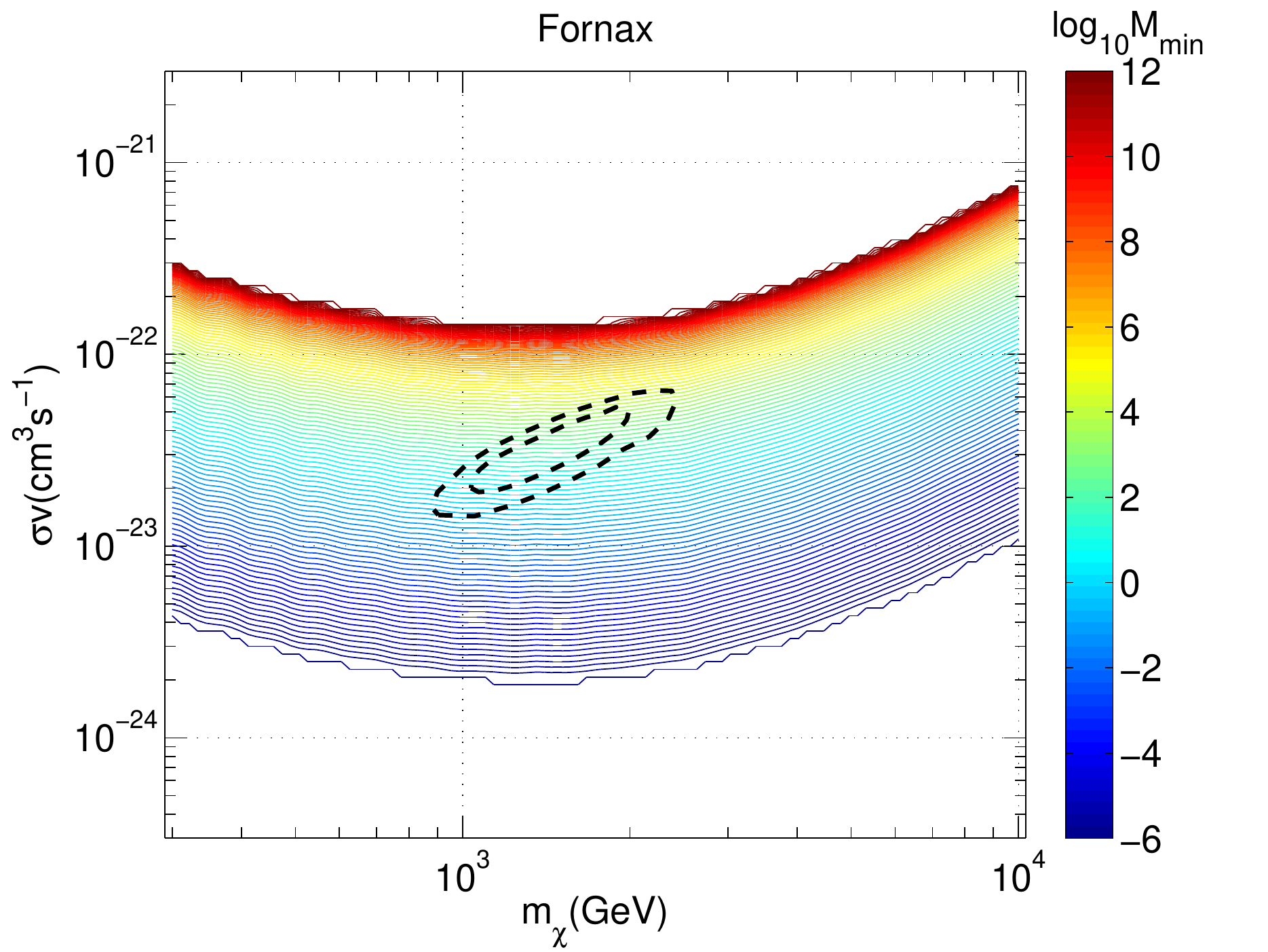}
\caption{The 95\% upper limits of DM annihilation cross section to
$\mu^+\mu$ from Fermi-LAT observations of galaxy cluster Fornax.
Different colors show the results for different values of the minimal
subhalo mass. The circles show the ``required'' parameter region to
fit the Milky Way $e^{\pm}$ excesses \cite{Meade:2009iu}.
From \cite{Yuan:2010gn}.
}
\label{fig:cluster}
\end{figure}

For the DM annihilation in the galaxy clusters, a large uncertainty comes from
the substructures in the clusters. We know there are substructures at
least down to the mass scale of dwarf galaxies, but it is not clear
whether the substructures can be extrapolated down to very low masses
(e.g., $10^{-6}$ M$_{\odot}$ as expected for typical CDM) which
are beyond the resolution limit of observations and numerical
simulations. Based on the direct extrapolation of simulation results
of CDM, the boost factor arises from substructures for typical
clusters can reach $10^{3}$ with significant variation for cutoff mass
$10^{-6}$ M$_{\odot}$ \cite{Gao:2011rf}. Fig. \ref{fig:cluster}
illustrates the uncertainties of the constraints on DM models 
from substructures \cite{Yuan:2010gn}. With conservative consideration of the
substructures down to dwarf galaxy mass, the constraint on DM annihilation
cross section is weaker than that derived from dwarf
galaxies \cite{Ackermann:2010rg}.

Another issue of searching for DM signal from galaxy clusters is
that the targets should be taken as extended sources. For some nearby
clusters, the virial radii reach several degrees, well beyond the
resolution angle of Fermi-LAT detector for $E>$GeV. In \cite{Han:2012au}
it was found that the spatial extension of the clusters would play
a crucial role in searching for the potential signals. A marginal
detection of extended $\gamma$-ray emission from Virgo cluster was
found assuming a spatial template of DM annihilation with significant
boost of substructures \cite{Han:2012au}. Although it was then
suggested that new point sources which were not included in the Fermi
catalog would contribute a fraction to the ``signal''
\cite{MaciasRamirez:2012mk,Han:2012uw}, it is still necessary to
pay attention to the spatial extension of the sources when doing
similar searches.

Finally we comment that the galaxy clusters are powerful to probe
decaying DM \cite{Dugger:2010ys,Ke:2011xw,Huang:2011xr,Combet:2012tt}.
Since the $\gamma$-ray flux induced by decaying DM is proportional to
the total mass of DM in the cluster, the uncertainty of the
constraint on decaying DM scenario arising from the DM substructure model is much smaller, and robust results
could be derived. It was shown that the constraint on the DM lifetime
from galaxy clusters was generally stronger than that from dwarf
galaxies and nearby galaxies \cite{Dugger:2010ys}.

\subsubsection{Star clusters}

The Milky Way globular clusters, defined as spherical ensembles of
stars that orbit the Galaxy as satellites, are potential targets
for the indirect detection of DM. Although observationally there is
in general no significant amount of DM in the globular clusters
\cite{Baumgardt:2009hv,Lane:2010rd,Conroy:2010bs}, the adiabatic
contraction process in the cosmological formation context
\cite{Peebles:1984az} due to the infall of baryons will give birth
to a high density spike of DM and can in principle result in a high
annihilation rate of DM.

In \cite{Feng:2011ab} the authors used the Fermi-LAT data of two
globular clusters, NGC 6388 and M 15 which favor the astrophysical
origin, to search for the DM signals. Strong $\gamma$-ray emission
from NGC 6388 was reported, which was generally thought to come from
the population of millisecond pulsars \cite{Abdo:2010bb}. No
$\gamma$-ray emission from M 15 was found. The constraints on
the DM annihilation cross section were derived \cite{Feng:2011ab}.
Compared with the constraints from dwarf galaxies, the globular
clusters could give even stronger constraints. However, there are
very large systematic uncertainties for such a study, from the
hypothesis of the origin of the globular clusters, the interaction
between DM and baryons during the cluster evolution and the modification
of DM profile due to adiabatic growth of the intermediate mass black
hole in the center of the clusters.

\subsubsection{Galactic center}

\begin{figure*}[!htb]
\centering
\includegraphics[width=\columnwidth]{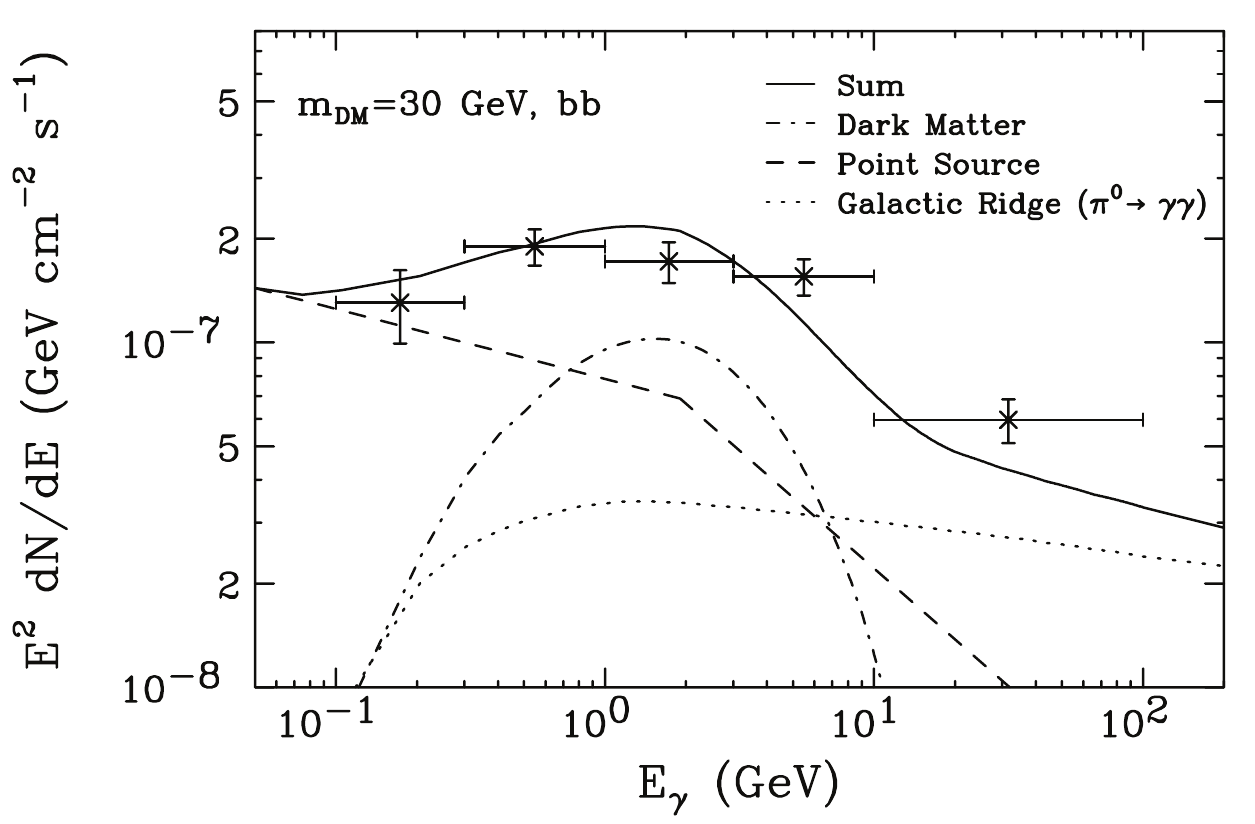}
\includegraphics[width=\columnwidth]{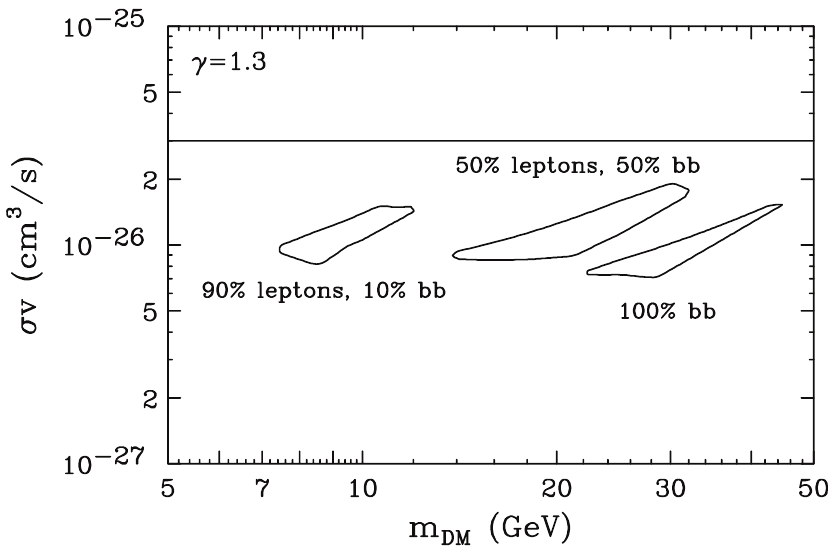}
\caption{Left: spectrum of the excess $\gamma$-rays in the Galactic
center and an illustration to employ a point source component, diffuse
component from the Galactic ridge together with a DM component to
explain the data. Right: fitting parameter space for DM annihilation
scenario to account for the excess for different assumptions of DM
annihilation channels. From \cite{Hooper:2011ti}.
}
\label{fig:gc}
\end{figure*}

The advantage of the Galactic center as the target for $\gamma$-ray
detection of DM is that it is nearby with expected high density of DM.
Although the density profile in small region around the Galactic center
is not clear, the average annihilation $J$-factor, which is defined as
the line-of-sight integral of DM density square averaged within a
solid angle, in a relatively large region of the Galactic center is
generally much higher than those of dwarf galaxies and cluster of
galaxies \cite{Cirelli:2009vg,Ackermann:2011wa,Ackermann:2010rg}.
However, the disadvantage is that the Galactic center is a so complex
astrophysical laboratory that the background $\gamma$-ray emission is
very strong and far from a clear understanding. Thus the problem is to
distinguish the potential signal, if any, from the background emission.

Through analyzing the Fermi data, one group reported the extended
$\gamma$-ray excess in the Galactic center which could be consistent
with a DM annihilation origin with mass $10$s GeV \cite{Goodenough:2009gk,
Hooper:2010mq,Hooper:2011ti}. Such a claim was confirmed by other groups
with independent analyses\footnote{Note in \cite{Boyarsky:2010dr} the
conclusion was against the existence of the extra ``signal'' from DM
annihilation, but in their analysis they did find that including an
additional source consistent with DM annihilation the fit to the data
was significantly improved.} \cite{Boyarsky:2010dr,Abazajian:2012pn}.
Fig. \ref{fig:gc} shows the extracted spectrum of the extended
$\gamma$-ray excess in the Galactic center and the required parameter
region if DM annihilation is adopted to explain it \cite{Hooper:2011ti}.
The DM density profile is adopted to be a generic Navarro-Frenk-White
(NFW, \cite{Navarro:1996gj}) profile with inner slope $\alpha=1.3$
in order to be consistent with the spatial distribution of the excess
$\gamma$-rays. It is interesting to note that the mass of the DM particle,
which is fitted to be about tens of GeV, may be consistent with the suspected
signals from direct detection experiments DAMA/LIBRA \cite{Bernabei:2010mq},
CoGeNT \cite{Aalseth:2010vx} and CRESST \cite{Angloher:2011uu}.

Nevertheless, because the Galactic center is so complicated, it is still
very difficult to distinguish the DM scenario of such an excess from the
potential astrophysical sources such as cosmic rays from the supermassive
black hole \cite{Goodenough:2009gk,Hooper:2010mq}, or a population of
millisecond pulsars \cite{Abazajian:2010zy}. It should be also
cautious that the understanding of the diffuse background emission
close to the Galactic center is poor and the contamination of the
diffuse background to this excess is still possible.

Conservatively one can set upper limits on the DM model parameters
with the observational data \cite{Hooper:2011ti,Bernal:2010ip,
Ellis:2011du,Huang:2012yf,Hooper:2012sr}. For DM mass less than
$\sim100$ GeV, the upper limit of the annihilation cross section was
found to be close to or lower than the natural value $3\times10^{-26}$
cm$^3$ s$^{-1}$ \cite{Hooper:2011ti,Huang:2012yf}. Such a limit is at
least comparable to that derived from other observations such as the
dwarf galaxies.

\subsubsection{Milky Way halo and subhalos}

More works are trying to search for the DM signal from $\gamma$-ray
observations of the Milky Way halo
\cite{Zhang:2009kp,Cirelli:2009dv,Papucci:2009gd,Zhang:2009ut,
Zaharijas:2010ca,Baxter:2011rc,Ackermann:2012qk,Ackermann:2012rg},
since the signal-to-noise ratio will be even higher in the halo than
that in the Galactic center \cite{Serpico:2008ga}. Furthermore,
the result from the Milky Way halo is less sensitive to the DM density
profile which is highly uncertain in the inner Galaxy.
Using 2-year Fermi-LAT data of the whole Milky Way halo excluding the
Galactic plane, the LAT collaboration derived constraints on the DM
annihilation cross section or decay lifetime for a wide range of final
states, conservatively through comparing the data with the expected
signal from DM \cite{Ackermann:2012qk}. The constraints are relatively
weak, however. With optimizing analysis regions, and improving diffuse
backgrounds, the LAT collaboration updated the results and gave more
stringent constraints on the DM model parameters \cite{Ackermann:2012rg}.
The uncertainties of the astrophysical diffuse background which were
poorly constrained, e.g., the diffusive halo height, the cosmic ray
source distribution, the injection spectrum index of electrons, and the
dust to gas ratio of the interstellar medium, were also taken into account
using a profile likelihood method \cite{Ackermann:2012rg}. The results
were found to be competitive with other probes like dwarf galaxies and
galaxy clusters.

Subhalos are expected to widely exist in the Milky Way halo, in the
CDM structure formation pattern. Less affected by the tidal
stripping effect of the main halo, subhalos could be more abundant
in the large halo away from the Galactic center. The dwarf galaxies
are part of the subhalos in the Milky Way. Besides the dwarf galaxies
we might expect the existence of DM-only subhalos which are not
visible with current astronomical observations. Such DM-only subhalos
could be $\gamma$-ray emitters if the flux is high enough
\cite{Kuhlen:2008aw,Springel:2008zz}. Several works tried to investigate
the possible connection between the Fermi unassociated sources and the
DM subhalos \cite{Buckley:2010vg,Zechlin:2011kk,Belikov:2011pu,
Ackermann:2012nb,Zechlin:2012by,Belotsky:2012ua}. The $\gamma$-ray
emission from DM subhalos should be non-variable, spatially extended and
spectrally hard. Applying these criteria to the unassociated Fermi sources,
the LAT collaboration found that most of the sources did not pass the
cuts, except two candidates \cite{Ackermann:2012nb}. However, further
analyses identified the rest two sources to a pulsar and two active
galactic nuclei respectively \cite{Ackermann:2012nb}. Such a result
could be interpreted in the context of structure formation of the
$\Lambda$CDM scenario. Using slightly different critera together
with the multi-wavelength observations, another group reached a similar
conclusion that no DM subhalo in the Fermi unassociated source
catalog could be identified now \cite{Zechlin:2011kk,Zechlin:2012by}.

The population of low mass subhalos, unresolved in the current numerical
simulations, could contribute to the diffuse $\gamma$-ray emission of
DM annihilation in the halo \cite{Zhang:2009kp,Baxter:2010fr,
Blanchet:2012vq}. The contribution of subhalos suffers from large
uncertainties of the structure parameters extrapolated according
to the numerical simulations. Roughly speaking for typical CDM scenario
the boost factors of the $\gamma$-ray signal could be several to several
tens, depending on the directions \cite{Blanchet:2012vq}. Thus the
constraints from the Milky Way halo could be even stronger considering
the contribution of subhalos.

Besides the flux of the $\gamma$-ray emission, the spatial morphology
can also provide useful identification of the DM signals. This
includes the large scale morphology \cite{Dixon:1998fk,Hooper:2007be,
Berezinsky:2006qm,Pieri:2009je,Ibarra:2009nw} and the statistical
properties at small scales \cite{SiegalGaskins:2008ge,Fornasa:2009qh,
Ando:2009fp,Cuoco:2010jb,Malyshev:2010zzc,Zhang:2010mg,Fornasa:2012gu}.
It was shown that even the DM annihilation contributed only a small
fraction of the diffuse background, the anisotropy detection could
have the potential to identify it from the background (e.g.,
\cite{Ando:2009fp}).

\subsubsection{Extragalactic gamma-ray background}

The extragalactic $\gamma$-ray background (EGRB) is a probe to explore
the accumulation of the DM evolving in the whole history of the Universe.
Measurement of the EGRB by Fermi-LAT showed a structureless power-law
from $200$ MeV to $100$ GeV \cite{Abdo:2010nz}. Most recently the
analysis extended up to $400$ GeV with result being basically consistent
with the previous published one \cite{Ackermann:2012}. The DM model
is in general difficult to produce such a single power-law spectrum
in a very wide energy range. Therefore the EGRB is widely employed to
constrain the DM model parameters \cite{Yuan:2009xq,Abdo:2010dk,
Abazajian:2010sq,Hutsi:2010ai,Arina:2010rb,Abazajian:2010zb,Zavala:2011tt,
Yuan:2011yb,Calore:2011bt,Yang:2011eg,Murase:2012xs,Cirelli:2012ut}.

The major uncertainty of the expectation of DM contribution to the EGRB
flux is the clumpiness enhancement factor for annihilating DM scenario.
For different assumptions of the density profile of each halo, the halo
mass/luminosity function, the cutoff mass of the minimal halo and/or the
concentration-mass relation, the clumpiness enhancement factor can differ
by several orders of magnitude (e.g., \cite{Abdo:2010dk}). The most
conservative limits can barely reach some models interested in the
community such as those proposed to explain the positron/electron
excesses \cite{Abdo:2010dk}. The EGRB has the potential to probe
larger range of the parameter space, but it depends on more precise
knowledge about the DM structure formation and the astrophysical
contribution to the EGRB.

\begin{figure}[!htb]
\centering
\includegraphics[width=\columnwidth]{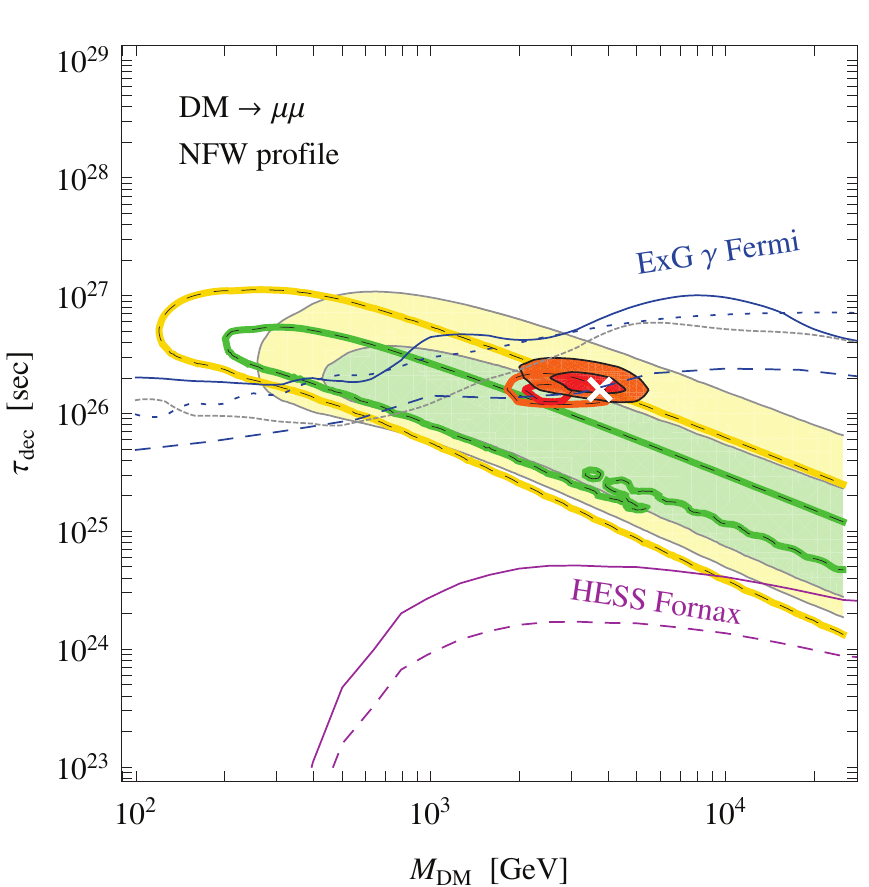}
\caption{Constraints on the DM lifetime with Fermi EGRB data and
HESS observation of galaxy cluster Draco. From \cite{Cirelli:2012ut}.
}
\label{fig:eg_mu_decay}
\end{figure}

For decaying DM scenario, the constraints are more robust due to the
less effect of structures \cite{Hutsi:2010ai,Cirelli:2012ut}. It was
shown that the EGRB measured by Fermi-LAT could exclude almost all
the parameter regions of DM models with two-body channels to account
for the cosmic ray positron/electron excesses (Fig. \ref{fig:eg_mu_decay},
\cite{Cirelli:2012ut}). For other decaying channels such as $b\bar{b}$,
the constraint from EGRB is also among the most stringent ones
\cite{Cirelli:2012ut}.

The statistical anisotropy of extragalactic DM annihilation may have
distinct behavior from the astrophysical sources and can be used to
detect the DM signal from EGRB \cite{Ando:2006cr,Cuoco:2007sh,
Taoso:2008qz}. It was proposed that the energy dependence of the
anisotropy of EGRB could be another interesting observable which might
be more sensitive to identify the DM signal \cite{SiegalGaskins:2009ux,
Hensley:2009gh,Hensley:2012xj}. With the intensity and anisotropy energy
spectra, different components of the diffuse background can be decomposed
model-independently \cite{Hensley:2012xj}. A positive detection of the
angular power for $155\leq l\leq504$ with Fermi-LAT data was reported
\cite{Ackermann:2012uf}. The measured angular power is approximately
independent with scale $l$, which implies that it originates from the
contribution of one or more unclustered source populations. Furthermore,
the lack of strong energy dependence of the amplitude of the angular
power normalized to the mean intensity in each energy bin indicates that
a single source class may dominate the contribution to the anisotropy
and it provides a constant fraction to the intensity of the EGRB
\cite{Ackermann:2012uf}. The result is consistent with the blazar
origin of the EGRB \cite{Ackermann:2012uf}, which implies the lack
of a signal from DM.

\subsubsection{Line emission}

The monochromatic $\gamma$-ray emission is usually called as the ``smoking
gun'' diagnostic of the DM signal \cite{Bergstrom:1988fp,Rudaz:1989ij,
Bergstrom:1997fh,Bern:1997ng,Ullio:1997ke}. The DM particles may annihilate
into a pair of photons or a photon plus another gauge boson or Higgs
boson, resulting quasi-monochromatic $\gamma$-ray emissions. In addition
the internal bremsstrahlung photons produced when DM annihilating into
charged particles can also give prominent spectral features which will
mimic the line emission \cite{Bergstrom:2004cy,Birkedal:2005ep,
Bergstrom:2005ss,Bringmann:2007nk}.

\begin{figure}[!htb]
\centering
\includegraphics[width=1.02\columnwidth]{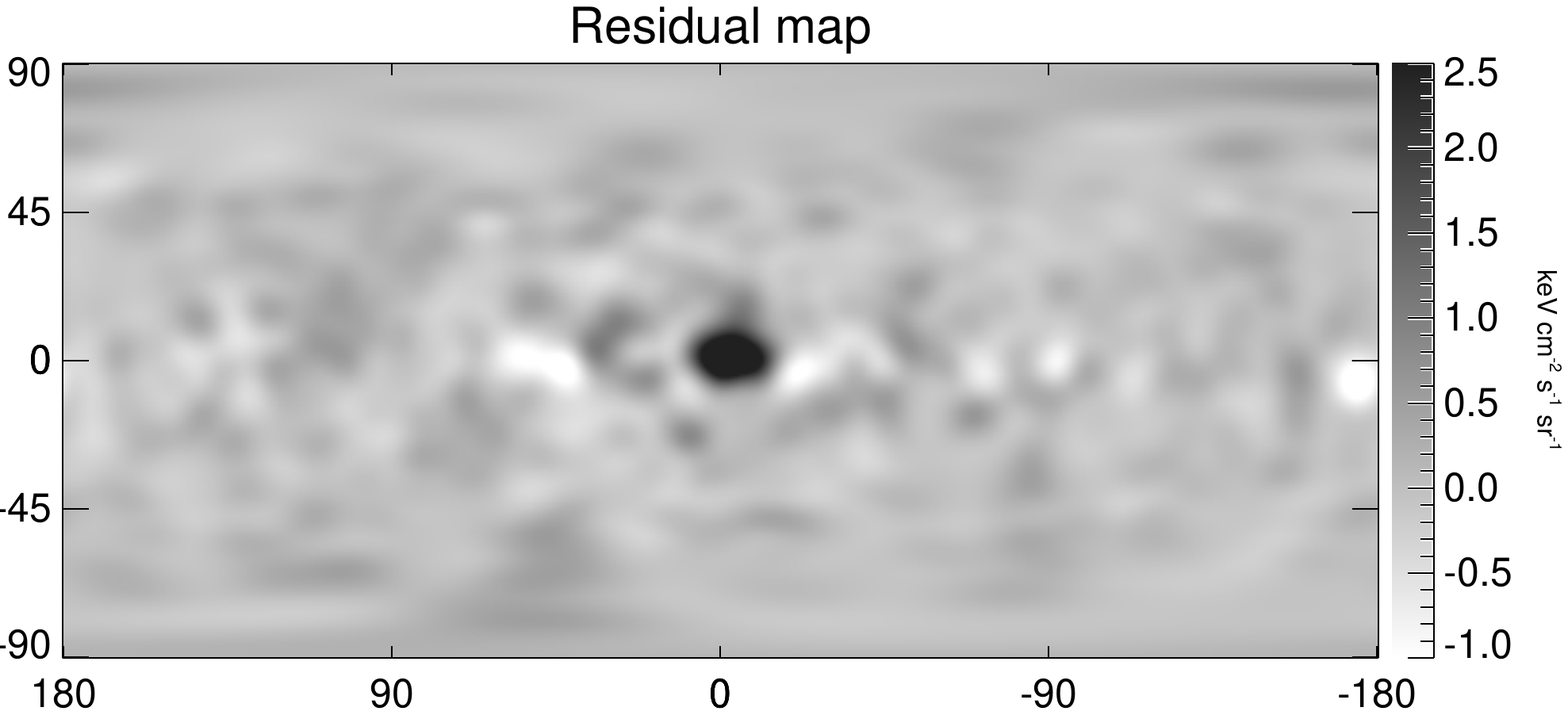}
\includegraphics[width=0.95\columnwidth]{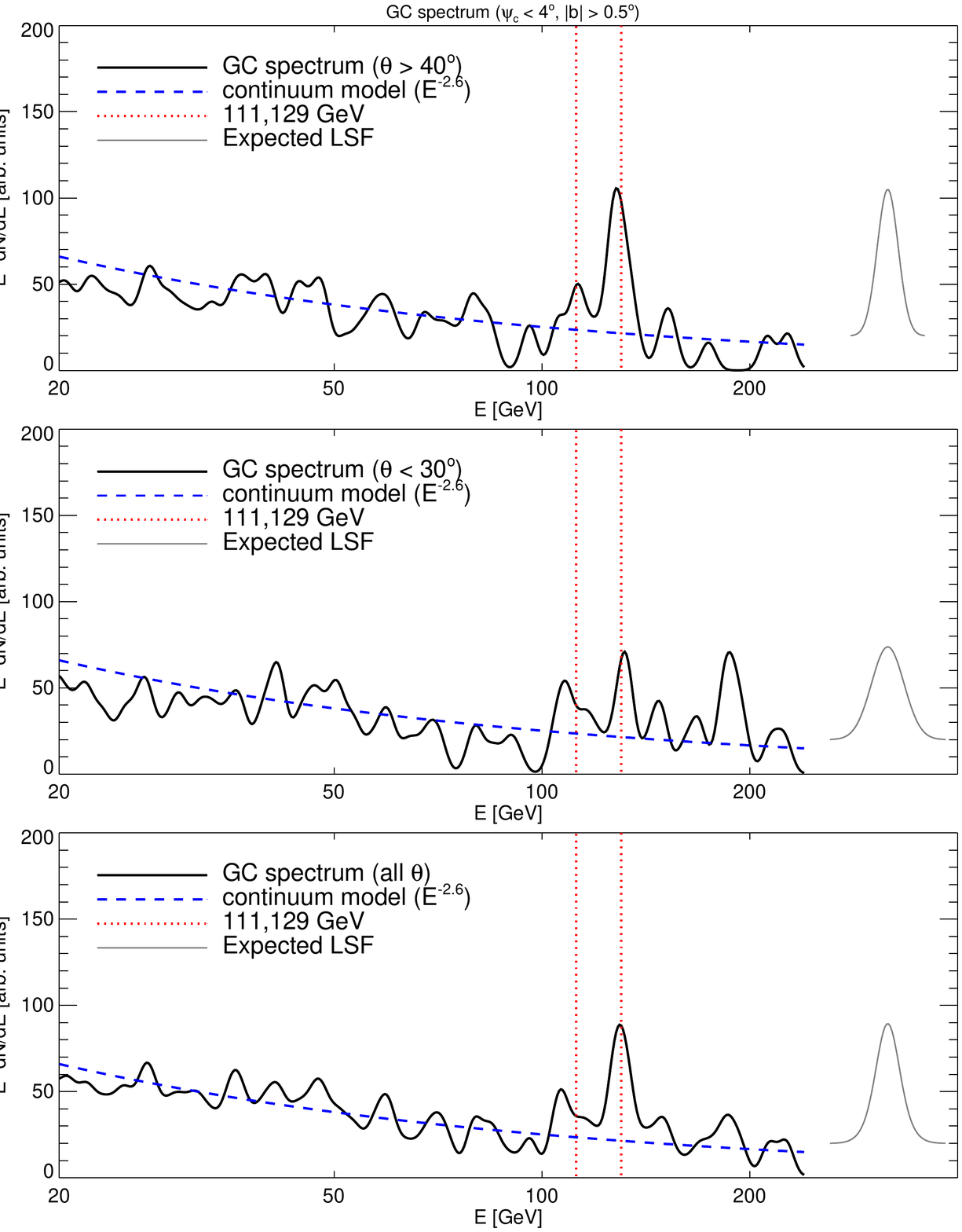}
\caption{Upper: the residual map of $120-140$ GeV photons with
subtraction of the maps of nearby energies: $80-100$ GeV, $100-120$
GeV, $140-160$ GeV and $160-180$ GeV. Lower three panels: spectra
of the emission within $4^{\circ}$ of the cusp center $(l,b)=
(-1.5^{\circ},0)$ excluding $|b|<0.5^{\circ}$, for large incidence
angle (second), small incidence angle (third) and total events
(bottom). From \cite{Su:2012ft}.
}
\label{fig:line}
\end{figure}

With the purpose of searching for sharp spectral features, one group
found a weak indication of the ``signal'' around $\sim130$ GeV
in the public Fermi-LAT data with global significance $\sim3\sigma$,
which may be consistent with either an internal bremsstrahlung like
signal or a $\gamma$-ray line \cite{Bringmann:2012vr,Weniger:2012tx}.
This ``signal'' was confirmed in the following studies
\cite{Tempel:2012ey,Su:2012ft}, and the significance could become
higher if a $\sim1.5^{\circ}$ offset of the ``signal'' region to the
Galactic center was included \cite{Su:2012ft}. The basic features
of the tentative ``signal'' are:
\begin{itemize}

\item Spatially: extended; concentrated in the Galactic center with
peak position slightly offset from the central supermassive black
hole; consistent with Einasto or cuspy NFW (inner slope
$\alpha\approx1.2$) profile.

\item Spectrally: sharp feature at $\sim130$ GeV with no significant
spread; possibly a second line at $\sim111$ GeV; sharper for larger
incidence angle events whose energy resolution is higher.

\end{itemize}
Fig. \ref{fig:line} shows the spatial and spectral results of the
line emission derived in \cite{Su:2012ft}.

Note there are still some discrepancies of the results of both the
spatial and spectral properties. The spatial distribution of the
$120-140$ GeV photons was found elsewhere other than the Galactic
center, with lower significance \cite{Tempel:2012ey,Boyarsky:2012ca}.
Similar spectral features at other energies along the Galactic plane
were also shown \cite{Boyarsky:2012ca}. The most recent results
reported by the LAT collaboration, with improvement of the energy
resolution and reprocessed data with corrected energy scale, confirmed
these complex features and found lower global significance $<2\sigma$
of the ``signal'' around $135$ GeV \cite{Albert:2012}.

It was further reported that there might be another hint of $130$ GeV
line emission from the accumulation of data from several nearby galaxy
clusters \cite{Hektor:2012kc}. However, the arrival directions of
the photons with highest significance are within $5^{\circ}$ or
$6^{\circ}$ cone around each cluster, which are beyond the virial
radii of all these clusters. The likelihood fit based on the DM
density profile (together with substructures) of each cluster actually
found no significant signal \cite{Huang:2012yf}. The analysis of
dwarf galaxies showed no signal of line emission either
\cite{GeringerSameth:2012sr}.

The analysis of the unassociated sources in the second Fermi-LAT
catalog (2FGL, \cite{Nolan:2011bm}) suggested a local $\sim3\sigma$
evidence of the existence of double line with energies consistent
with that found in the inner Galaxy \cite{Su:2012zg}. However, the
overall $\gamma$-ray spectra of most of these unassociated sources
which have potential line photons show distinct shape from that expected
from DM annihilation \cite{Hooper:2012qc}. It turns out that most of
those unassociated sources should not be DM subhalos. Another
independent study argued that the double lines from unassociated sources
could not be identified in the present statistics and energy resolution
of Fermi-LAT and the ``signal'' should be an artifact of the applied
selection criteria \cite{Hektor:2012jc}.

It is very important to consider the possible instrument systematics
of this line ``signal'' \cite{Whiteson:2012hr,Hektor:2012ev,
Finkbeiner:2012ez}. The detailed study with currently available
information of the detectors showed no systematical differences
between the $130$ GeV photons and photons with other energies
\cite{Whiteson:2012hr}. The photons from the Galactic center region
also show no systematical differences from those of other sky regions
\cite{Finkbeiner:2012ez}. A marginally significant line feature at
$E\sim130$ GeV in the photons of the Earth limb, which is produced
by the collisions between cosmic rays and the atmosphere, was found
within a limited range of detector incidence angles
\cite{Hektor:2012ev,Finkbeiner:2012ez,Albert:2012}. Such a result
raises concerns about the line signal found in the inner Galaxy.
However, it is not easy to be understood with any plausible cause
of the instrument behavior \cite{Finkbeiner:2012ez}.

Described above are the current observational status of the $130$
GeV line emission. We are not clear whether it is real or not at
present from the Fermi-LAT data only. Nevertheless, there are
many discussions on the theoretical implication of such a line in
the sky, most of which focus on the DM models
\cite{Aharonian:2012cs,Profumo:2012tr,Dudas:2012pb,Cline:2012nw,
Choi:2012ap,Kyae:2012vi,Lee:2012bq,Rajaraman:2012db,Acharya:2012dz,
Buckley:2012ws,Chu:2012qy,Das:2012ys,Kang:2012bq,Weiner:2012cb,
Oda:2012fy,Park:2012xq,Tulin:2012uq,Li:2012jf,Cline:2012bz,Bai:2012qy,
Bergstrom:2012bd,Wang:2012ts,Fan:2012gr,Lee:2012wz,Baek:2012ub,
Shakya:2012fj,D'Eramo:2012rr,SchmidtHoberg:2012ip,Farzan:2012kk,
Chalons:2012xf,Rajaraman:2012fu,Bai:2012yq,Zhang:2012da,Bai:2012nv,
Lee:2012ph}. If DM annihilation is responsible for the
line ``signal'', the required annihilation cross section is
estimated to be $\sim10^{-27}$ cm$^3$ s$^{-1}$ for NFW or Einasto
profile \cite{Weniger:2012tx}. Such a cross section seems
too large for typical DM annihilation into a pair of photons
through a loop, due to the lack of strong continuous $\gamma$-ray
emission as expected from the tree level contribution
\cite{Buchmuller:2012rc,Cohen:2012me,Cholis:2012fb,Huang:2012yf}.
On the other hand if the tree level annihilation is suppressed or
forbidden, then we may need to finely tune the model parameters to
reconcile with the relic density of DM \cite{Buckley:2012ws,
Tulin:2012uq}. Other studies to constrain the DM models of the
monochromatic line emission include the electron/positron spectra
\cite{Feng:2012gs}, radio data \cite{Laha:2012fg} and antiprotons
\cite{Asano:2012zv}. Note that these constraints are not directly
applicable on the line emission.

The offset of the $\gamma$-ray peak from the central black hole serves
another challenge to the DM interpretation. However, in the case of low
statistics, the fluctuation could naturally explain such an offset
\cite{Yang:2012ha,Rao:2012fh}. Numerical simulation also suggest
that an offset of several hundred parsec is generally plausible
\cite{Kuhlen:2012qw}. Note in \cite{Gorbunov:2012sk} it was pointed
out that the density cusp of DM could not survive the tidal force
of the Milky Way given an off set of $\sim1.5^{\circ}$. A caveat
is that the above estimate is based on the assumption of static
equilibrium of the DM density profile.

Due to the potential importance of the $\gamma$-ray line for
physics and astrophysics, it is very important to test it with
other independent measurements. The currently operating experiment
on the International Space Station (ISS), Alpha Magnetic Spectrometer
2 (AMS-02), can measure the electrons and photons up to TeV with
an energy resolution of $2-3\%$ \cite{ams02}. There is another on-going
mission, CALorimetric Electron Telescope (CALET), which is planned
to be placed on the ISS around 2014, has an energy resolution of
$2\%$ for photons with energies higher than $100$ GeV \cite{calet}.
The geometry factors of AMS-02 and CALET for photons are much smaller
than that of Fermi-LAT, and it will need much longer time to have
enough statistics to test the line emission. A Chinese spatial
mission called DArk Matter Particle Explorer (DAMPE) which is
planned to be launched in 2015, may have large enough geometry
factor ($\sim0.6$ m$^2$ sr) and high enough energy resolution
($1-1.5\%$ at $100$ GeV) \cite{Chang:DAMPE2011}. It is possible
for DAMPE to test this line emission with one to two year
operation \cite{Li:2012qg}. A recently available test may come
from the ground based Cerenkov telescopes HESS II \cite{Becherini:2009zz}.
For $50$ hour exposure of the Galactic center region by HESS II
a $5\sigma$ detection of the Fermi-LAT 130 GeV line can be reached
\cite{Bergstrom:2012vd}. The detectability or exclusion power will
be much higher for the Cerenkov Telescope Array (CTA,
\cite{Consortium:2010bc}) project \cite{Bergstrom:2012vd}.

\subsection{Ground based telescopes}

In this subsection we briefly compile the results (limits) of DM searches
with the ground based VHE $\gamma$-ray detectors, especially from Cerenkov
telescopes. The threshold detection energy of the ground based atmospheric
Cerenkov telescopes is about tens to hundreds GeV, and they are most
sensitive for TeV photons. Therefore it will be more effective to probe
the heavy DM using the ground based telescopes.

The search for DM signal has been carried out by Whipple \cite{Wood:2008hx},
HESS \cite{Aharonian:2007km,Aharonian:2008wt,Aharonian:2008dm,
Abramowski:2010aa,Abramowski:2011hc,Abramowski:2011hh}, MAGIC
\cite{Aliu:2008ny,Aleksic:2011jx} and VERITAS \cite{Aliu:2012ga}.
The primary search targets are dwarf galaxies. Up to now no $\gamma$-ray
emission was found from the dwarf galaxies, even for the very deep
observations, and stringent upper limits of the $\gamma$-ray emission
could be set \cite{Aharonian:2007km,Aharonian:2008dm,
Abramowski:2010aa,Aliu:2008ny,Aleksic:2011jx,Aliu:2012ga}. The upper
limit of the DM annihilation cross section derived by the Cerenkov
telescopes can reach $10^{-24}$ cm$^3$ s$^{-1}$ for neutralino DM
\cite{Aharonian:2007km}, and will be better than that given by Fermi-LAT
for $m_{\chi}>$TeV. Better constraint comes from the observations
of the Galactic center region \cite{Abramowski:2011hc}. However, the
uncertainty from the density profile becomes larger.

The future experiment CTA will improve the sensitivity of VHE
$\gamma$-ray detection by an order of magnitude compared with the
current Cerenkov telescope arrays. It is expected to significantly
improve the capability of searching  for heavy DM \cite{Doro:2012xx}.

\section{Status of indirect detection -- neutrinos}

\subsection{High energy neutrino telescopes}

Unlike other products induced by DM, neutrinos have less trajectory defection and energy loss
during the propagation due to the weak interaction. Therefore neutrinos may carry
the information of the property and distribution of DM. For the same reason, neutrinos are more difficult to be
detected compared with charged particles and photons. For a review of the high energy neutrino telescopes, see Ref. \cite{Hoffman:2008yu}.

Neutrinos can only be observed indirectly through the charged leptons induced by neutrinos interacting
with nuclei inside/outside the detector. These secondary leptons, such as electrons or muons, with
high energy will emit Cerenkov radiation when they penetrate in the detector. Since the
secondary lepton carries almost all the energy of neutrino and only has small trajectory
defection from the original direction of neutrino, the information of neutrino can be
well reconstructed via the Cerenkov emissions. The telescope can also detect the cascade showers induced by
electron neutrinos and tau neutrinos, and by neutrino-nucleon scatterings
via neutral current interactions \cite{:2012zk}. However, the efficiency of such detection is much lower than
detecting Cerenkov emissions.

The high energy neutrino telescopes, such as Super-Kamiokande (Super-K) \cite{Tanaka:2011uf}, ANTARES \cite{Lambard:2011zza} and IceCube \cite{Aartsen:2012ef,Abbasi:2012ws}, are located in the deep underground, water and ice to be shielded from high energy cosmic ray backgrounds. The water or ice can be used as Cerenkov radiator for high energy muons. In order to improve the detection capability, the volume of telescope should be very large. Because high energy muons can propagate a long distance, the telescope may observe the muons produced outside the detector. Such effect enlarges the effective volume of the detector. For the same reason, in order to reduce high energy atmospheric muon background, the telescope observes the up-going muons induced by the up-going neutrinos which travel through the earth.

The final muon event rate at the detector can be given by \cite{Halzen:2009vu} (for the calculation considering the energy dependent muon flux, see Ref. \cite{Erkoca:2009by,Wikstrom:2009kw})
\begin{eqnarray}
\phi_\mu & \simeq &\int^{m_\chi}_{E_{th}} dE_\mu \int dE_{\nu_\mu} \frac{dN_{\nu_\mu}}{dE_{\nu_\mu}} \left[ \frac{d\sigma_{cc}^{\nu p}}{dE_{\nu_\mu}} n_p + \frac{d\sigma_{cc}^{\nu n}}{dE_{\nu_\mu}} n_n \right]  \nonumber \\
&\times& (R(E_\mu)+L) A_{eff} + (\nu \to \bar{\nu})
\end{eqnarray}
where $n_p$($n_n$) is the number density of protons(neutrons) in matter around the detector, the muon range $R(E_\mu)$ denotes
the distance that a muon could travel in matter before its energy drops below the detector's threshold energy $E_{th}$, $L$ is the
depth of the detector, $A_{eff}$ is the detector's effective area for muons which depends on the muon energy, the notation $(\nu \to \bar{\nu})$ denotes that the anti-neutrino flux is also taken into account. $d\sigma^{\nu p}_{cc}/dE_\mu$ is the the cross section of deep inelastic neutrino-nucleon scattering which produces muons via charged current
interactions. $dN_{\nu_\mu}/dE_{\nu_\mu}$ is the flux of the neutrinos induced by DM.

The irreducible backgrounds are the up-going atmospheric neutrinos
which are produced by the cosmic rays interacting with nuclei in
the atmosphere \cite{Honda:2006qj}. In fact, almost all the high
energy neutrinos observed at the neutrino telescopes are atmospheric
neutrinos \cite{Sullivan:2012hf}. The atmospheric neutrinos are
almost isotropic, while the neutrinos from DM are produced from
particular direction. The flux of atmospheric neutrinos decreases
as $\sim E_\nu^{-3.7}$, while the neutrinos from DM may have a
harder spectrum. Therefore, high angular and energy resolutions
of the telescope are essential to
extract the signals from the smooth backgrounds.

\subsection{Solar neutrinos from DM}
\label{solarneu}

When DM particles travel through a massive astrophysical object, such as the Sun (Earth), they may be gravitationally trapped and continuously lose energy by collisions with nuclei \cite{Faulkner:1985rm,Press:1985ug}. Once captured, DM particles may have large annihilation rate due to high number density \cite{Silk:1985ax}. The time evolution of the DM population in the Sun can be given by
\begin{equation}
\dot{N}=C_\odot- C_A N^2-C_E N
\end{equation}
where $C_\odot$ is the capture rate, $C_A \def \langle \sigma v \rangle /V_{eff}$ is the thermally averaged DM annihilation cross
section per volume, $C_E$ is the evaporation rate which is only significant for light DM. If the capture process and annihilation process reach equilibrium over a long time scale, the annihilation
rate is determined by capture rate as $\Gamma=\frac{1}{2} C_\odot$. The capture rate can be approximately given by \cite{Gould:1991hx} ( general discussions of DM capture rate for the Earth and the Sun can be found in Ref. \cite{Gould:1987ir,Jungman:1995df,Wikstrom:2009kw,Menon:2009qj})
\begin{eqnarray}
C_\odot & \sim & 10^{20} {\rm s}^{-1} \left( \frac{\rho_\chi}{0.3 \rm{GeV} \; {\rm cm}^{-3}} \right) \left( \frac{270 {\rm km}\; {\rm s}^{-1}}{\bar{v}} \right)^3 \nonumber \\
&\times&   \left( \frac{100 {\rm GeV}}{m_\chi} \right)^2 \left( \frac{\sigma_{SD}^{\chi H}+ \sigma_{SI}^{\chi H}+ \sum_i \xi_i \sigma^{\chi N_i} }{10^{-42} {\rm cm}^2}  \right)
\end{eqnarray}
where $\rho_\chi$ and $\bar{v}$ are the mass density and RMS velocity of DM in the solar system respectively. The contribution of the i-th nuclear species depends on the elastic scattering cross section between DM and the i-th nucleus $\sigma^{\chi N_i}$. 
It also depends on the mass fraction and distribution of the i-th nuclei in the Sun and the properties of the scattering
which can be presented by a numerical factor $\xi_i$ \cite{Gould:1991hx,Jungman:1995df}. Since the SI cross section between DM and nucleon has been stringent constrained by the direct detections, the most important contribution for the capture rate and thus for the annihilation rate may be from the SD scattering between the DM and hydrogen in the Sun (for the discussions of inelastic DM, see Ref. \cite{Nussinov:2009ft,Menon:2009qj,Shu:2010ta}).

If the products of DM annihilations are $e^+e^-$ or $\mu^+\mu^-$, they will not contribute to neutrino signals. For muons, the reason is they always lose most of energy before decay in the center of the Sun. For annihilation
channels into $\tau^+\tau^-$, $W^+W^-$, $ZZ$, $t\bar{t}$, they produce neutrinos via cascade decays
and the neutrino spectra for such channels are hard. For quark channels, since neutrinos are induced via hadron decays after hadronization process, the neutrino spectra are soft. Moreover, the light mesons lose energy easily before decay, therefore the contributions from light quarks to the neutrino signals are always small.

The high energy neutrinos produced at the solar center will interact with the nuclei before they escape from the Sun.
The effects include the neutral current interaction, the charged current interaction and tau neutrino $\nu_\tau$
re-injection from secondary tau decays. The other important effects are neutrino oscillations including the vacuum
mixing and the MSW matter effects. The comprehensive discussions can be found in Ref. \cite{Cirelli:2005gh,Blennow:2007tw,Barger:2007xf}.

The final neutrino flux arrived at the Earth is
\begin{equation}
\frac{dN_\nu}{dE_\nu} \simeq \frac{C_\odot}{2}\frac{1}{4\pi R_{SE}^2} \sum_i Br_i \left( \frac{dN_\nu}{dE_\nu} \right)_i
\end{equation}
where $i$ runs over all the DM annihilation channels contributing to neutrino signals with branching fractions $Br_i$,
$\left( \frac{dN_\nu}{dE_\nu} \right)_i$ is the neutrino energy spectrum after propagation for the i-th channel, and $R_{SE}$ is the
Sun-Earth distance. In fact, the high energy solar neutrino detections search for the combinations of $\sigma^{\chi p}\cdot Br_i$.

\begin{figure}[!htbp]
\begin{center}
\includegraphics[width=0.45\textwidth]{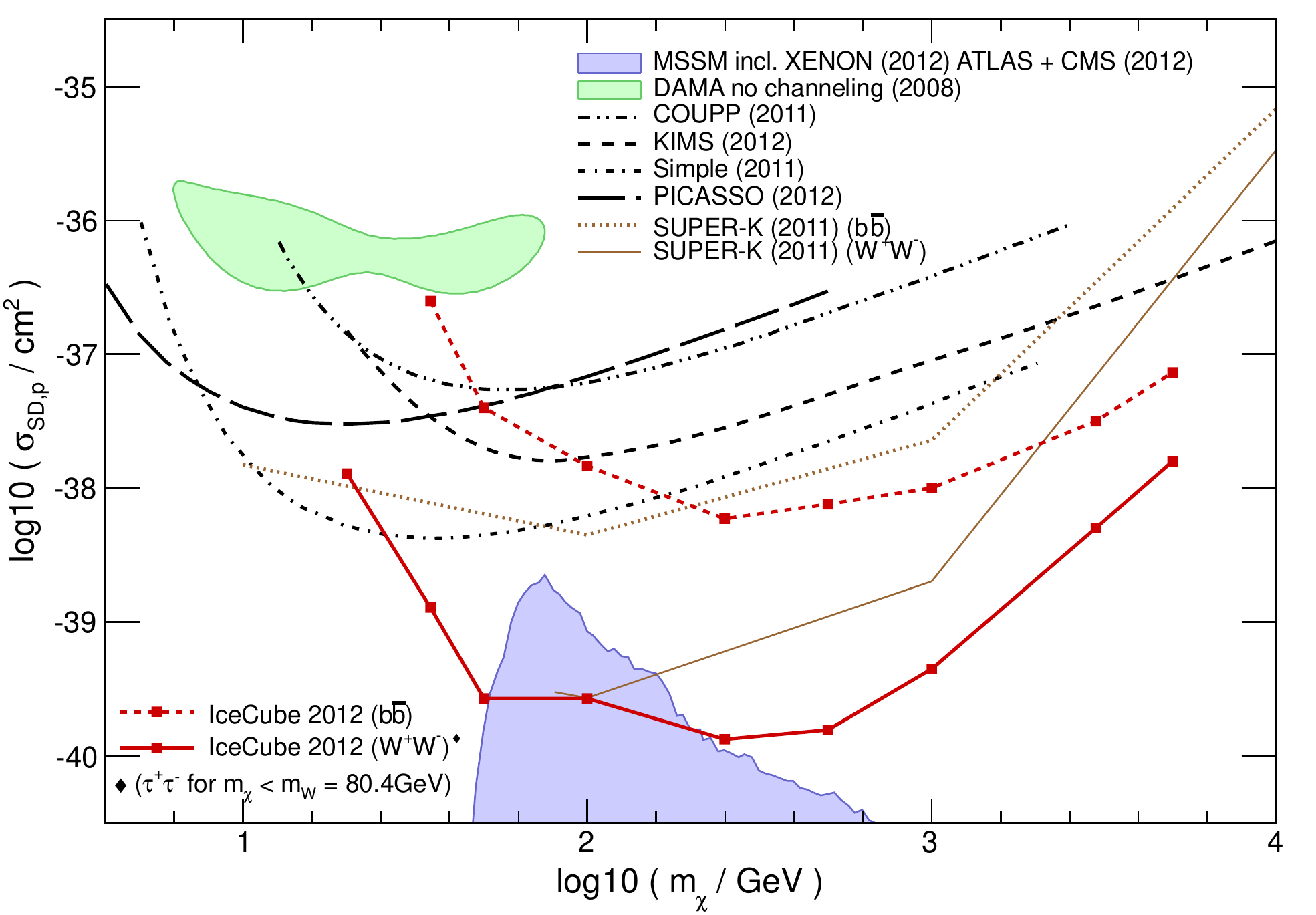}
\caption{Constraints on SD DM-nucleon scattering cross section for $W^+W^-$ and $b\bar{b}$ annihilation channels by IceCube 79 strings. For comparison, other results from  Super-K \cite{Tanaka:2011uf}, DAMA \cite{Bernabei:2008yi,Savage:2008er}, COUPP \cite{Behnke:2010xt}, PICASSO \cite{Archambault:2012pm}, KIMS \cite{Lee.:2007qn}, SIMPLE \cite{Felizardo:2011uw}, are also shown, together with the preferred regions in MSSM \cite{Silverwood:2012tp}. Figure from \cite{Aartsen:2012ef}.
\label{icecos}}
\end{center}
\end{figure}

Recently, IceCube reported the results of the high energy solar
neutrinos with the 79-string configuration and 317 days running
\cite{Aartsen:2012ef}. Since no events from the DM are confirmed,
upper-limits are set on the SD and SI DM-proton scattering cross
sections for DM masses in the range of $20-5000$GeV. In the analysis,
two typical initial neutrino spectra from $b\bar{b}$(``soft'')
and $W^+W^-$(``hard'') channels with the branching fractions
Br$=$1 are adopted. For the SI cross section, the limits given
by IceCube are weaker than those from CDMS \cite{Ahmed:2010wy,Ahmed:2011gh} and XENON100 \cite{Aprile:2011hi}. The most stringent limits for DM masses of $O(100)$ GeV and $W^+W^-$ channel from IceCube have reached $10^{-43}$cm$^2$. For the SD cross section, the most stringent limits are given by IceCube for DM masses above 35GeV and $W^+W^-$ channel. The strict limits for lower DM masses are set by superheated liquid experiments, such as PICASSO \cite{Archambault:2012pm} and SIMPLE \cite{Felizardo:2011uw}. It is also worth noting that in a particular DM model, such as MSSM, the SD constraints set by IceCube can also exclude some parameter regions allowed by current CDMS and XENON100 results.

\subsection{Cosmic neutrinos from DM}

The DM annihilations or decays in the Galactic halo, Galactic Center(GC) \cite{Yuksel:2007ac,PalomaresRuiz:2007ry,Covi:2009xn,Erkoca:2010qx}, subhalos \cite{Yin:2008mv,Liu:2008ci}, dwarf satellite galaxies \cite{Sandick:2009bi} and galaxy clusters \cite{Yuan:2010gn,Dasgupta:2012bd,Murase:2012rd} can produce high energy neutrinos. There are almost no astrophysical high energy neutrino sources in these regions which can mimic the DM signals. The flux of neutrinos observed at the earth can be given by (see e.g. \cite{Bergstrom:1997fj})
\begin{eqnarray}
\left(\frac{dN_\nu}{dE_\nu}\right)^A &=& \frac{1}{4\pi} \frac{\langle \sigma v \rangle}{2 m_\chi^2} \times \left( \frac{dN_\nu}{dE_\nu} \right)^A_i \times J^A(\Delta \Omega) \label{neua} \; , \\
\left(\frac{dN_\nu}{dE_\nu}\right)^D &=& \frac{1}{4\pi} \frac{1}{\tau_\chi m_\chi} \times \left( \frac{dN_\nu}{dE_\nu} \right)^D_i \times J^D(\Delta \Omega)
\label{neud}
\end{eqnarray}
where the superscripts $A$ and $D$ denote annihilating and decaying DM respectively, $\tau_\chi$ is the lifetime of decaying DM, $\left( \frac{dN_\nu}{dE_\nu} \right)_i$ is the initial neutrino energy spectrum. The J-factors $J^A$ and $J^D$ are the line-of-sight (l.o.s) integrals of the DM density $\rho$ toward a direction of observation $\psi$ integrated over a solid angle $\Delta \Omega$, which can be written as
\begin{eqnarray}
J^A(\Delta \Omega) &=& \int d \Omega \int_{l.o.s}dl \;\; \rho^2[l(\psi)]  \; , \\
J^D(\Delta \Omega) &=& \int d \Omega \int_{l.o.s}dl \;\; \rho [l(\psi)] \ .
\end{eqnarray}

The GC is the best object to search the neutrino signals due to the high density of DM. For IceCube located at the south pole, the neutrinos from the GC located in the southern sky are down-going. Although the down-going atmospheric muon background is very large, the IceCube collaboration has developed some techniques to reduce such background efficiently \cite{Abbasi:2012ws}. Since the cosmic muons enter the detector from outside, only the muon events which are produced inside the detector are selected. Due to the large volume of IceCube,
the upper digital optical modules on each string and strings
in the outer layer can be used to veto the atmospheric muons.
Especially, the central strings of DeepCore with higher module
density will use the surrounding IceCube detector as an active veto,
and have strong capability to detect down-going neutrinos
\cite{Collaboration:2011ym}.

\begin{figure}[!htbp]
\begin{center}
\includegraphics[width=0.40\textwidth]{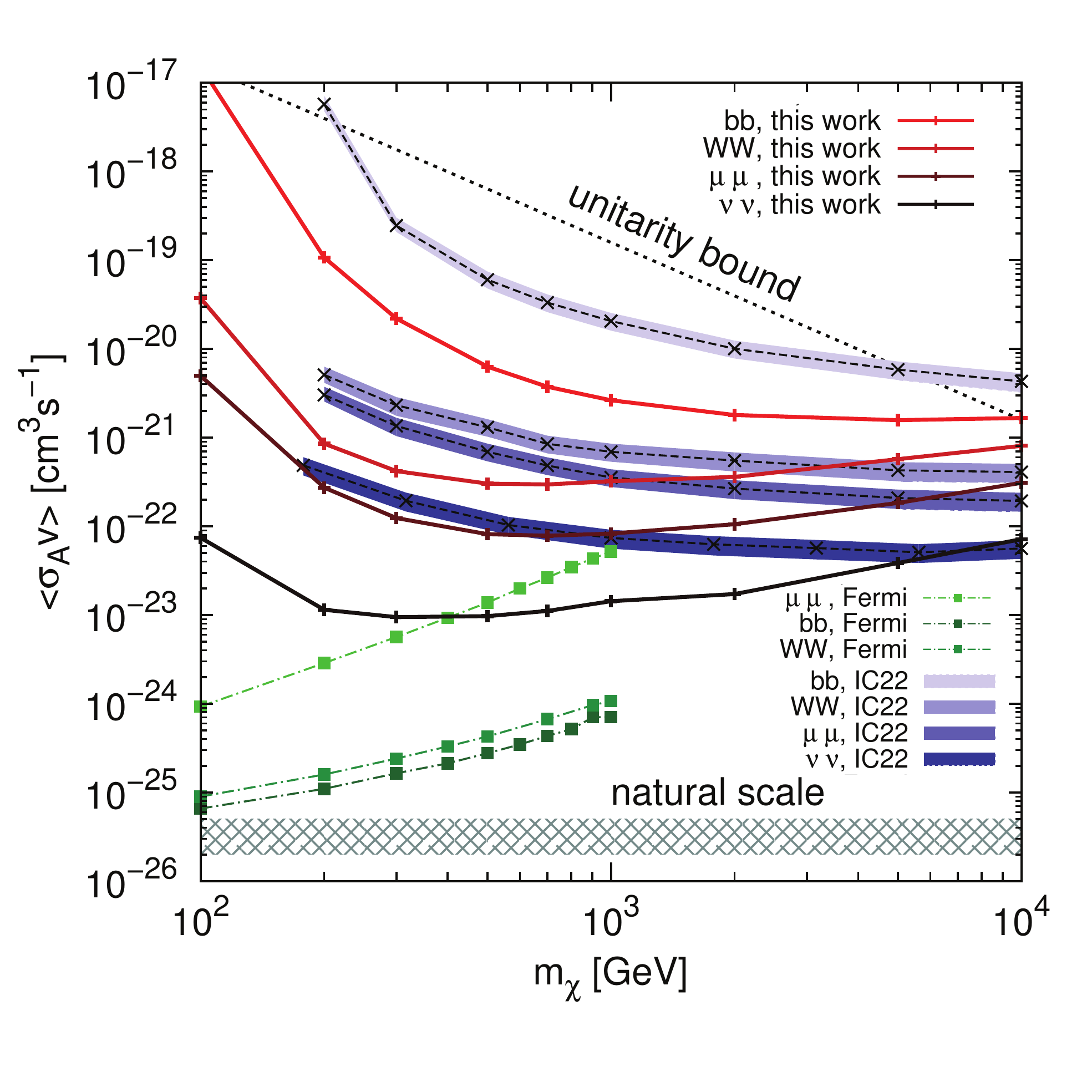}
\caption{Constraints on DM annihilation cross section for four channels by IceCube-40. For comparison, the results from the IceCube-22 outer Galactic halo analysis \cite{Abbasi:2011eq} and the Fermi-LAT dwarf galaxies observation \cite{Ackermann:2011wa} are also shown. Figure from \cite{Aartsen:2012ef}.
\label{icecos1}}
\end{center}
\end{figure}

IceCube collaboration has reported the constraints on the neutrino signals from DM annihilations in the GC based on the performance of 40 strings during 367 days \cite{Abbasi:2012ws}. No data from DeepCore strings are used in this analysis. The main constraints are shown in Fig. \ref{icecos1}. Since the neutrinos with higher energy are more easily reconstructed (due to large cross section and muon range) and suffer from smaller atmospheric neutrino background, the constraints for heavy DM are more stringent. If the dominant DM annihilation channel is $\nu \bar{\nu}$ ($W^+W^-$), the upper limit on the DM annihilation cross section for $m_\chi \sim 1$TeV is $\sim 10^{-23} $cm$^3$s$^{-1}$($\sim 10^{-22}$cm$^3$s$^{-1}$). In the future, the IceCube with 79 strings and DeepCore will significantly improve these results due to larger volume and lower threshold.

Since the dwarf satellite galaxies and galaxy clusters are far away from the earth, the constraints for DM annihilation in these regions are weaker than the GC. For instance, for $m_\chi \sim 1$TeV and $W^+ W^-$ channel,
the upper-limit given by dwarf satellite galaxy searches on
the DM annihilation cross section is about
$\sim 10^{-21}$cm$^3$s$^{-1}$ \cite{Icecosmicn}.
Note that the DM substructure models of e galaxy clusters have large uncertainties.
The constraints on DM annihilation cross sections can be improved by
some substructure models and parameter configurations \cite{Yuan:2010gn}.

\section{Summary}

The existence of DM has been firmly established for a long time
by widely astronomical observations. To detect DM particle and study
its properties is a fundamental problem in cosmology and particle physics.
Extreme efforts have been paid for DM detection and great progresses
have been achieved in recent years.

The direct detection experiments have been widely developed all
over the world. The sensitivity was improved rapidly in last years.
However, most experiments give null results and very strong constraints
on the interaction strength between DM and nucleon have been given.
DAMA, CoGeNT and CRESST have observed some anomalous events.
Interpreting these events as DM signal will lead to inconsistency with other
null results.

As the successful running of LHC, constraints on the nature of DM from
collider data are derived. It is shown that the collider search is
complementary to the direct detection. Especially for small DM mass the
LHC gives very strong constraints while the direct detection sensitivity
becomes worse as DM mass decreases.

The most important progress comes from the indirect detection.
PAMELA, ATIC and Fermi all observed positron and electron excesses in
cosmic rays. It is widely accepted that these data mean new sources
contributing to primary positrons and electrons. Both astrophysical origins
and DM origins are extensively studied in the literature. However,
if the anomalies are interpreted by DM annihilation the DM property is
highly non-trivial. Firstly, DM couples with leptons dominantly and the coupling
with quarks should be suppressed. Secondly, the annihilation rate should
be boosted with a very large BF at $O(10^3)$. Several mechanisms are proposed
to give the large BF. However, later careful studies show those
proposals can not work efficiently. The decay scenario is still working
quite well. There are also
many discussions trying to discriminate the astrophysical
and DM scenarios.

The Fermi-LAT has made great success in $\gamma$-ray detection.
However, all the observations are consistent with CR expectation and
thus give strong constraints on the DM annihilation rate.
Observations from dwarf galaxies, galaxy clusters, Galactic center and
Galactic halo all lead to strong constraints on the DM annihilation rate.
An interesting progress recently is that the line spectrum $\gamma$-ray
emission from the Galactic center is observed in Fermi data.
If such observation is finally confirmed it is certainly the first signal
from DM particles, as line spectrum is thought the smoking gun of
DM annihilation.

The detection of neutrinos is usually difficult. However, as the
running of IceCube constraints on DM from neutrino observation are
given. In some cases they can be stronger than the direct detection.
Especially the sensitivity of direct detection for the SD interaction
is weak. In this case the detection of neutrinos by DM annihilation from
the sun gives complementary constraints to direct detection.

In the near future we expect the sensitivity of DM detection
will be improved greatly.
In direct detection the upcoming experiments will improve the present
sensitivity by two orders of magnitude. In collider search LHC will upgrade
its center of mass energy to $13\sim 14$ TeV and improve the probe range
of DM mass. The AMS02 is accumulating data right now and will greatly
improve the CR spectrum measurement. DAMPE is expected to measure electron
spectrum up to 10 TeV precisely. We expect the AMS02 and DAMPE will solve
the anomaly in cosmic rays finally.

\begin{acknowledgments}
This work is supported by the Natural Science Foundation of China under
the grant NOs. 11075169, 11135009,11105155, 11105157 and 11175251.
\end{acknowledgments}

\bibliography{refs}
\bibliographystyle{apsrev}

\end{document}